\documentclass[preprint,3p,12pt]{elsarticle}
\usepackage{amsmath}
\usepackage{amssymb}
\usepackage{graphicx}
\usepackage{subfigure}
\usepackage{float}
\usepackage{threeparttable}
\usepackage{multirow}
\usepackage{dcolumn}
\usepackage{booktabs}
\usepackage{color}
\renewcommand{\vec}[1]{\boldsymbol{#1}}
\begin{document}

\begin{frontmatter}

\title{An Implicit Unified Gas-kinetic Scheme for Unsteady Flow in All Knudsen Regimes}	
	
\author[ad1]{Yajun Zhu}
\ead{zhuyajun@mail.nwpu.edu.cn}
\author[ad1]{Chengwen Zhong}
\ead{zhongcw@nwpu.edu.cn}
\author[ad2]{Kun Xu\corref{cor1}}
\ead{makxu@ust.hk}
\address[ad1]{National Key Laboratory of Science and Technology on Aerodynamic Design and Research, Northwestern Polytechnical University, Xi'an, Shaanxi 710072, China}
\address[ad2]{Department of Mathematics, Hong Kong University of Science and Technology, Hong Kong, China}
\cortext[cor1]{Corresponding author}

\begin{abstract}
The unified gas-kinetic scheme (UGKS) is a direct modeling method for multiple scale transports.
The modeling scale of the scheme is the mesh size and time step, and the ratios of mesh size over particle mean free path or the time step over the particle collision time determine the local evolution regime of flow physics.
For a multiscale flow problem, such as the hypersonic flow around a flying vehicle in near space,
the UGKS is able to capture the highly compressed Navier-Stokes solution in one region and fully expanded free molecular flow in another region, with the significant variations of the ratio of the time step over the local particle collision time around the vehicle.
For an explicit UGKS, the time step in the whole computational domain is determined by the CFL condition.
For steady state computation, in order to increase the efficiency of the scheme, the implicit and multigrid techniques have been implemented in UGKS \cite{zhu2016implicit,zhu2017multigrid}, where the efficiency can be improved by two orders of magnitude in comparison with the explicit scheme.
However, for unsteady flow computation, due to the CFL condition the time step in the explicit UGKS is limited by the smallest cell size in the computational domain.
As a result, for a largely stretched non-uniform mesh the global time step becomes very small and
the ratio of the time step over the local particle collision time may get a very small value.
Under such a circumstance, even though the UGKS is a multiscale method,
the physics in explicit UGKS may be constrained by the kinetic scale physics only, such as the overall time step being less than the
local particle collision time.
Therefore, for unsteady flow the multiscale UGKS may become a single scale discrete velocity method (DVM)
where the free transport is used for the flux evaluation. In order to keep the advantages of UGKS and
improve the efficiency for unsteady flow computation,
the restriction from global CFL condition on the time step has to be removed.
In this paper, we will develop an implicit UGKS (IUGKS) for unsteady flow by alternatively solving the macroscopic
and microscopic governing equations within a step iteratively.
With a pre-defined uniform large evolution time step, the local CFL number varies greatly in different region,
such as on the order $1$ in the large numerical cell size region, and $100$ in the small cell size region.
In order to preserve the coherent flow evolution and multiscale nature, the numerical flux across a cell interface is still evaluated by the explicit UGKS with the local CFL condition.
Therefore, the multiscale property of the UGKS modeling is preserved over non-uniform meshes.
With improved temporal discretization, the current IUGKS can automatically go back to the explicit UGKS and obtain identical solutions when the time step of the implicit scheme gets to that of an explicit scheme.
Many numerical examples are included to validate the scheme for both continuum and rarefied flows with a large variation of artificially generated mesh size.
The IUGKS has a second order accuracy and presents reasonably good results for unsteady flow computation and its efficiency has been
improved  by dozen of times in comparison with the explicit UGKS.
\end{abstract}


\end{frontmatter}
\section{Introduction}\label{sec:introduction}
The unified gas-kinetic scheme (UGKS) is a direct modeling method on the scale of mesh size and time step \cite{xu2010unified,huang2012unified,xu2015book}, which is able to capture multiscale flow physics in all Knudsen number regimes.
In UGKS, a time evolution solution based on the kinetic relaxation model is used at a cell interface for the flux evaluation,
where the ratio of the time step over the particle collision time determines the local flow regime.
Different from the gas-kinetic scheme (GKS) \cite{xu2001gas} for the update of macroscopic flow varibales only for the continuum flow,
the UGKS updates both the macroscopic flow variables and the gas distribution function.
Therefore, highly non-equilibrium effects can be properly captured by UGKS through the update of the gas distribution function, which
has much more degrees of freedom to describe the non-equilibrium property.
Due to its unified treatment for the continuum and rarefied, as well as high and low speed flows,
the UGKS shows excellent performance in many flow problems, such as the studies of hypersonic flows \cite{chen2012adaptive,liu2014diatomic}, rarefied flows \cite{jiang2016thesis}, and micro-flows \cite{huang2013unified,liu2015micro}.
In addition, the methodology of direct modeling \cite{xu2015book,xu2017paradigm} can be extended to other multi-scale transport processes,
such as plasma \cite{liu2017plasma}, radiative transfer \cite{sun2015radiative}, and phonon heat transfer \cite{guo2016discrete}.
The UGKS provides a useful tool to solve complex multi-scale problems, and has great potentials in the engineering applications, such as micro-electro-mechanical system and spacecraft design.

One of distinguishable features of the UGKS is to use the scale-dependent flow evolution in the numerical flux evaluation, where the ratio of time step over particle mean collision time ($\Delta t /\tau$) identifies the continuum ($\Delta t/\tau >> 1 $)and rarefied ($\Delta t/\tau \sim 1 $) flow regimes and this ratio can be changed by several orders of magnitude continuously over the computational domain for a multiscale flow problem.
For a single scale method, such as the direct simulation of Monte Carlo (DSMC) method and the discrete velocity method (DVM), it always
requires to resolve the flow physics on the scale of ($\Delta t /\tau \leq 1 $), where the mesh size and time step are less than the mean free path and mean collision time.
These constraints on the cell size and time step are not enforced by UGKS because the time evolution solution with coupled collision and transport
is used in the interface flux evaluation instead of free transport (upwind) only for the single scale method \cite{chen-ap}.
The truly multi-scale nature makes the UGKS very efficient in the near continuum flow regime.
However, for unsteady flow simulations the explicit UGKS uses a uniform time step which is determined by the Courant-Friedrichs-Lewy (CFL) condition and the smallest cell size may determine the time step.
For hypersonic flow with a wide range of cell size, such as the cell size around the corner or the tip of a wing of a flying vehicle, the overall time step can be very small.
So, for the explicit scheme the global time step $\Delta t$ may go to the order of collision time $\tau$.
As a result, the advantage of UGKS may be demolished when using a time step everywhere with the order $\Delta t /\tau \sim 1$,
even though in most of the computational domain a much large cell size is used and the flow is in the near equilibrium one,
such as those far away from the vehicle surface.
The above permissible uniform CFL time step for stability is much smaller than the preferred time step for a reasonable accuracy for unsteady
flow computation \cite{jameson1991time}. Therefore, it is necessary to develop an implicit UGKS (IUGKS) that can break the barrier
from the CFL stability condition from these very small cell sizes,  and preserve the multi-scale nature of the UGKS
for simulating different flow regime physics in different locations at the same time.

For unsteady flow simulation in continuum regime, implicit treatments to solve the Navier-Stokes (NS) equations have already been fully developed  using dual-time stepping methods \cite{jameson1991time,pulliam1993time,tomaro1997implicit}.
Recently, the time-accurate implicit GKS has also been constructed \cite{tan2017time,li2017implementation}
to accelerate computational efficiency for high-speed unsteady and turbulent flows.
In rarefied flow study, there are several implicit schemes that have been constructed based on the DVM \cite{yang1995rarefied,mieussens2000discrete,peng2016implicit} and the UGKS \cite{mao2015implicit,zhu2016implicit} for steady state solutions only.
In the study of radiative transfer \cite{sun2017implicit}, implicit techniques have been used in UGKS to remove the restriction of speed of light on the numerical time step for unsteady radiative transport.
In order to fully utilize the multi-scale nature of UGKS for unsteady flow simulations, the construction of an implicit UGKS is necessary,
especially for mutiscale flow problem with both continuum and rarefied flow regimes and a highly stretched mesh distributions.

The current paper presents an implicit UGKS (IUGKS) for unsteady flows in all Knudsen number regimes.
Starting from discretized conservation laws for conservative flow variables and the gas distribution function,
we employ the semi-discrete formulation to construct the implicit scheme.
To maintain the multiscale transport property, the time derivatives in the semi-discrete UGKS for the flow variables are
evaluated from the averaged transport according to the local time step.
The temporal derivatives involve the cell size effect on the flow evolution with respect to the local particle mean free path.
Based on the implicit discretization for the macroscopic flow variables and the gas distribution function,
a matrix connecting the variables in the whole computational domain can be easily solved through the lower-upper symmetric Gauss-Seidel (LU-SGS)  method \cite{jameson1987lower,yoon1988lower}, point-relaxation (PR) method \cite{yuan2002comparison},
and the generalized minimal residual (GMRES) algorithm \cite{saad1986gmres}.
During the inner iterative process within each time-marching step,
the implicit governing equations for macroscopic flow variables and microscopic distribution are solved alternatively
with high efficiency for both continuum and rarefied flows.

This paper is organized as follows. Section \ref{sec:numericalMethod}  gives a brief introduction to the explicit UGKS
and presents details in the construction of IUGKS. In Section \ref{sec:validation},
numerical validations and discussions are carried out. A conclusion is given in the last section.

\section{Implicit unified gas-kinetic scheme}\label{sec:numericalMethod}
\subsection{Unified gas-kinetic scheme}\label{sec:EUGKS}
The UGKS is constructed based on a direct modeling of the flow evolution on the scale of time step and cell size.
For a discrete finite volume cell $i$ and discretized time step $\Delta t = t^{n+1} - t^{n}$,
the discretized governing equation for the gas distribution function is
\begin{equation}\label{eq:directMicro}
f_{i}^{n+1} = f_{i}^{n} - \dfrac{1}{V_i} \sum_{j \in N(i)} {\int_{0}^{\Delta t} u_{n} f_{ij}(t) S_{ij} dt} + \int_{0}^{\Delta t}{\Omega (f,f)dt},
\end{equation}
where $N(i)$ denotes the set of neighboring cells of $i$, and cell $j$ is one of the neighbors. The interface connecting the cells $i$ and $j$ is referred to as $ij$. $V_i$ is volume of cell $i$, and $S_{ij}$ is area of the cell interface $ij$. Here $u_n$ is the normal component of the particles' microscopic velocity $\vec{u}$, which is perpendicular to the cell interface. $f_{ij}(t)$ is a time-dependent distribution function at the cell interface $ij$, which describes the time evolution of flow physics around the cell interface. The collision term $\Omega (f,f)$ takes into account the particles' interaction in one time step.
Eq.~(\ref{eq:directMicro}) is an exact physical law  for the discrete cell $i$,
which simply describes the evolution of the gas distribution function due to particles' transport and collision on the discretized level.
The corresponding macroscopic governing equations are
\begin{equation}\label{eq:directMacro}
\vec{W}_i^{n+1} = \vec{W}_i^{n} - \dfrac{1}{V_i} \sum_{j \in N(i)} { \int {\int_{0}^{\Delta t} u_{n} f_{ij}(t) \vec{\psi}(\vec{u}) S_{ij} dt} d\vec{u} },
\end{equation}
where $\vec{\psi}(\vec{u}) = (1,u,v,\frac{1}{2} (u^2+v^2+\xi^2))^T$ for two dimensional cases. $u$ and $v$ are the particles' velocities along $x$- and $y$-direction, respectively. $\xi$ denotes the particles' random motion for the internal degrees of freedom. $\vec{W}$ is a vector of the conservative flow variables, i.e., $(\rho, \rho U, \rho V, \rho E)^T$ representing the density of mass, momentum and energy.
Eq.~(\ref{eq:directMacro}) describes the conservation of macroscopic variables, which is exact as well.

The above two governing equations are the physical laws on a discretized level.
The accuracy of the solutions from Eq.(\ref{eq:directMicro}) and (\ref{eq:directMacro}) depends on the construction of the collision term and the flux transport, which is similar to
the constitutive relationship provided in the NS equations. The special ingredient of the UGKS is to construct the collision term and the
flux function according to the scale of cell size and time step.
According to the scale of time step $\Delta t$ relative to the particle collision time $\tau$,
the full Boltzmann collision term can be replaced by the BGK-type kinetic model in the Eq.(\ref{eq:directMicro}),
especially for the cases of $\Delta t \ge 4 \tau$ \cite{liu2016boltzmann}. In the following, the kinetic equation with the BGK-type model will be used only.
The multi-scale nature of the UGKS comes from the construction of the time-dependent distribution function at the cell interface, $f_{ij}(t)$, which is the evolution solution of the BGK model.
The analytic solution gives
\begin{equation}\label{eq:analyticSolution}
\begin{aligned}
f_{ij}(t)
=&  (1- e^{-t/\tau}) g_0 + e^{-t/\tau} f^{l,r}\\
&+ \tau ( e^{-t/\tau} -1 + \frac{t}{\tau} e^{-t/\tau})  ( \vec{u} \cdot \bar{\vec{a}}^{l,r}) g_0 + \tau (\dfrac{t}{\tau}+e^{-t/\tau}-1) \bar{A} g_0 \\
 & - \tau (\frac{t}{\tau} e^{-t/\tau})  (\vec{u} \cdot \vec{\sigma}^{l,r}),
\end{aligned}
\end{equation}
from the reconstructed initial gas distribution function
\begin{equation}\label{eq:initialDistributionfunction}
f_0(x) = \left\{
\begin{array}{c}
f^l + \vec{\sigma}^l \cdot \vec{x}, \quad x_n \le 0,\\
f^r + \vec{\sigma}^r \cdot \vec{x}, \quad x_n > 0,
\end{array}\right.
\end{equation}
and an equilibrium state distributed in time and space
\begin{equation}\label{eq:equilibriumState}
g(x,t) = g_0\left( 1 + \bar{\vec{a}}^{l,r} \cdot \vec{x} + \bar{A} t\right),
\end{equation}
where $\vec{\sigma}^{l}$ and $\vec{\sigma}^{r}$ are the spatial derivatives of the initial distribution functions for the left and right hand sides of the interface. $\bar{\vec{a}}^{l,r}$ and $\bar{A}$ are the spatial and temporal derivatives of the equilibrium state, respectively. $g_0$ is the Maxwellian distribution, corresponding to the local conservative flow variables.

\begin{figure}[H]
\centering
\includegraphics[width=0.6\textwidth]{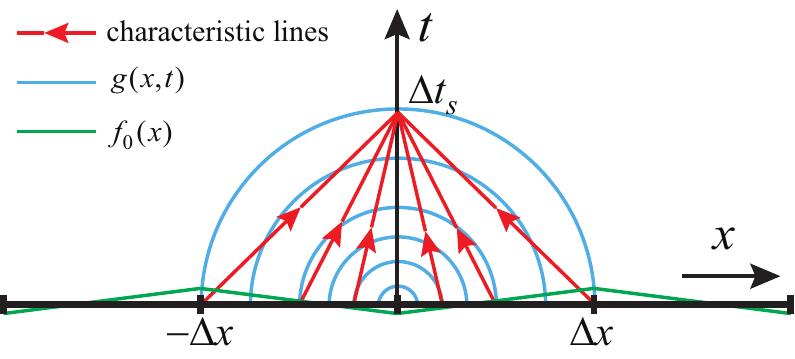}
\caption{\label{fig:flowPhysics}Physical picture illustrating the time evolution of flow physics around a cell interface in GKS and UGKS.}
\end{figure}

From the first two terms in Eq.~(\ref{eq:analyticSolution}), we can observe a transition from the initial non-equilibrium
distribution function to the equilibrium one with the increment of time scale,
which denotes a flow transition from kinetic scale to hydrodynamic one.
It describes an accumulating effect from particle collision in the evolution process.
The physical picture is illustrated in Fig.~\ref{fig:flowPhysics}, where a time evolution of flow physics around the interface within a cell-size-determined time step $\Delta t_s$ is presented.
The automatic scale variation in the physical solution depends on the ratio of $t/\tau$, which is the key ingredient to construct a multi-scale scheme.  Details about the implementation of the explicit UGKS can be found in \cite{xu2010unified} and \cite{xu2015book}.

With a local constant $\tau$ and the assumed local distribution functions in Eq.~(\ref{eq:initialDistributionfunction}) and Eq.~(\ref{eq:equilibriumState}),
the analytic solution (\ref{eq:analyticSolution}) is valid in a local small region on a cell size scale $\Delta x$
and a small time step $\Delta t_s$ from the CFL condition, as illustrated in Fig.~\ref{fig:flowPhysics}.
Eq.~(\ref{eq:directMicro}), Eq.~(\ref{eq:directMacro}) and Eq.~(\ref{eq:analyticSolution}) are fundamental physical laws,
which presents a local flow evolution in an explicit way. Due to its equations-based formulation instead of particle-based,  Eq.~(\ref{eq:directMicro}) and Eq.~(\ref{eq:directMacro}) can be transformed into semi-discrete forms
for the construction of an implicit scheme. The accelerating techniques in CFD can be easily employed here in the UGKS.

\subsection{Time-accurate implicit UGKS}\label{sec:IUGKS}
From Eq.~(\ref{eq:directMicro}) and Eq.~(\ref{eq:directMacro}), the semi-discrete macroscopic governing equations for conservative variables can be written as
\begin{equation}\label{eq:semiMacro}
\dfrac{\partial \vec{W}_i}{\partial t}+\dfrac{1}{V_i}\sum_{j \in N(i)}{\hat{\vec{F}}_{ij}S_{ij}} = \vec{0},
\end{equation}
where $\hat{\vec{F}}_{ij}$ is the macroscopic flux across the cell interface.
The governing equation for the gas distribution function is
\begin{equation}\label{eq:semiMicro}
\dfrac{\partial f_{i}}{\partial t} + \dfrac{1}{V_i} \sum_{j \in N(i)}{u_{n} \hat{f}_{ij} S_{ij}} = \dfrac{g_i-f_i}{\tau_i},
\end{equation}
where $u_n \hat{f}_{ij}$ is the interface flux of the gas distribution function.
The above semi-discrete formulations describe the instantaneous variation  of the flow field.
It seems to present the flow evolution in a time scale $t \to 0$.
Although Eq.~(\ref{eq:semiMacro}) and Eq.~(\ref{eq:semiMicro}) are still consistent with the UGKS formulation,
it reduces the time-dependent UGKS fluxes into a particle free transport, which makes UGKS as
a single scale method within the mean free path and mean collision time.
Therefore, in order to incorporate the multi-scale property of UGKS into the above formulation,
it is better to reconstruct the interface fluxes in Eq.~(\ref{eq:semiMacro}) and Eq.~(\ref{eq:semiMicro})
as a time-averaged quantity instead of instantaneous one.
Specifically, the above fluxes are averaged over a cell-size-determined time step, which is given in Fig.~\ref{fig:flowPhysics}.
Therefore,  a cell-size-determined time scale related flow physics evolution is implicitly included in the semi-discrete governing equations.
In order to distinguish this cell-size-determined time step from the numerical marching time step $\Delta t$ for an implicit scheme,
this time step is given by $\Delta t_s$ which determines the local flow physics.
$\Delta t_s$ is the same as the explicit local time step, which is computed by the local CFL number less than one.

For a large numerical marching time step $\Delta t = t^{n+1} - t^n$, we have the discrete macroscopic governing equations
\begin{equation}\label{eq:discretizeMacro}
\dfrac{\vec{W}_i^{n+1}-\vec{W}_i^n}{\Delta t} +
\dfrac{1}{V_i}\sum_{j \in N(i)}{\left(\epsilon \hat{\vec{F}}_{ij}^{n+1}+(1-\epsilon)\hat{\vec{F}}_{ij}^{n}\right)S_{ij}} = \vec{0},
\end{equation}
and the discrete microscopic equations
\begin{equation}\label{eq:discretizeMicro}
\dfrac{f_{i}^{n+1}-f_{i}^n}{\Delta t} +
\dfrac{1}{V_i}\sum_{j \in N(i)}{u_{n}\left(\epsilon \hat{f}_{ij}^{n+1}+(1-\epsilon) \hat{f}_{ij}^n\right)S_{ij}}
= \epsilon \dfrac{g_i^{n+1}-f_i^{n+1}}{\tau_i^{n+1}}+(1-\epsilon)\dfrac{g_i^n-f_i^n}{\tau_i^n},
\end{equation}
where $\hat{\vec{F}}_{ij}^{n+1}$ and $\hat{\vec{F}}_{ij}^{n}$ denote the macroscopic fluxes computed by taking moments of the microscopic flux of a gas distribution function.
Here $u_n \hat{f}_{ij}^{n+1}$ and $u_n \hat{f}_{ij}^n$ are the time-averaged fluxes over
$\Delta t_s$ from the explicit UGKS formulation.
For $\epsilon = 0.5$ in Eq.~(\ref{eq:discretizeMacro}) and Eq.~(\ref{eq:discretizeMicro}), it corresponds to the Crank-Nicolson method with second-order accuracy in time, and for $\epsilon=1$ it is a backward Euler scheme with first-order temporal accuracy.

Eq.~(\ref{eq:discretizeMacro}) and Eq.~(\ref{eq:discretizeMicro}) are fully coupled and $g_i^{n+1}$ in Eq.~(\ref{eq:discretizeMicro})
has one-to-one corresponding with $\vec{W}_{i}^{n+1}$ in Eq.~(\ref{eq:discretizeMacro}).
It is difficult to simultaneously solve these two implicit systems.
Instead we can solve Eq.~(\ref{eq:discretizeMacro}) and Eq.~(\ref{eq:discretizeMicro}) in an alternative way.
Specifically, we first get an approximate equilibrium state $\tilde{g}^{n+1}$ from Eq.~(\ref{eq:discretizeMacro}) from an assumed approximate solution $f^{n+1}$, then obtain a more precise solution of $f^{n+1}$ from Eq.~(\ref{eq:discretizeMicro}) using the previously obtained $\tilde{g}^{n+1}$. After several alternations, both $f^{n+1}$ and $g^{n+1}$ can be solved.
However it isn't straightforward to get solutions from Eq.~(\ref{eq:discretizeMacro}) and Eq.~(\ref{eq:discretizeMicro})
due to the implicit fluxes on the left hand side of these equations. These equations will result in a large nonlinear system, especially for a second-order accurate numerical fluxes.
Therefore, implicit schemes usually adopt a delta-form
with implicit parts on left hand side and explicit parts on right hand side of an equation \cite{pulliam1993time}.
Based on the delta-form equations, the implicit parts on the left hand side can be much simplified using a first-order flux function,
and it becomes much easier to take iterative processes  to solve the implicit system \cite{tomaro1997implicit}.
Details are given in the following.

Given an intermediate approximate solution $\vec{W}^{(s)}$ and $f^{(s)}$, the delta-form governing equations of Eq.~(\ref{eq:discretizeMacro}) and Eq.~(\ref{eq:discretizeMicro}) can be rewritten as
\begin{equation}\label{eq:implicitMacro}
\dfrac{1}{\Delta t}{\Delta \vec{W}_i^{(s)}}+\dfrac{\epsilon}{V_i}\sum_{j \in N(i)}{{\Delta \hat{\vec{F}}_{ij}^{(s)}}S_{ij}} = \vec{R}_i^{(s)},
\end{equation}
and
\begin{equation}\label{eq:implicitMicro}
\left(\dfrac{\epsilon}{\tau_i^{(s+1)}}+\dfrac{1}{\Delta t}\right) {\Delta f_i^{(s)}} +\dfrac{\epsilon}{V_i}\sum_{j \in N(i)}{u_{n}S_{ij} {\Delta \hat{f}_{ij}^{(s)}}} =r_i^{(s)},
\end{equation}
where
\begin{equation}\label{eq:residualMacro}
\vec{R}_i^{(s)}=\dfrac{\vec{W}_i^{n}-\vec{W}_i^{(s)}}{\Delta t}
-\dfrac{1}{V_i}\sum_{j \in N(i)}{\left((1-\epsilon)\hat{\vec{F}}_{ij}^{n}+\epsilon \hat{\vec{F}}_{ij}^{(s)}\right)S_{ij}},
\end{equation}
and
\begin{equation}\label{eq:residualMicro}
\begin{aligned}
r_i^{(s)}=\dfrac{f_i^n-f_i^{(s)}}{\Delta t}
&-\dfrac{1}{V_i}\sum_{j \in N(i)}{u_{n}\left(\epsilon \hat{f}_{ij}^{(s)}+(1-\epsilon)\hat{f}_{ij}^n\right)S_{ij}}\\
&+\epsilon \dfrac{g_i^{(s+1)}-f_i^{(s)}}{\tau_i^{(s+1)}} +(1-\epsilon)\dfrac{g_i^n-f_i^n}{\tau_i^n},
\end{aligned}
\end{equation}
where $\Delta \vec{W}_i^{(s)} =  \vec{W}_i^{n+1}-\vec{W}_i^{(s)}$ and $\Delta f_i^{(s)} = f_i^{n+1}-f_i^{(s)}$.
The fluxes in the residuals $r_i^{(s)}$ and $\vec{R}_i^{(s)}$ are fully evaluated by the UGKS with an averaging over a time step $ \Delta t_s$, i.e.,
\begin{equation}\label{eq:macroUGKSflux}
\hat{\vec{F}}_{ij} = \dfrac{1}{\Delta t_s}{ \int {\int_{0}^{\Delta t_s} u_{n} f_{ij}(t) \vec{\psi}(\vec{u}) dt} d\vec{u} },
\end{equation}
and
\begin{equation}\label{eq:microUGKSflux}
u_n \hat{f}_{ij}= \dfrac{1}{\Delta t_s}{\int_{0}^{\Delta t_s} u_{n} f_{ij}(t) dt},
\end{equation}
where $f_{ij}(t)$ is given in Eq.~(\ref{eq:analyticSolution}).

For the terms on the left hand side of Eq.~(\ref{eq:implicitMacro}), the Euler equations-based flux
\begin{equation}
\Delta \hat{\vec{F}}_{ij}^{(s)} = \dfrac{1}{2} \left[ \Delta \vec{T}_{i}^{(s)} + \Delta \vec{T}_{j}^{(s)} +
\Gamma_{ij} \left(\Delta \vec{W}_{i}^{(s)}-\Delta \vec{W}_{j}^{(s)}\right) \right]
\end{equation}
is adopted to simplify the implicit macroscopic fluxes, where $\vec{T}$ is the Euler fluxes and $\Gamma_{ij}$ is the spectral radius of the Euler flux Jacobian with an additional stable factor related to kinematic viscosity \cite{chen2000fast}, i.e.,
\begin{equation}
\Gamma_{ij} = |\vec{U}_{ij} \cdot \vec{n}_{ij}|+ a_s + \dfrac{2 \mu}{\rho |\vec{n}_{ij} \cdot (\vec{x}_j - \vec{x}_i)|},
\end{equation}
where $\vec{n}_{ij}$ is the normal vector of cell interface $ij$, $\vec{U}_{ij}$ is macroscopic velocity and $a_s$ is the speed of sound. For simplifying the numerical fluxes on the left hand side of Eq.~(\ref{eq:implicitMicro}), we use an upwind approach
\begin{equation}
\Delta \hat{f}_{ij}^{(s)} = \dfrac{1}{2}\left[1+{\rm sign}(u_{n})\right] \Delta f_{i}^{(s)} + \dfrac{1}{2}\left[1-{\rm sign}(u_{n})\right]  \Delta f_{j}^{(s)}.
\end{equation}
Substituting Eq.~(\ref{eq:macroUGKSflux}) and Eq.~(\ref{eq:microUGKSflux}) into Eq.~(\ref{eq:implicitMacro}) and Eq.~(\ref{eq:implicitMicro}), the implicit governing equations for conservative flow variables become
\begin{equation}\label{eq:implicitMacroGov}
C_i {\Delta \vec{W}_i^{(s)}}
+ \dfrac{1}{2 V_i} \sum_{j \in N(i)} {\epsilon_{ij} S_{ij} \left[T(\vec{W}_j^{(s)}+\Delta \vec{W}_j^{(s)}) - T(\vec{W}_j^{(s)}) - \Gamma_{ij} {\Delta \vec{W}_j^{(s)}}\right]}=\vec{R}_i^{(s)},
\end{equation}
where
\begin{equation}
C_i = \dfrac{1}{\Delta t} + \dfrac{1}{2 V_i}\sum{\epsilon_{ij} \Gamma_{ij} S_{ij}}.
\end{equation}
The implicit governing equation for gas distribution function is
\begin{equation}\label{eq:implicitMicroGov}
D_i {\Delta f_i^{(s)}}
+\sum_{j \in N(i)}{ D_j {\Delta {f}_{j}^{(s)}}} =r_i^{(s)},
\end{equation}
where
\begin{equation}
\begin{aligned}
D_i =& \dfrac{\epsilon}{\tau_i^{(s+1)}}+\dfrac{1}{\Delta t}
+\dfrac{1}{2 V_i}\sum_{j \in N(i)}{u_{n} \epsilon_{ij} S_{ij} \left[1+{\rm sign}(u_{n})\right]},\\
D_j =&  \dfrac{1}{2 V_i} u_{n} \epsilon_{ij} S_{ij}  \left[1-{\rm sign}(u_{n})\right].
\end{aligned}
\end{equation}
For the above delta-form implicit equations, the LU-SGS method \cite{jameson1987lower,yoon1988lower}, the point-relaxation (PR) method \cite{yuan2002comparison}, or the GMRES algorithm \cite{saad1986gmres} can be applied to solve the variation of conservative variables and the gas distribution function. In the current paper, we employ the PR(2) scheme described in Ref.~\cite{yuan2002comparison}, for which each inner iterative process is dealt with twice forward and backward sweeps. The details for solving the matrix system will not be further discussed, see \cite{zhu2016implicit,yuan2002comparison}. Therefore, starting with $\vec{W}^{(s=0)} = \vec{W}^n$ and $f^{(s=0)} = f^n$, Eq.~(\ref{eq:implicitMacroGov}) and Eq.~(\ref{eq:implicitMicroGov}) can be alternatively solved within several inner iterations.

In summary, the time evolution of the implicit UGKS for unsteady flow simulations from $t^n$ to $t^{n+1}$ can be described as
\begin{description}
\item[Step 1\label{item:start}] Calculate the numerical fluxes $\hat{\vec{F}}_{ij}^n$ and $u_{n} \hat{f}_{ij}^n$ averaged over $\Delta t_s$ using the gas distribution function $f^n$ and equilibrium state $g^n$ by the explicit UGKS fluxes.
\item[Step 2\label{item:recomputeFlux}] Compute the intermediate fluxes $\vec{F}_{ij}^{(s)}$ and $u_{n} \hat{f}_{ij}^{(s)}$ averaged over $\Delta t_s$ using the intermediate solution $f^{(s)}$ and $\vec{W}^{(s)}$ by Eq.~(\ref{eq:macroUGKSflux}) and Eq.~(\ref{eq:microUGKSflux}). For the first inner step, $f^{(s=0)} = f^n$ and $g^{(s=0)} = g^n$ are adopted.
\item[Step 3] Evaluate the macroscopic residual $\vec{R}_i^{(s)}$ in Eq.~(\ref{eq:residualMacro}) using the numerical macroscopic fluxes obtained in Step 1 and Step 2.
\item[Step 4\label{item:prediction}] Solve Eq.~(\ref{eq:implicitMacroGov}) to obtain the equilibrium state $\tilde{g}_i^{(s+1)}$.
\item[Step 5] Evaluate the microscopic residual $r^{(s)}$ in Eq.~(\ref{eq:residualMicro}) using the equilibrium state $\tilde{g}_i^{(s+1)}$ recently obtained in Step 4 and the micro-fluxes $u_{n} \hat{f}_{ij}^n$ and $u_{n} \hat{f}_{ij}^{(s)}$ evaluated in Step 1 and Step 2.
\item[Step 6] Solve Eq.~(\ref{eq:implicitMicroGov}) to obtain the gas distribution function $f_i^{(s+1)}$, and update the equilibrium state $g_i^{(s+1)}$ by compatibility condition.
\item[Step 7] Check the convergence through $L_2$ norm of the macroscopic residual $\vec{R}^{(s)}$
\begin{itemize}
	\item if convergent state is reached for the current time step, update $f^n$ and $\vec{W}^n$ into $f^{n+1}$ and $\vec{W}^{n+1}$ by $f^{(s+1)}$ and $\vec{W}^{(s+1)}$
	\item if not, go to Step 2 and continue the inner iterations.
\end{itemize}
\end{description}

\subsection{Modification for temporal discretization}\label{sec:modification}
\begin{figure}[H]
	\centering
	\includegraphics[width=0.6\textwidth]{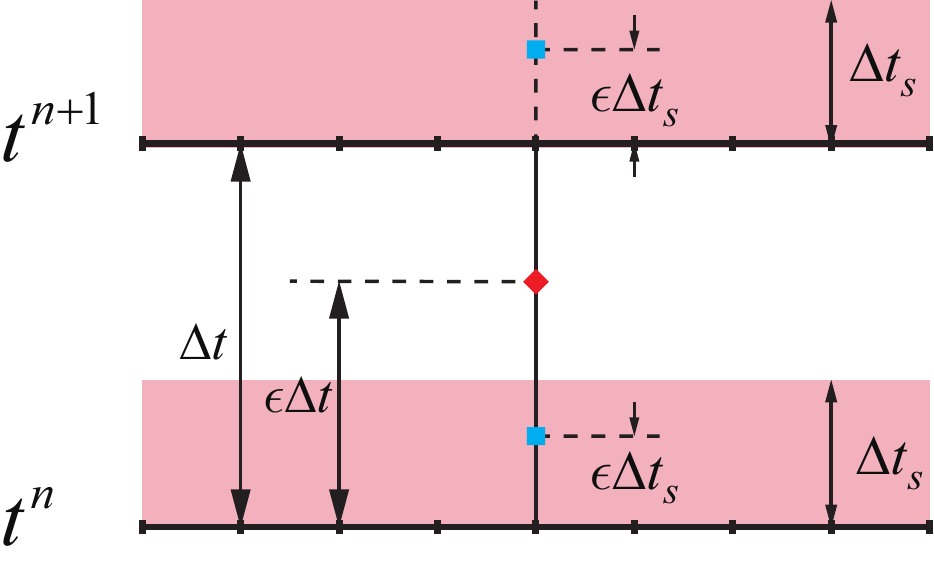}
	\caption{\label{fig:epsilonModified}Structure of temporal discretization.}
\end{figure}

The time-averaged fluxes are evaluated over a finite time step $\Delta t_s$ instead of at the instant of $t^{n+1}$ and $t^n$.
When the evolution time step $\Delta t$ is comparable with $\Delta t_s$, the size of $\Delta t_s$ should be used in the temporal discretization.
As illustrated in Fig.\ref{fig:epsilonModified}, the fluxes at time $t^n+\epsilon \Delta t$ can be approximated by those at time $t^n + \epsilon \Delta t_s$ and $t^{n+1} + \epsilon \Delta t_s$ with weighted coefficients
\begin{equation}
\begin{aligned}
\epsilon^{\prime} &= \dfrac{\epsilon \Delta t - \epsilon \Delta t_s}{\Delta t},\\
1-\epsilon^{\prime} &= \dfrac{(1-\epsilon) \Delta t + \epsilon \Delta t_s}{\Delta t}.
\end{aligned}
\end{equation}
Then Eq.~(\ref{eq:discretizeMacro}) and Eq.~(\ref{eq:discretizeMicro}) become
\begin{equation}\label{eq:modifiedMacro}
\dfrac{\vec{W}_i^{n+1}-\vec{W}_i^n}{\Delta t} +
\dfrac{1}{V_i}\sum_{j \in N(i)}{\left(\epsilon^{\prime} \hat{\vec{F}}_{ij}^{n+1}+(1-\epsilon^{\prime})\hat{\vec{F}}_{ij}^{n}\right)S_{ij}} = \vec{0},
\end{equation}
and
\begin{equation}\label{eq:modifiedMicro}
\dfrac{f_{i}^{n+1}-f_{i}^n}{\Delta t} +
\dfrac{1}{V_i}\sum_{j \in N(i)}{u_{n}\left(\epsilon^{\prime} \hat{f}_{ij}^{n+1}+(1-\epsilon^{\prime}) \hat{f}_{ij}^n\right)S_{ij}}
= \epsilon \dfrac{g_i^{n+1}-f_i^{n+1}}{\tau_i^{n+1}}+(1-\epsilon)\dfrac{g_i^n-f_i^n}{\tau_i^n},
\end{equation}
where the modification is only applied to the flux evaluation terms.
Then the delta-form of governing equations can be written as
\begin{equation}
\dfrac{1}{\Delta t}{\Delta \vec{W}_i^{(s)}}+\dfrac{\epsilon^{\prime}}{V_i}\sum_{j \in N(i)}{{\Delta \hat{\vec{F}}_{ij}^{(s)}}S_{ij}} = \vec{R}_i^{(s)},
\end{equation}
and
\begin{equation}
\left(\dfrac{\epsilon}{\tau_i^{(s+1)}}+\dfrac{1}{\Delta t}\right) {\Delta f_i^{(s)}} +\dfrac{\epsilon^{\prime}}{V_i}\sum_{j \in N(i)}{u_{n}S_{ij} {\Delta \hat{f}_{ij}^{(s)}}} =r_i^{(s)},
\end{equation}
where
\begin{equation}\nonumber
\vec{R}_i^{(s)}=\dfrac{\vec{W}_i^{n}-\vec{W}_i^{(s)}}{\Delta t}
-\dfrac{1}{V_i}\sum_{j \in N(i)}{\left((1-\epsilon^{\prime})\hat{\vec{F}}_{ij}^{n}+\epsilon^{\prime} \hat{\vec{F}}_{ij}^{(s)}\right)S_{ij}},
\end{equation}
and
\begin{equation}\nonumber
\begin{aligned}
r_i^{(s)}=\dfrac{f_i^n-f_i^{(s)}}{\Delta t}
&-\dfrac{1}{V_i}\sum_{j \in N(i)}{u_{n}\left(\epsilon^{\prime} \hat{f}_{ij}^{(s)}+(1-\epsilon^{\prime})\hat{f}_{ij}^n\right)S_{ij}}\\
&+\epsilon \dfrac{g_i^{(s+1)}-f_i^{(s)}}{\tau_i^{(s+1)}} +(1-\epsilon)\dfrac{g_i^n-f_i^n}{\tau_i^n}.
\end{aligned}
\end{equation}

Particularly, when $\Delta t = \Delta t_s$, $\epsilon^{\prime}$ equals zero. Then we have
\begin{equation}\nonumber
\dfrac{\vec{W}_i^{n+1}-\vec{W}_i^{(s)}}{\Delta t} =
\dfrac{\vec{W}_i^{n}-\vec{W}_i^{(s)}}{\Delta t}
-\dfrac{1}{V_i}\sum_{j \in N(i)}{\hat{\vec{F}}_{ij}^{n} S_{ij}}
\end{equation}
and
\begin{equation}\nonumber
\left(\dfrac{\epsilon}{\tau_i^{(s+1)}}+\dfrac{1}{\Delta t}\right) {\Delta f_i^{(s)}} =\dfrac{f_i^n-f_i^{(s)}}{\Delta t}
-\dfrac{1}{V_i}\sum_{j \in N(i)}{u_{n}\hat{f}_{ij}^n S_{ij}}
+\epsilon \dfrac{g_i^{(s+1)}-f_i^{(s)}}{\tau_i^{(s+1)}} +(1-\epsilon)\dfrac{g_i^n-f_i^n}{\tau_i^n}
\end{equation}
which automatically reduce to the explicit UGKS where both macroscopic variables and distribution function are updated within one inner iteration only. For the C-N scheme with $\epsilon = 1/2$, the collision term is approximated by trapezoidal rules.
It should be noted that without the modified coefficient $\epsilon^{\prime}$ the early implicit UGKS can not directly go to the explicit scheme. Even though it can still get identical solutions, it may take several inner iterations to obtain the convergent solutions.

\section{Numerical validation}\label{sec:validation}
\subsection{Sod's shock tube}\label{sec:SodCase}
Sod's shock tube is computed to validate the IUGKS for unsteady flows in one-dimensional case. The initial condition is given by
\begin{equation}\nonumber
\begin{aligned}
\left(\rho_l, \rho_l U_l, p_l\right) &= \left(1.0,0.0,1.0\right),    &\quad x \le 0;\\
\left(\rho_r, \rho_r U_r, p_r\right) &= \left(0.125,0.0,0.1\right),  & \quad x > 0.
\end{aligned}
\end{equation}
The time steps are denoted by CFL numbers,
\begin{equation}\label{eq:cflCondition}
\Delta t = {\rm CFL} \dfrac{\Delta x}{\max{\left|u_k\right|}}.
\end{equation}
For the explicit UGKS, ${\rm CFL}=0.5$ is adopted, which is also used to give the scale-determining time step $\Delta t_s$  for evaluating the time-averaged fluxes in the implicit scheme.

\begin{figure}[H]
\centering
\subfigure[Density]{\includegraphics[width=0.48\textwidth]{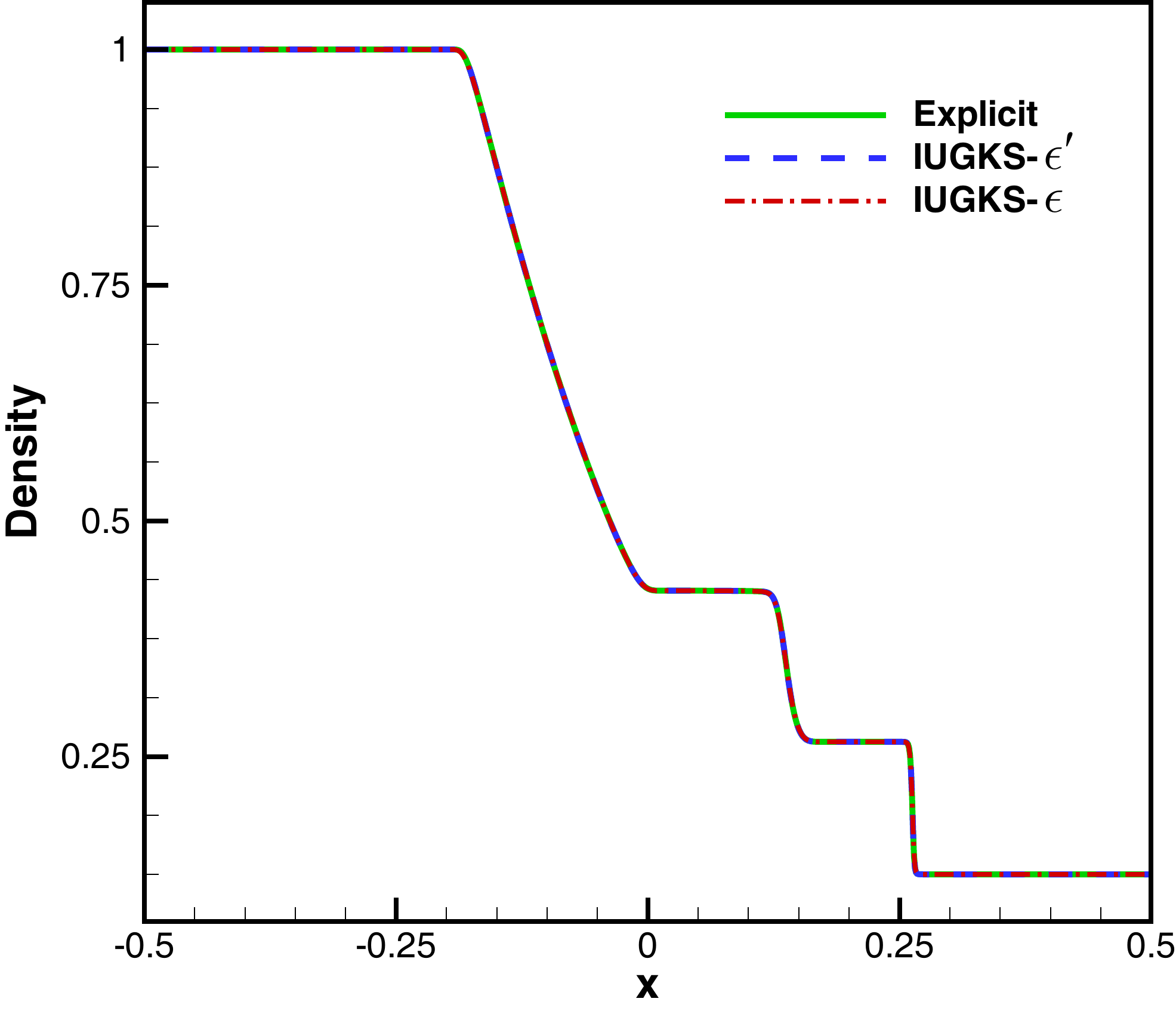}}
\subfigure[Velocity]{\includegraphics[width=0.48\textwidth]{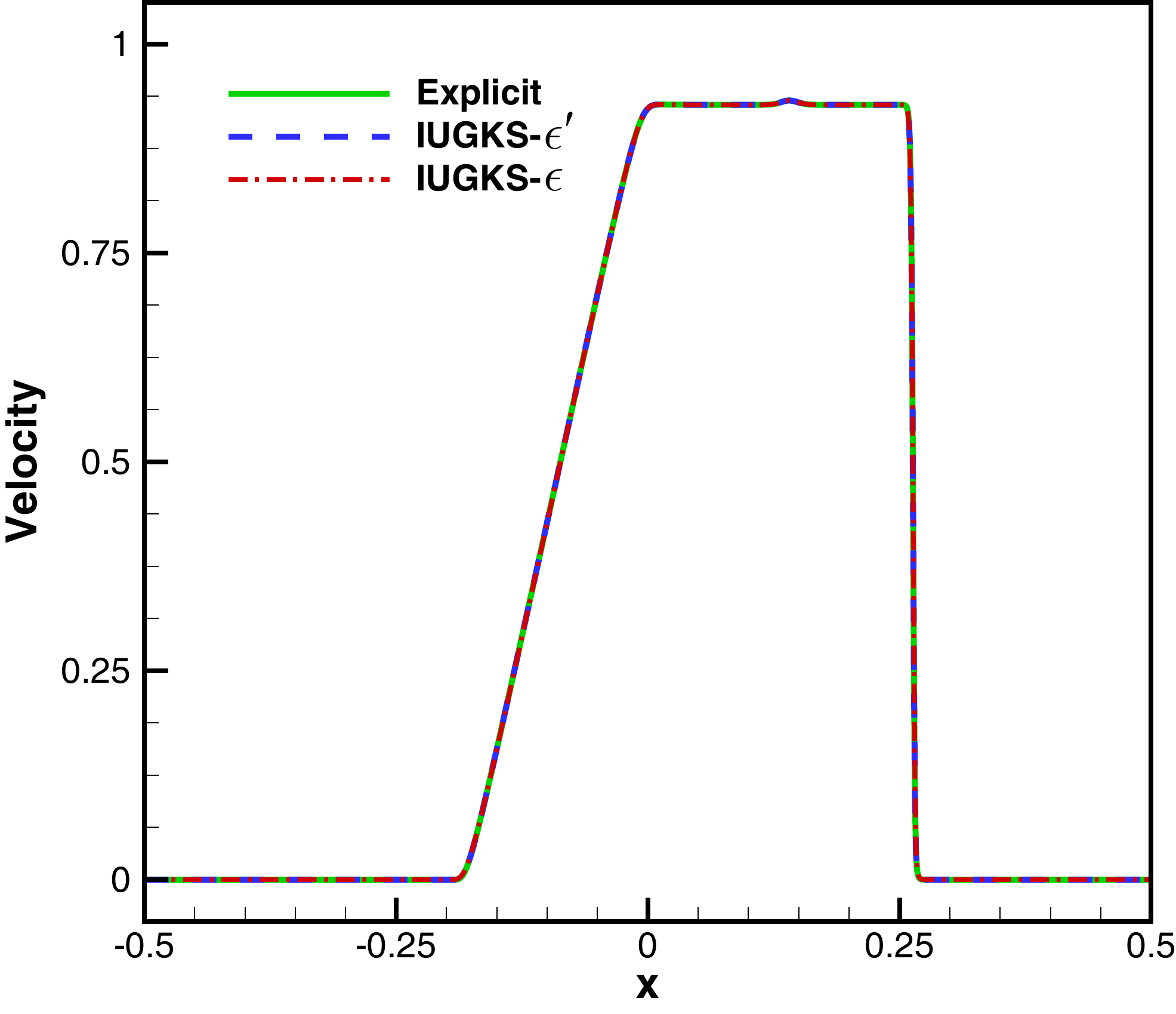}}\\
\subfigure[Temperature]{\includegraphics[width=0.48\textwidth]{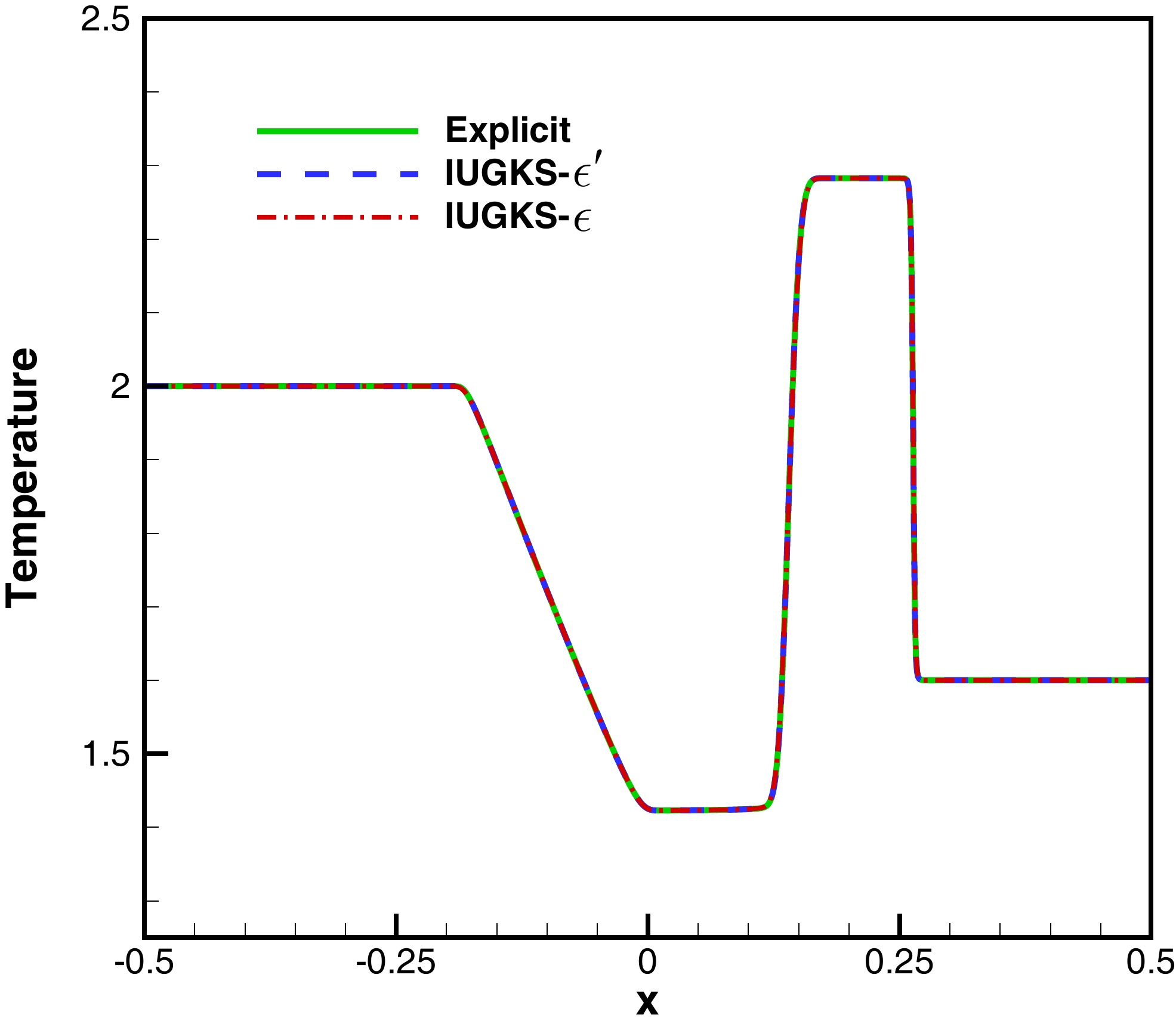}}
\subfigure[CPU time\label{fig:SodContinuumEpsilonCPU}] {\includegraphics[width=0.48\textwidth]{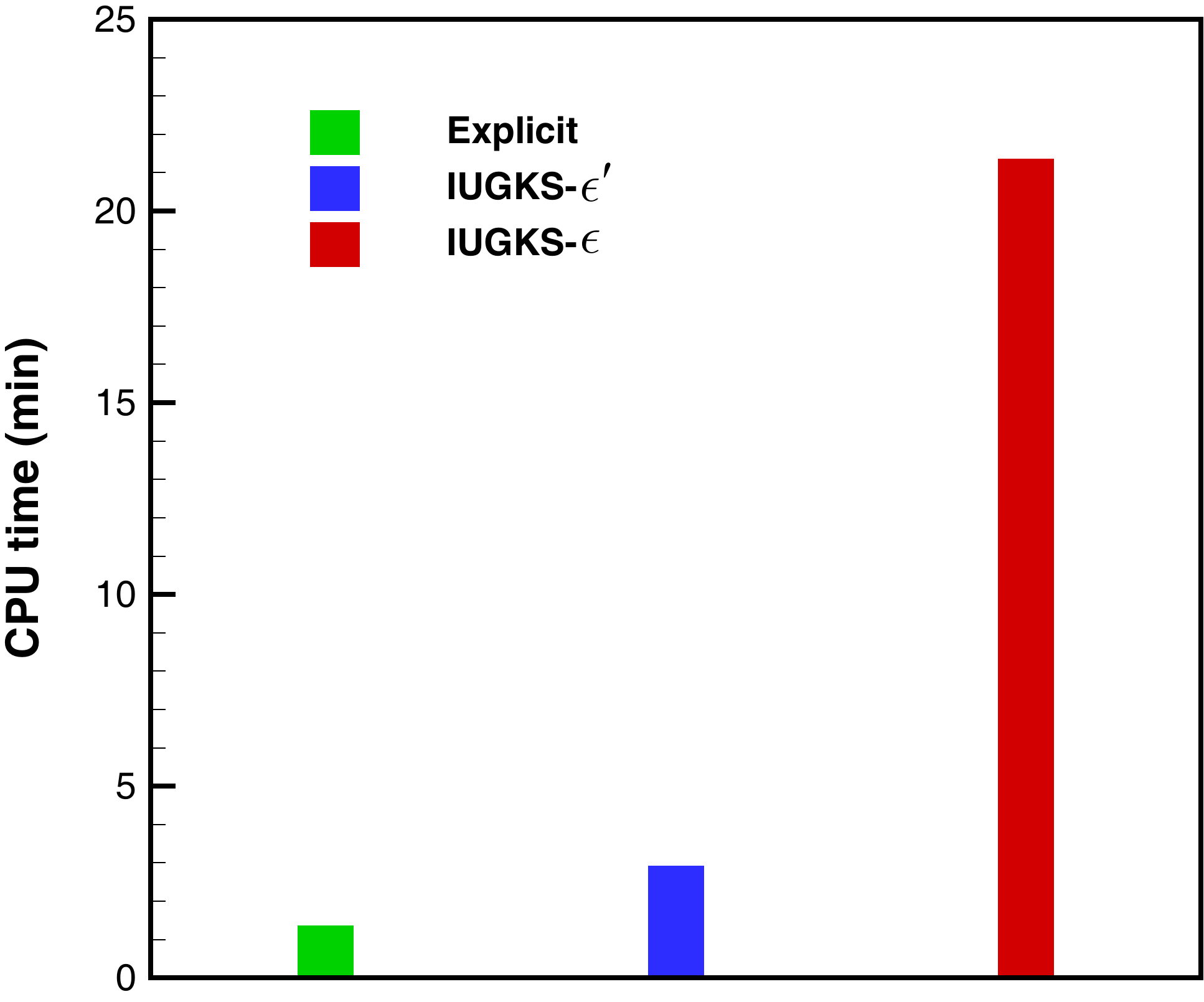}}
\caption{\label{fig:SodContinuumEpsilon}Numerical results and computational cost of the Sod's shock tube problems at $Kn=1.0\times 10^{-4}$ obtained from the explicit UGKS, IUGKS-$\epsilon$ and IUGKS-$\epsilon^{\prime}$ with $\Delta t = \Delta t_s$. $\epsilon=0.5$ is adopted for the implicit schemes and it gives $\epsilon^{\prime} = 0$ for IUGKS-$\epsilon^{\prime}$.}
\end{figure}

First of all, we compute the Sod's shock tube problem at $Kn=10^{-4}$ with a small time step of $\Delta t = \Delta t_s$.
The computational domain is discretized into 1000 uniform cells. Comparison between the implicit schemes with the original $\epsilon$ and the modified $\epsilon^{\prime}$ is carried out when the time step becomes as small as that used in the explicit scheme. For simplification, in the following the IUGKS with $\epsilon$ will be denoted by IUGKS-$\epsilon$, and the other one with $\epsilon^{\prime}$ will be IUGKS-$\epsilon^{\prime}$. In this case, $\epsilon = 0.5$ for the C-N method is chosen to give a second-order accuracy in time, and $\epsilon^{\prime}$ will become $0$ due to $\Delta t = \Delta t_s$.  Distributions of density, velocity and temperature at the time $t=0.15$ are given in Fig.~\ref{fig:SodContinuumEpsilon}. It can be observed that both implicit schemes can give identical solutions to that of  explicit UGKS.
As to the computational cost,  from Fig.\ref{fig:SodContinuumEpsilonCPU} we can find that IUGKS-$\epsilon^{\prime}$ is comparable with the explicit scheme and is much cheaper than IUGKS-$\epsilon$. This is due to the reason given in Section \ref{sec:modification},
where IUGKS-$\epsilon^{\prime}$ with $\Delta t = \Delta t_s$ automatically reduces to the explicit scheme and one inner iteration is needed for each time step. However, for IUGKS-$\epsilon$ several inner iterations are still needed to get a convergent solution in each time step.

\begin{figure}[H]
	\centering
	 \subfigure[Density]{\includegraphics[width=0.48\textwidth]{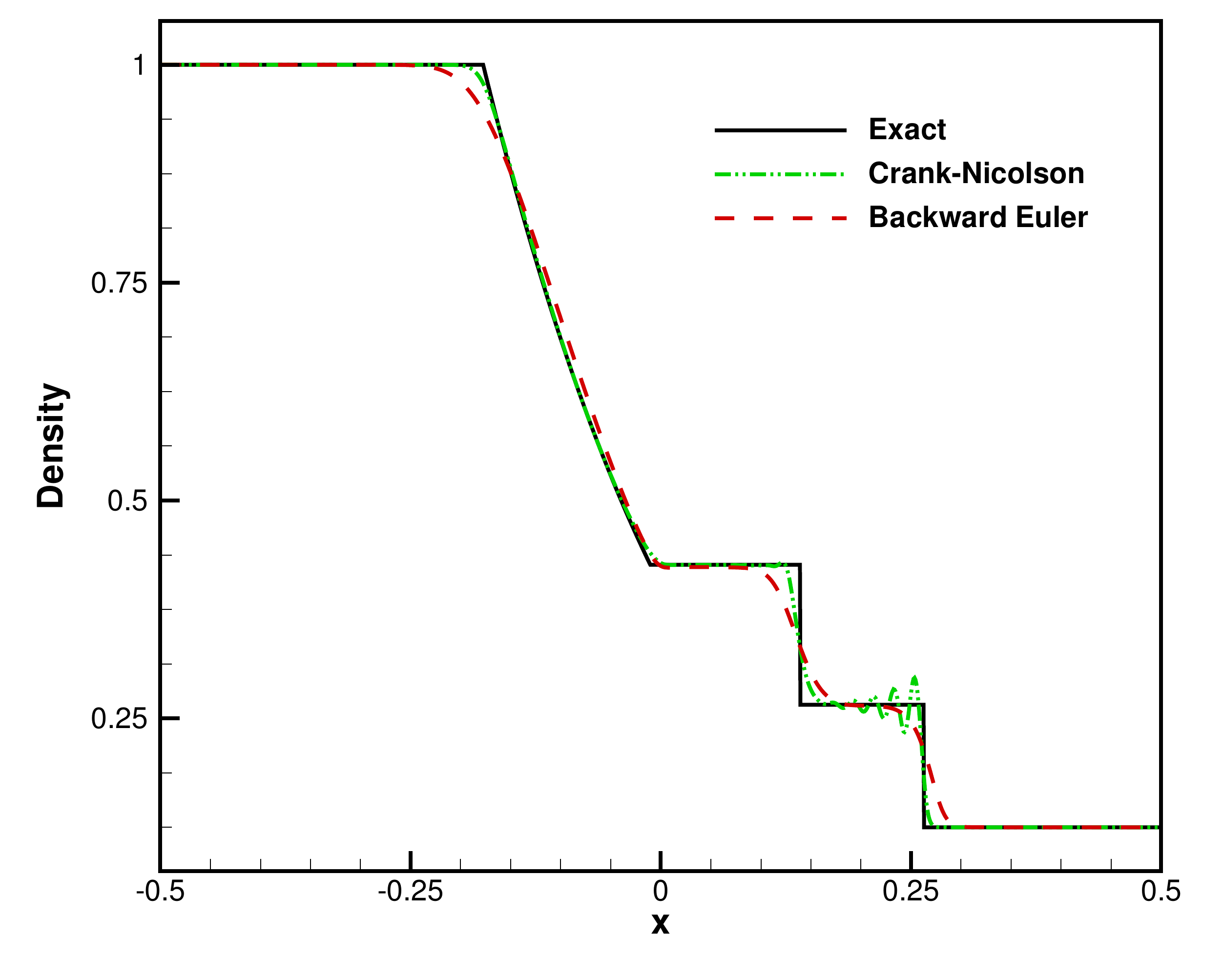}}
	 \subfigure[Velocity]{\includegraphics[width=0.48\textwidth]{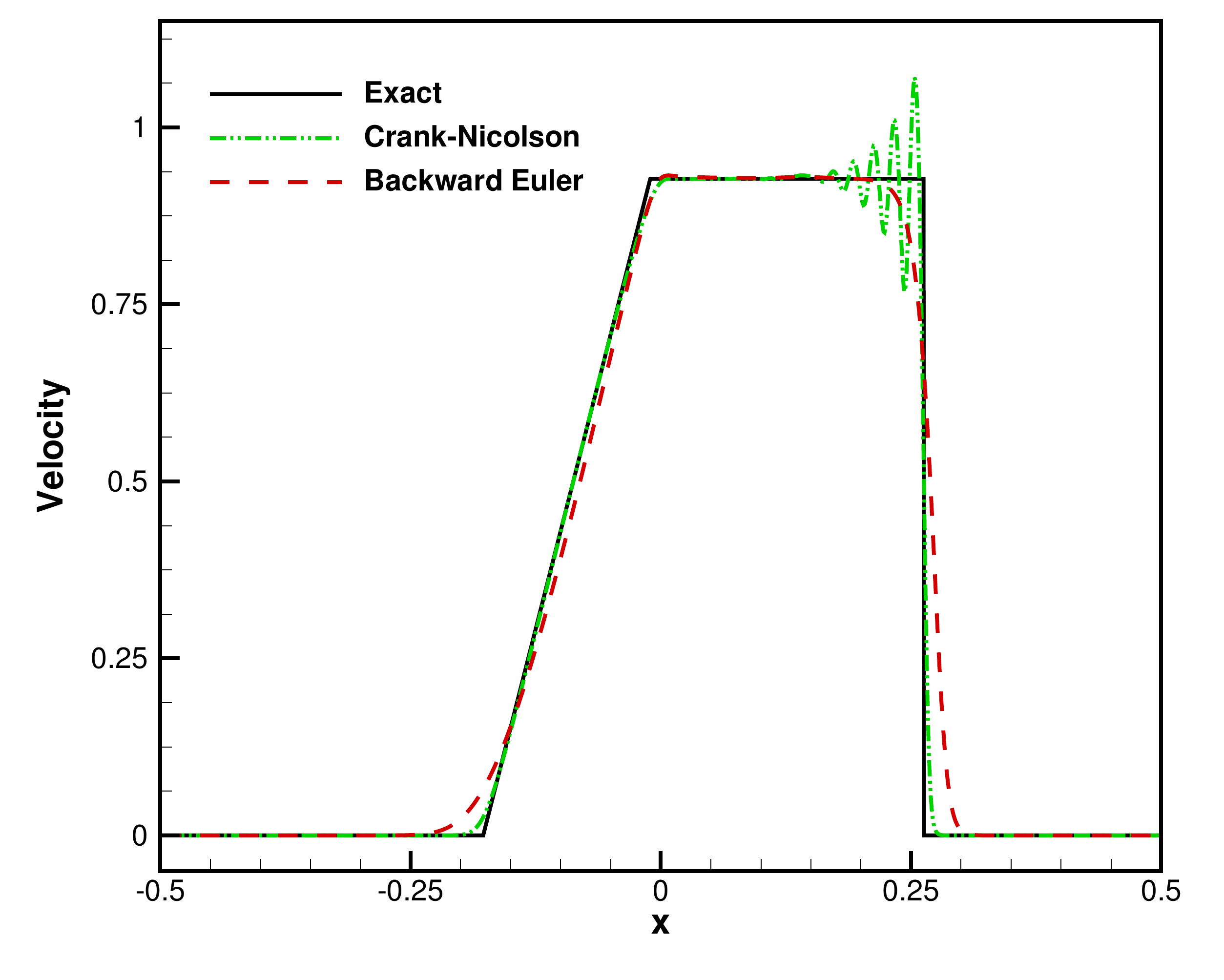}}\\
	 \subfigure[Pressure]{\includegraphics[width=0.48\textwidth]{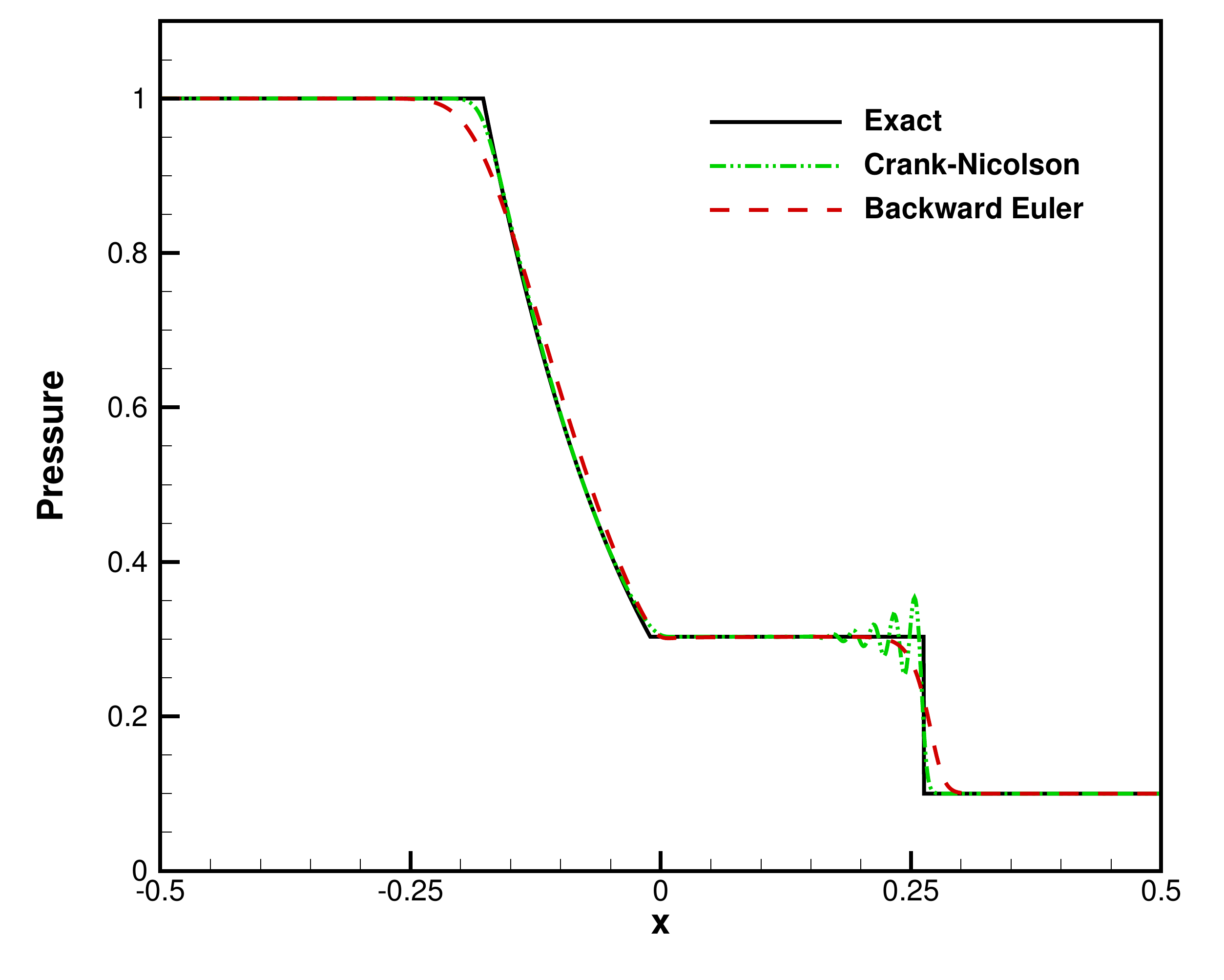}}
	 \subfigure[Temperature]{\includegraphics[width=0.48\textwidth]{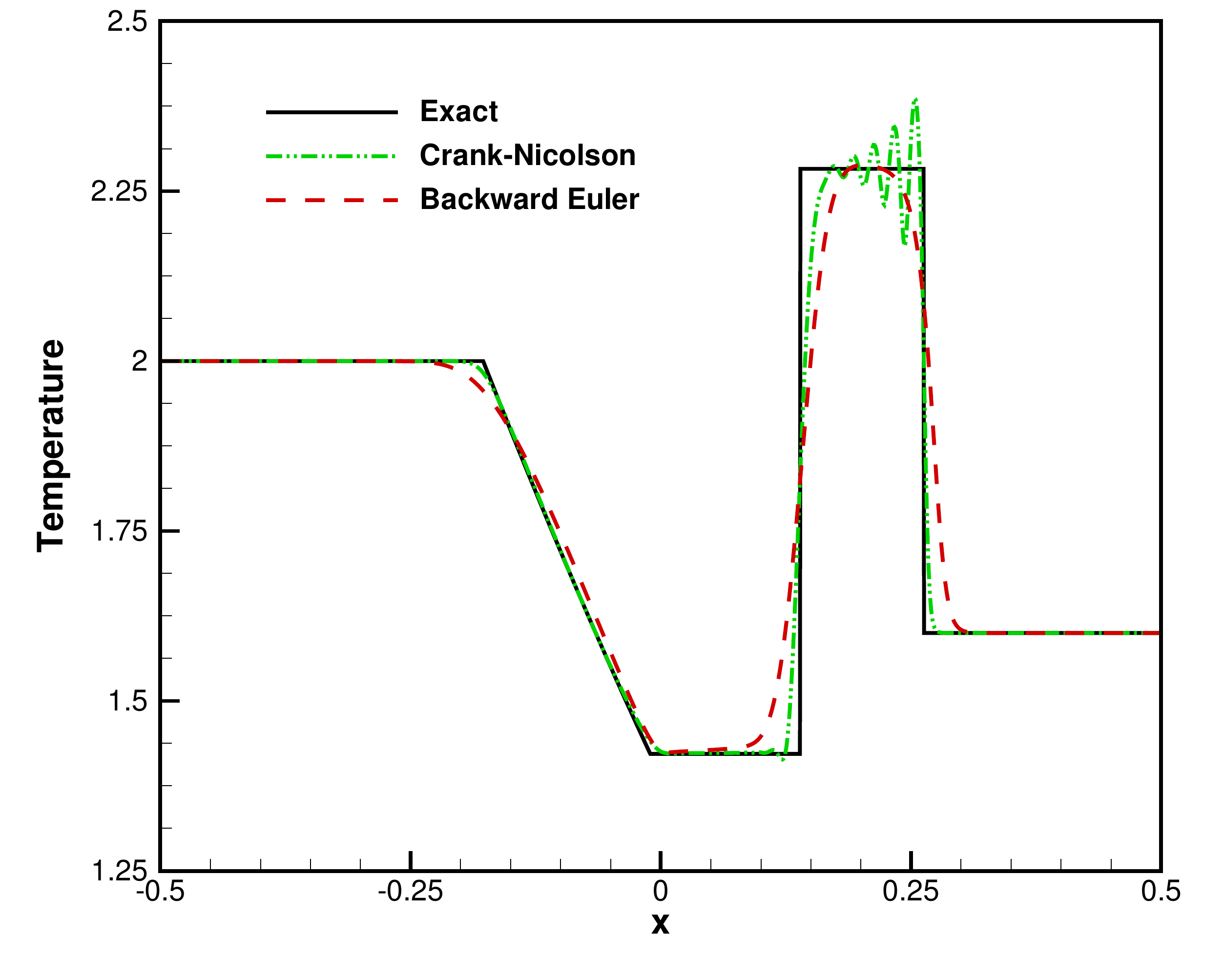}}
	\caption{\label{fig:SodContinuumDispersion}Comparison of two different temporal discretization methods, i.e., the Crank-Nicolson method and the backward Euler scheme for IUGKS-$\epsilon^{\prime}$ in the Sod test case with ${\rm CFL}=50$.}
\end{figure}

Different values of $\epsilon$ correspond to different temporal discretization methods.
Two typical discretization methods for the IUGKS-$\epsilon^{\prime}$ will be presented in  the Sod's test with  ${\rm CFL} = 50$.
Distributions of density, velocity, pressure, and temperature obtained from IUGKS-$\epsilon^{\prime}$ with $\epsilon = 0.5 $ of the C-N scheme,
and $\epsilon=1$ of the backward Euler method are given in Fig.~\ref{fig:SodContinuumDispersion}. As observed in the papers \cite{luo2001accurate,tan2017time}, due to the dispersion characteristics of the C-N scheme, numerical oscillations will appear around the normal shock wave in the large time step evolution. Regardless of numerical oscillations, the C-N scheme can give a sharp discontinuity, while the backward Euler scheme can effectively suppress the numerical oscillations around the smeared normal shock wave.
Therefore, for a balance of capturing discontinuity and accuracy preserving, a selection of proper $\epsilon$ in the interval of $\left[0.5,1\right]$ is necessary.

\begin{figure}[H]
\centering
\subfigure[Density]{\includegraphics[width=0.48\textwidth]{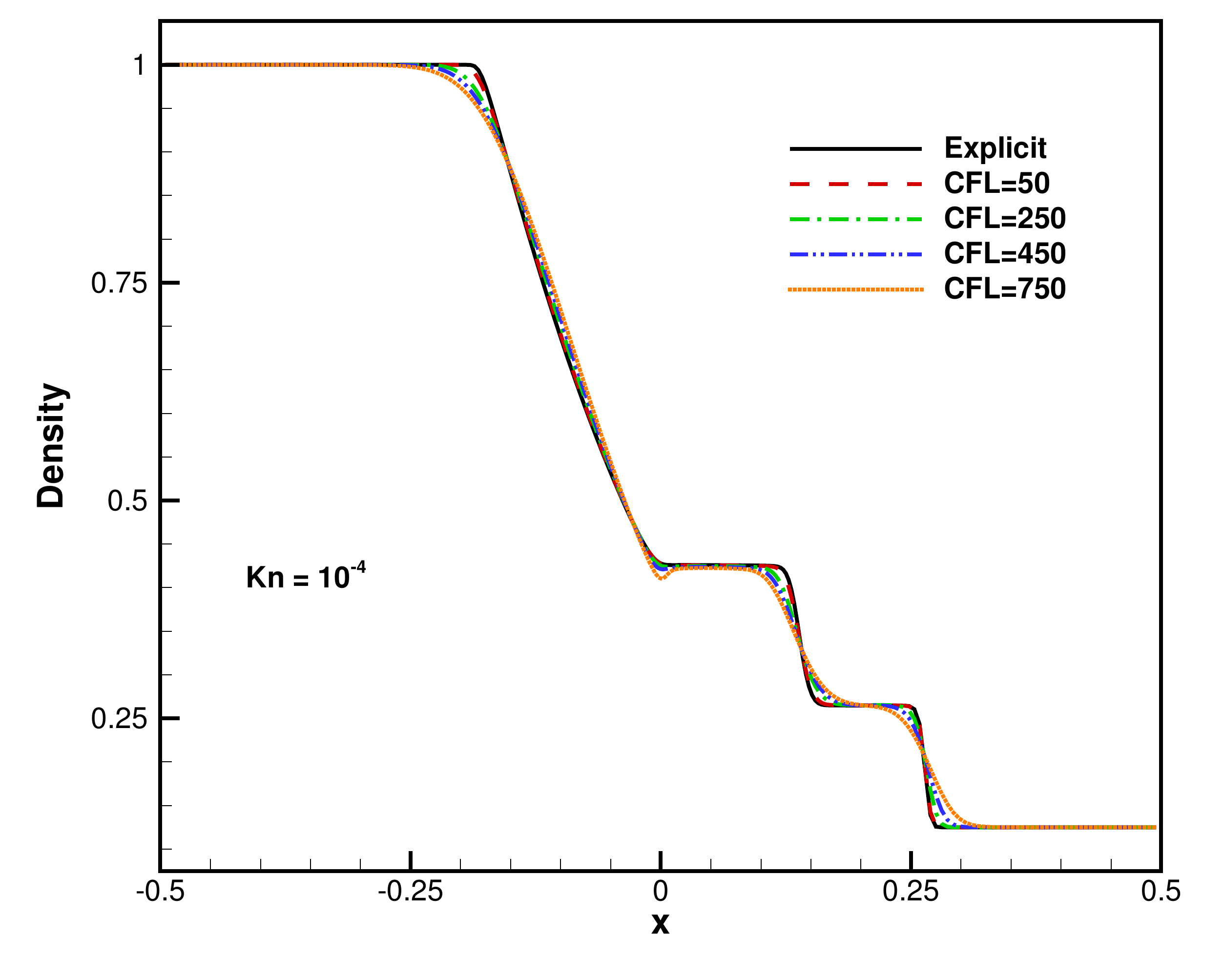}}
\subfigure[Density]{\includegraphics[width=0.48\textwidth]{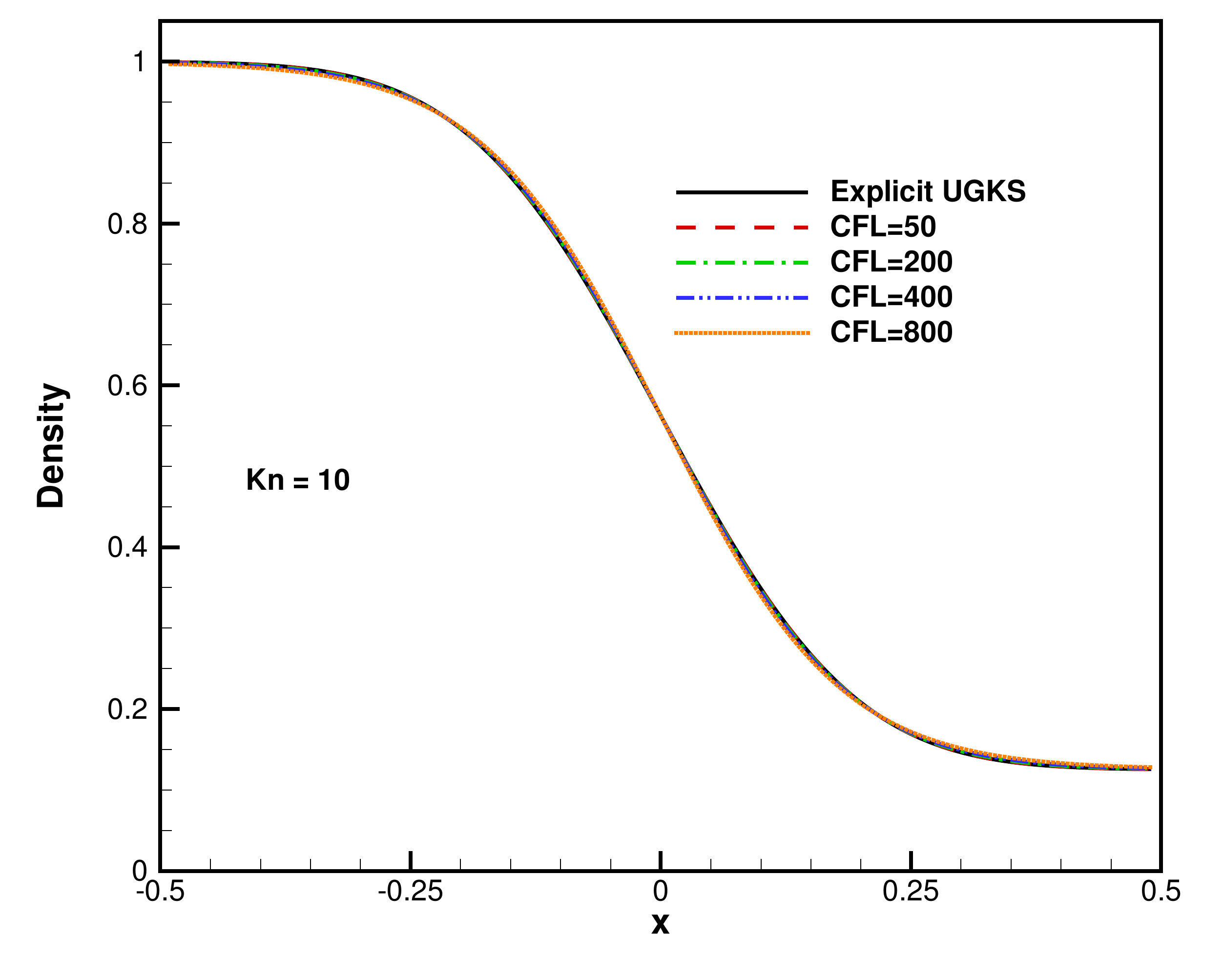}}\\
\subfigure[Velocity]{\includegraphics[width=0.48\textwidth]{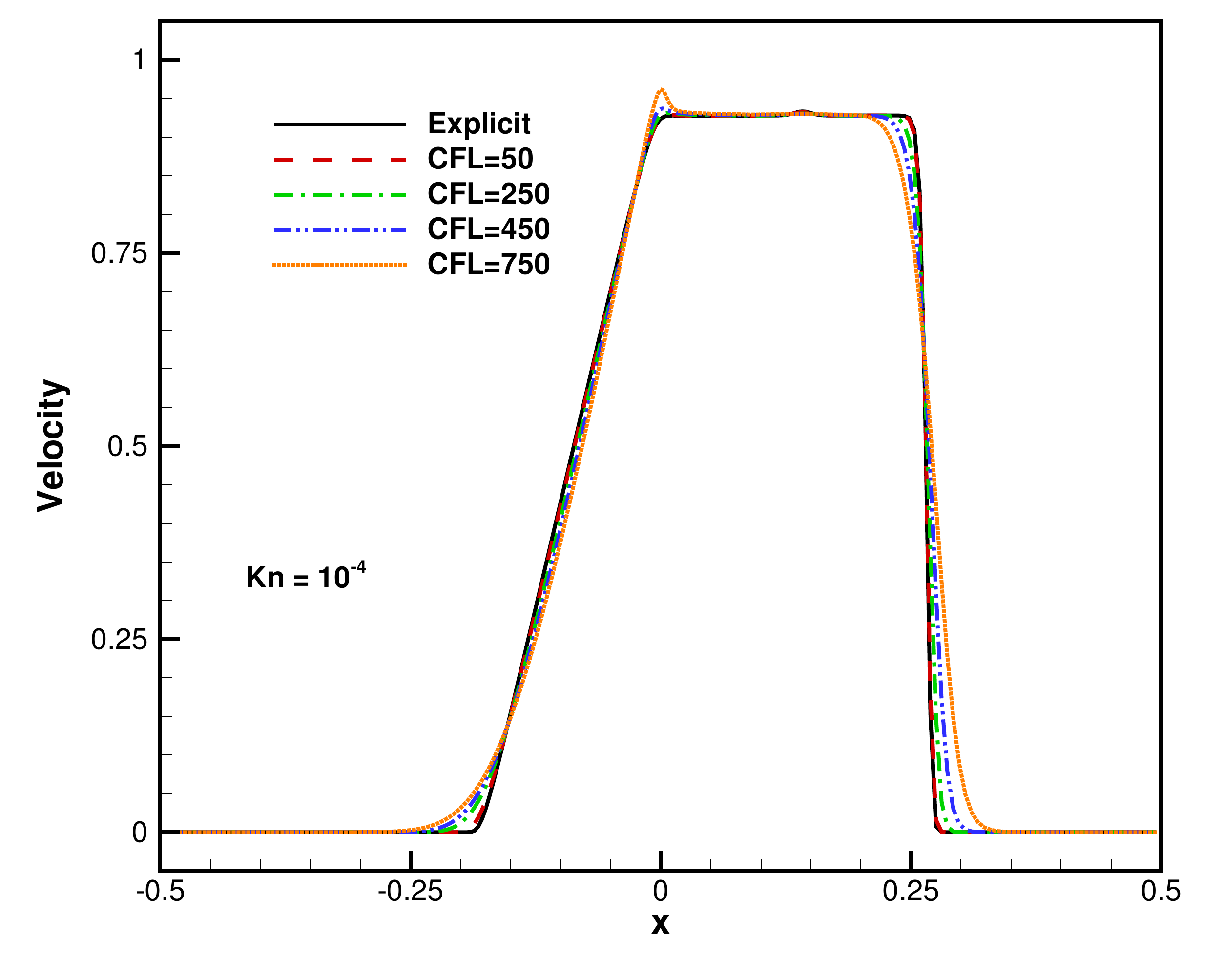}}
\subfigure[Velocity]{\includegraphics[width=0.48\textwidth]{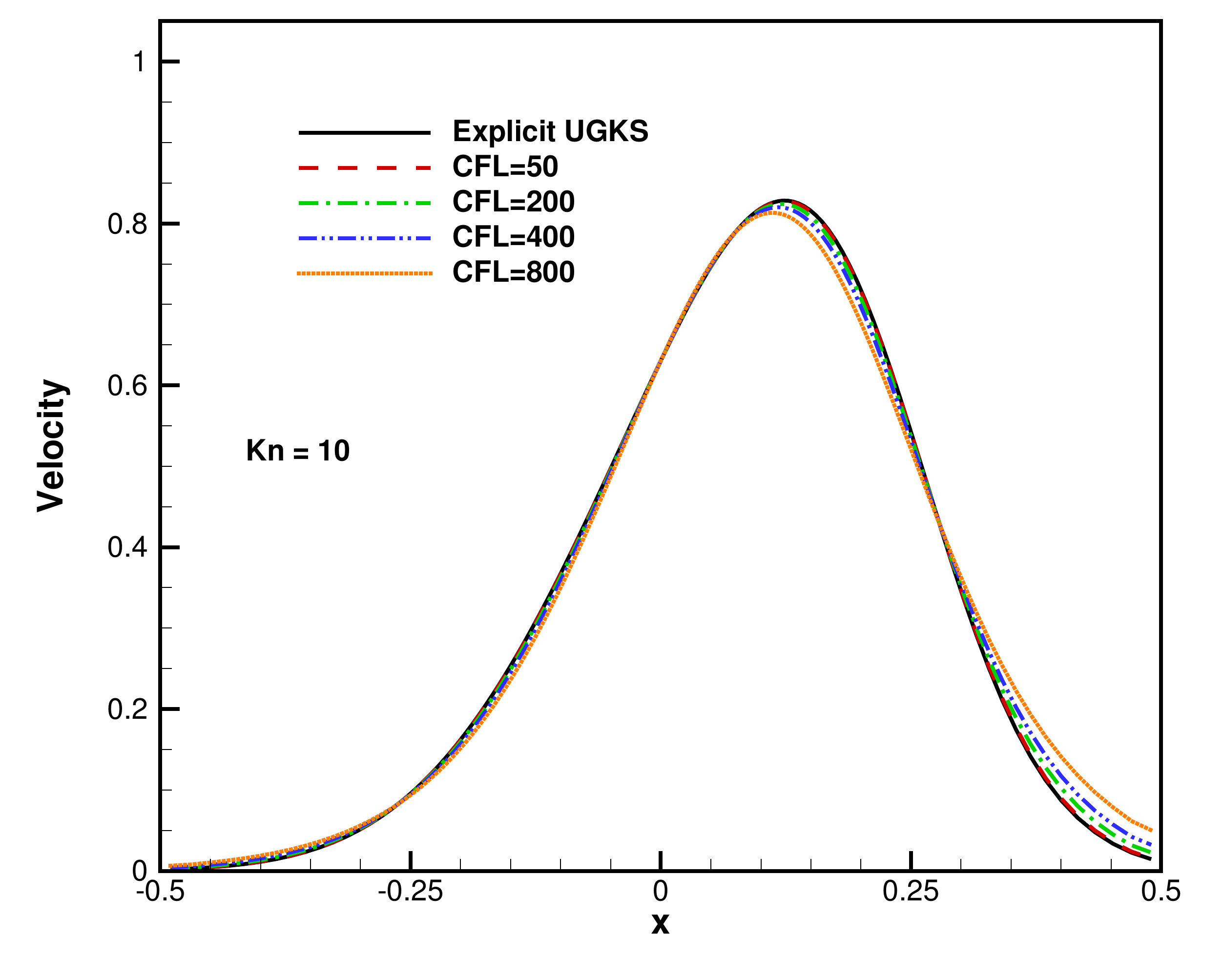}}\\
\subfigure[Temperature]{\includegraphics[width=0.48\textwidth]{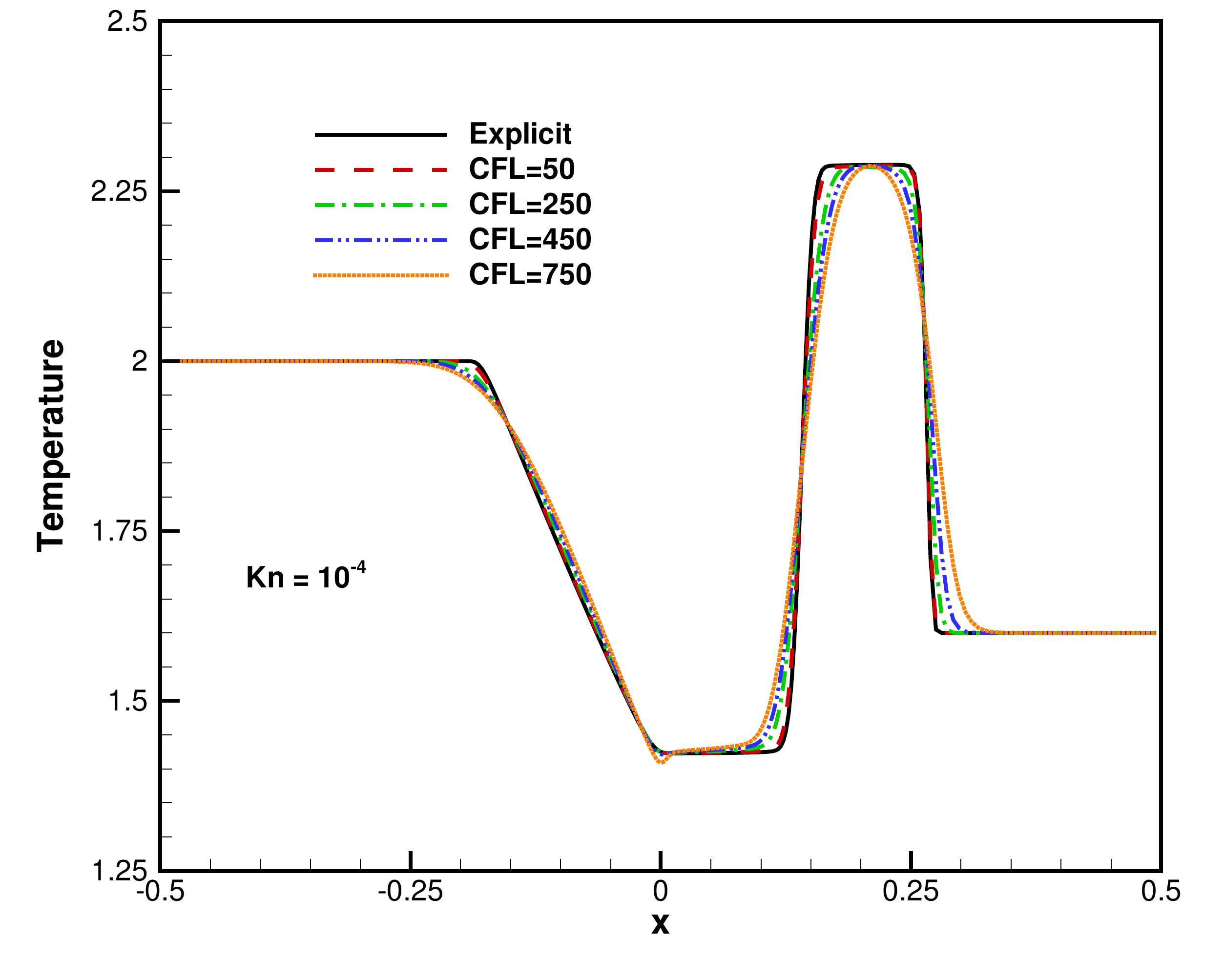}}		 \subfigure[Temperature]{\includegraphics[width=0.48\textwidth]{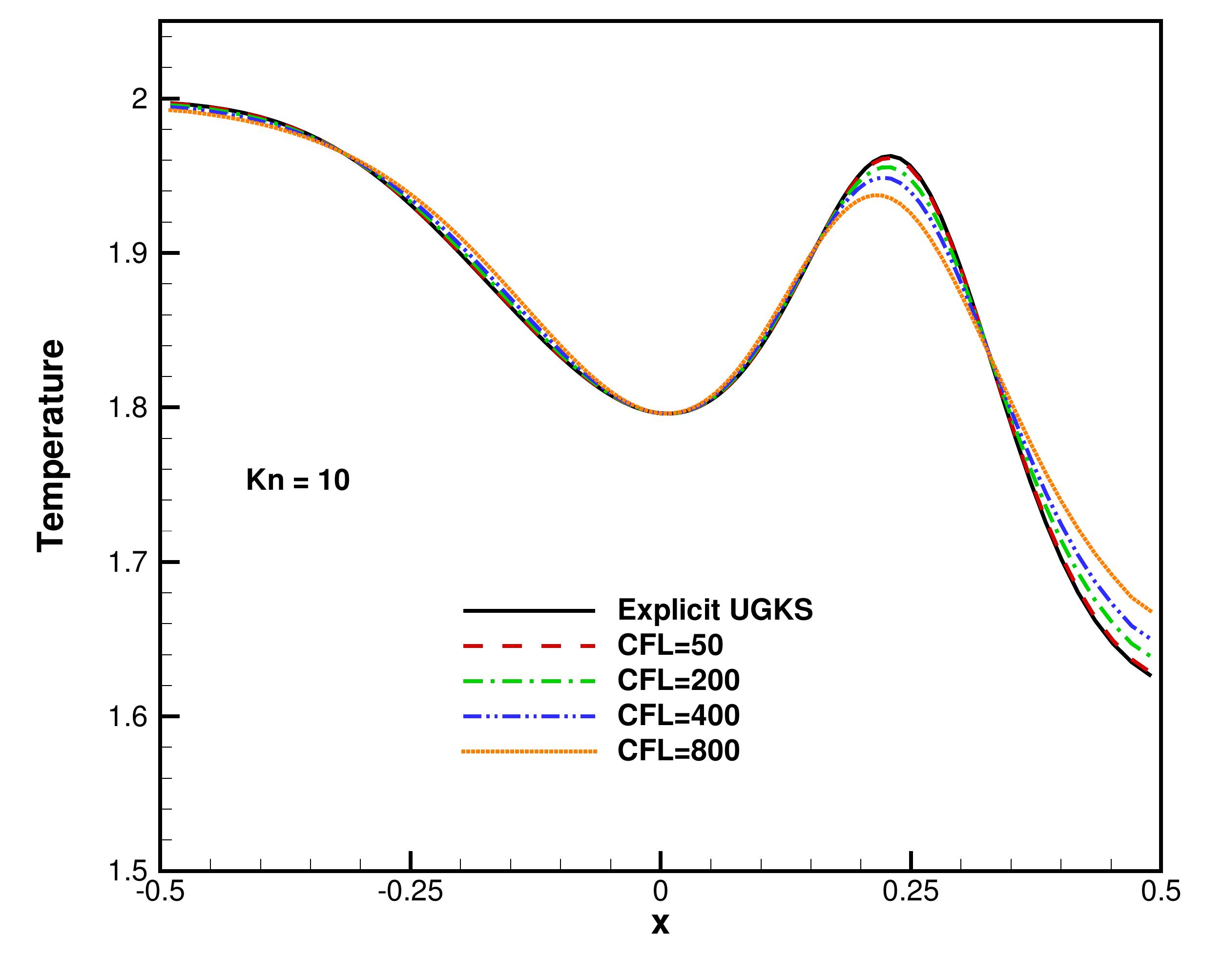}}
\caption{\label{fig:SodEfficiency}Numerical results of the Sod test case from IUGKS-$\epsilon^{\prime}$ with different choices of time step.
The left ones are the results at $Kn=10^{-4}$, and the right ones are those at $Kn=10$.}
\end{figure}

In order to figure out how much the implicit scheme can accelerate the numerical simulations, we compute the Sod case using IUGKS-$\epsilon^{\prime}$ with $\epsilon=0.75$ and different time steps. The computational domain is discretized into 400 nonuniform cells with a minimum cell size of $2\times 10^{-4}$ and a maximum cell size of $0.01$. The results in both continuum and rarefied flow regimes are given in Fig.~\ref{fig:SodEfficiency}, which are compared with the solutions from the exlicit UGKS. For the continuum case at $Kn=10^{-4}$, the velocity space is discretized into 200 velocity points, where the trapezoidal rule is used for the moments integration.
The computational cost are listed in Table.~\ref{tab:SodContinuumEfficiency}.
It can be seen that generally the IUGKS-$\epsilon^{\prime}$ is more than ten times faster than the explicit UGKS in the continuum regime.
For the rarefied flow at $Kn = 10$, we employ a non-uniform mesh with $200$ cells which has a smallest cell size of $0.0005$ around the initial discontinuity. In order to get rid of the ray effect and to obtain a smooth solution, we employ 20000 points for velocity space discretization. Actually, using uniform mesh and much fewer velocity points can also give acceptable results for this case, but here we use the much large number of grid points for testing the efficiency of the schemes. The computational cost and acceleration rate from IUGKS are listed in Table.~\ref{tab:SodRarefiedEfficiency} for the rarefied case. It can be seen that for a reasonably good result, e.g. the one with ${\rm CFL=50}$, the computational efficiency can increase by dozens of times.

\begin{table}[H]
\centering
\begin{threeparttable}[c]
\caption{\label{tab:SodContinuumEfficiency}Computational cost of the implicit UGKS and the explicit UGKS for the Sod test case at $Kn=10^{-4}$ with different time steps.}
\begin{tabular*}{0.9\textwidth}{@{\extracolsep{\fill}}lcrrrr}
\toprule
			     & Explicit UGKS & \multicolumn{4}{c}{IUGKS-$\epsilon^{\prime}$ with $\epsilon = 0.75$}\\
\midrule
	CFL number     & 0.5   & 50   & 250   & 450  & 750 \\
	CPU time (sec) & 159.6 & 23.9 & 9.2   & 6.6  & 5.6\\
    Speedup        & 1.0   & 6.7  & 17.4  & 24.3 & 28.7\\
\bottomrule
\end{tabular*}
\end{threeparttable}
\end{table}

\begin{table}[H]
	\centering
	\begin{threeparttable}[c]
		\caption{\label{tab:SodRarefiedEfficiency}Computational cost of the implicit UGKS and the explicit UGKS for Sod test case at $Kn=10$ with different time steps.}
		\begin{tabular*}{0.9\textwidth}{@{\extracolsep{\fill}}lcrrrr}
			\toprule
			& Explicit UGKS & \multicolumn{4}{c}{IUGKS-$\epsilon^{\prime}$ with $\epsilon = 0.75$}\\
			\midrule
			CFL number     & 0.5     & 50    & 200   & 400   & 800\\
			CPU time (sec) & 4405.4  & 270.4 & 77.7  & 46.0  & 29.8\\
			Speedup   & 1.0     & 16.3  & 56.7  & 95.8  & 147.7\\
			\bottomrule
		\end{tabular*}
	\end{threeparttable}
\end{table}

\subsection{Couette flow with a temperature gradient}\label{sec:CouetteCase}
In the incompressible limit, the Couette flow with a temperature gradient has a steady state analytic solution under the assumption of constant viscosity and heat conduction coefficients. Such a steady state problem provides a good test case to validate the spatial accuracy of the IUGKS,
where the influence on the time accuracy can be ignored. This case describes the flow between two parallel solid boundaries. The setup is as follows. The top solid boundary has a constant temperature $T_1$ and moves horizontally at a velocity $U$. The bottom wall is fixed and has a lower temperature $T_0$. If the distance between these two walls is $H$, it gives an analytic solution for the normalized temperature at steady state
\begin{equation}
\hat{T}_e(y) = \dfrac{T - T_0}{T_1 - T_0} = \dfrac{y}{H} + \dfrac{Pr Ec}{2} \dfrac{y}{H} \left(1-\dfrac{y}{H}\right)
\end{equation}
where $Ec = U^2/C_p(T_1-T_0)$ is the Eckert number, and $C_p$ is the specific heat ratio at constant pressure.

\begin{figure}[H]
	\centering
	\subfigure[Velocity]{\includegraphics[width=0.48\textwidth]{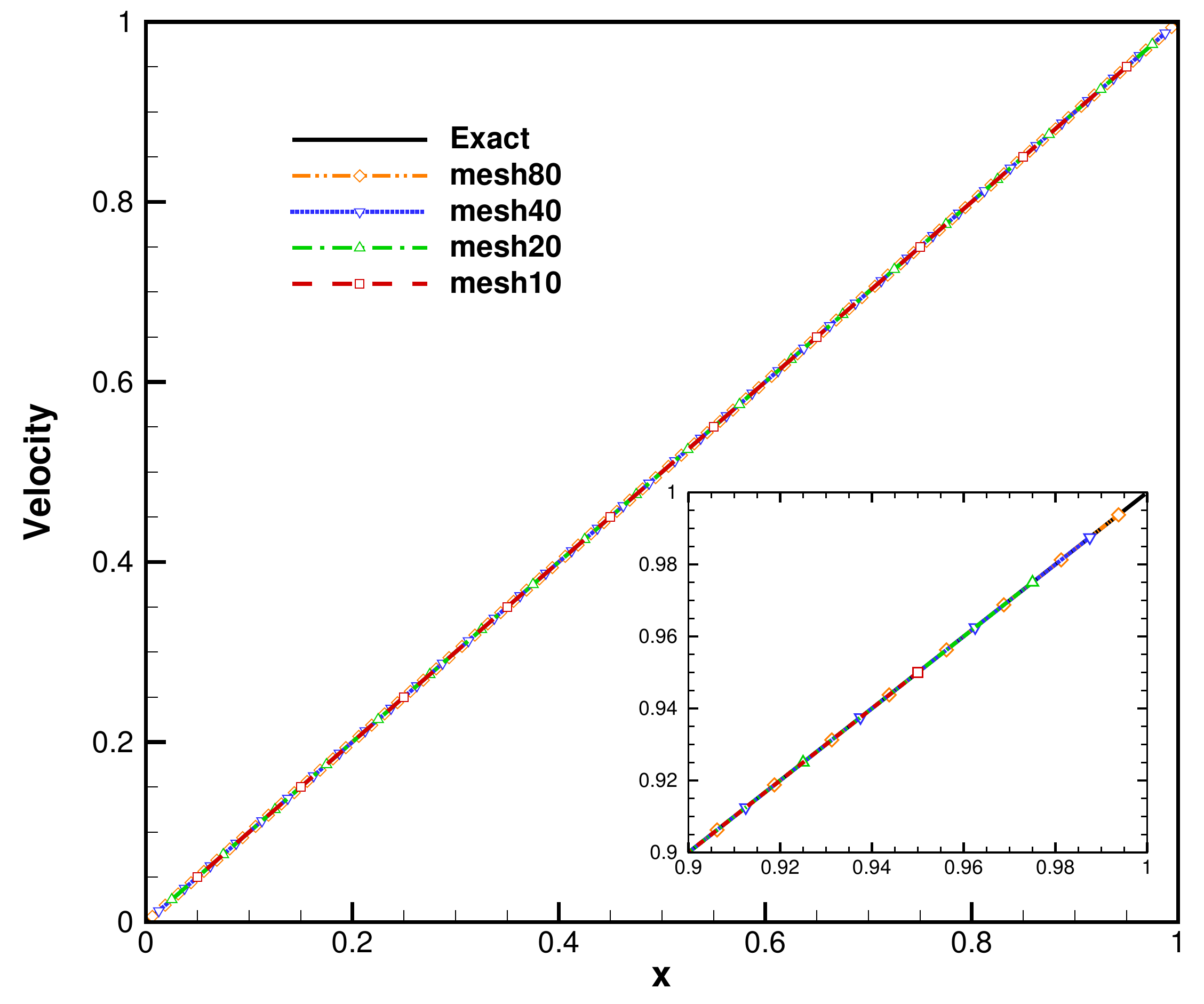}}
	\subfigure[Temperature]{\includegraphics[width=0.48\textwidth]{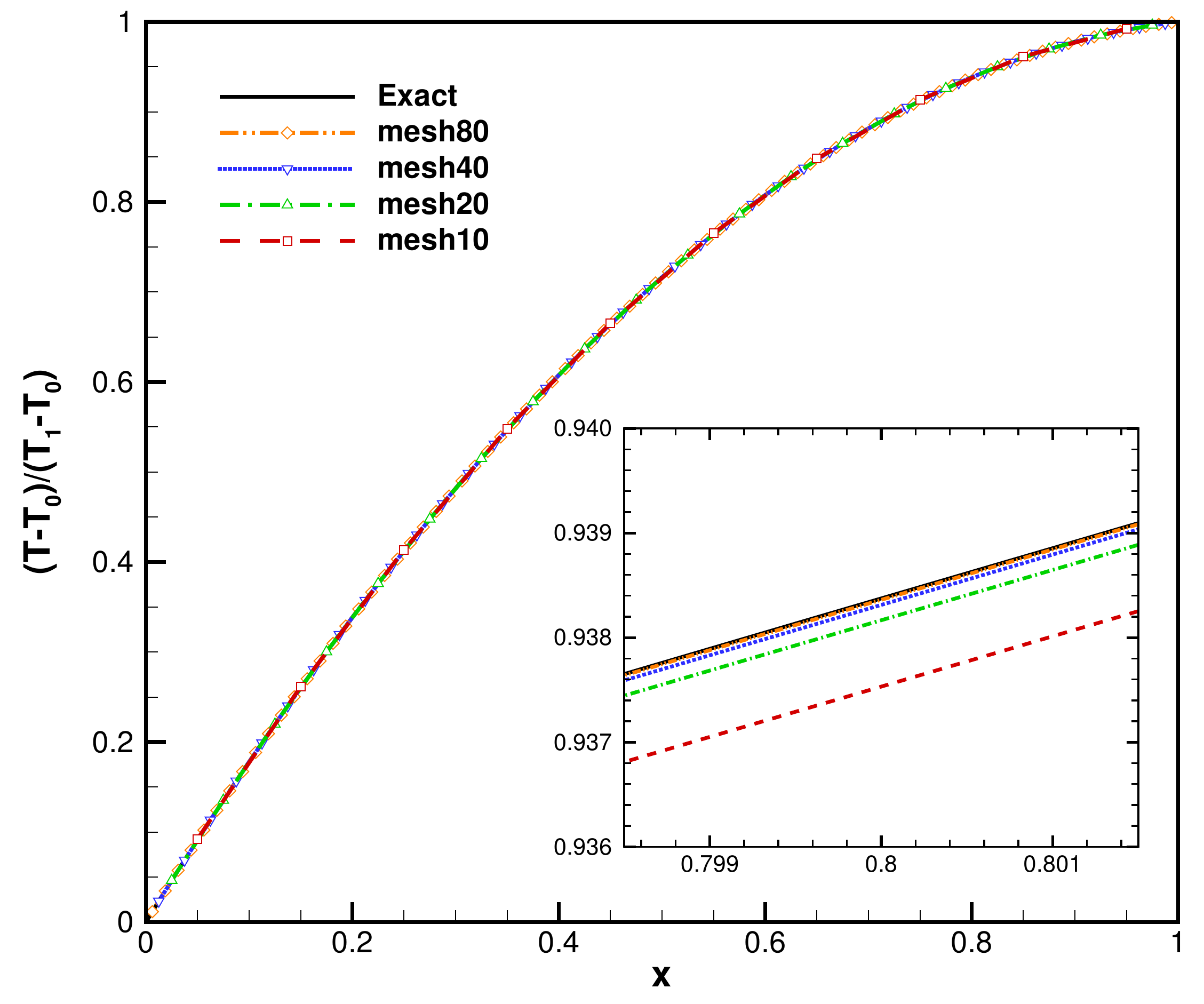}}
	\caption{\label{fig:CouetteContinuum}The normalized velocity and temperature distributions obtained on different meshes by IUGKS with $\tau = 0.001 \Delta t_s$.}
\end{figure}

This case has $T_1 = 274 {\rm K}$, $T_0 = 273{\rm K}$, and $U=30 {\rm m/s}$ for the incompressible limit. The argon gas with molecular mass  $m_0 = 6.63\times 10^{-26} {\rm kg}$ is employed, and $Ec \approx 1.73$. Uniform meshes with $10$, $20$, $40$ and $80$ discrete cells are used. The results obtained by the IUGKS with $\tau = 0.001 \Delta t_s$ in continuum regime are plotted in Fig.~\ref{fig:CouetteContinuum}.

\begin{figure}[H]
\centering
\subfigure[\label{fig:CouetteAccuracyContinuum}]{\includegraphics[width=0.46\textwidth]{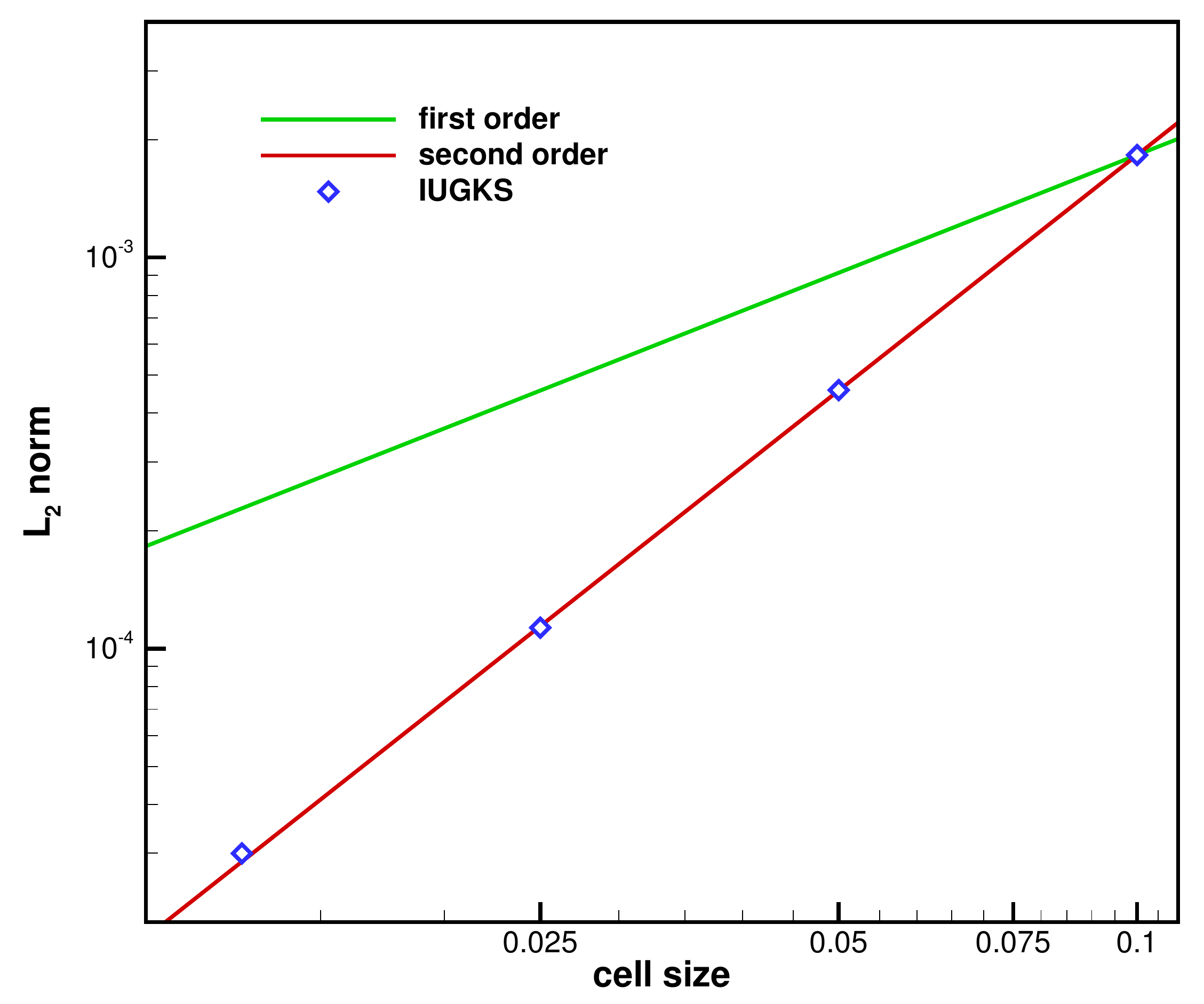}}	
\subfigure[\label{fig:CouetteAccuracyRarefied}]{\includegraphics[width=0.46\textwidth]{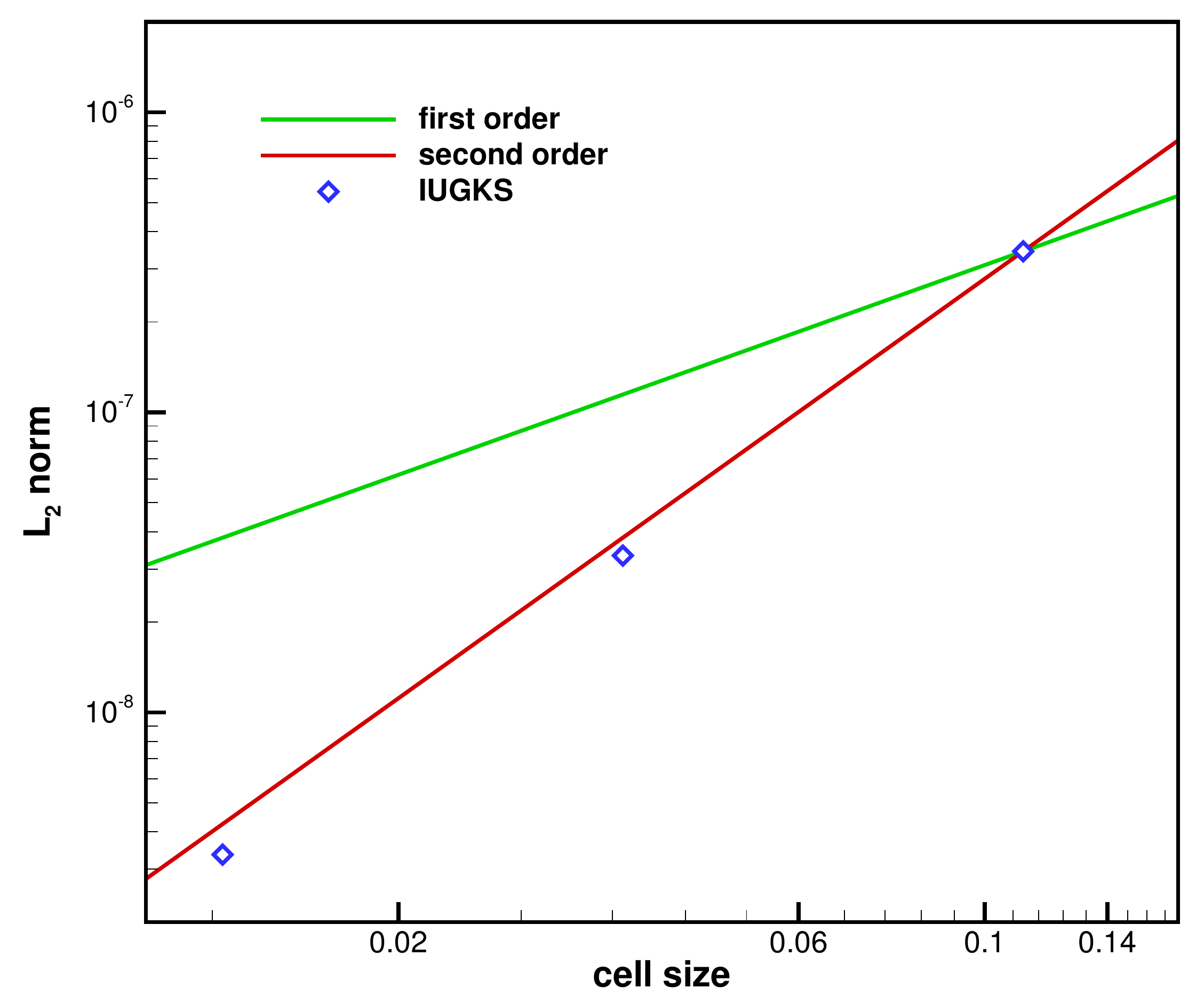}}
\caption{Spatial accuracy of the IUGKS for (a) $\tau = 0.001\Delta t_s $ in continuum flow, and (b) $Kn=10$ in rarefied flow.}
\end{figure}

The errors of $L_2$ norms are calculated by
\begin{equation}\label{eq:normA}
E_{L_2} = \dfrac{\|y(x) - y_e (x) \|_2}{\|y_e (x)\|_2},
\end{equation}
where $y(x)$ denotes numerical distribution of a specific flow variable, and $y_e (x)$ is the corresponding exact solution.
For this case, a normalized temperature distribution is used to evaluate the errors.
The $L_2$ norms with respect to mesh size are plotted in Fig.~\ref{fig:CouetteAccuracyContinuum}. It can be seen that the IUGKS has a second-order accuracy in space for the continuum flow.

\begin{figure}[H]
\centering
\subfigure[Velocity]{\includegraphics[width=0.46\textwidth]{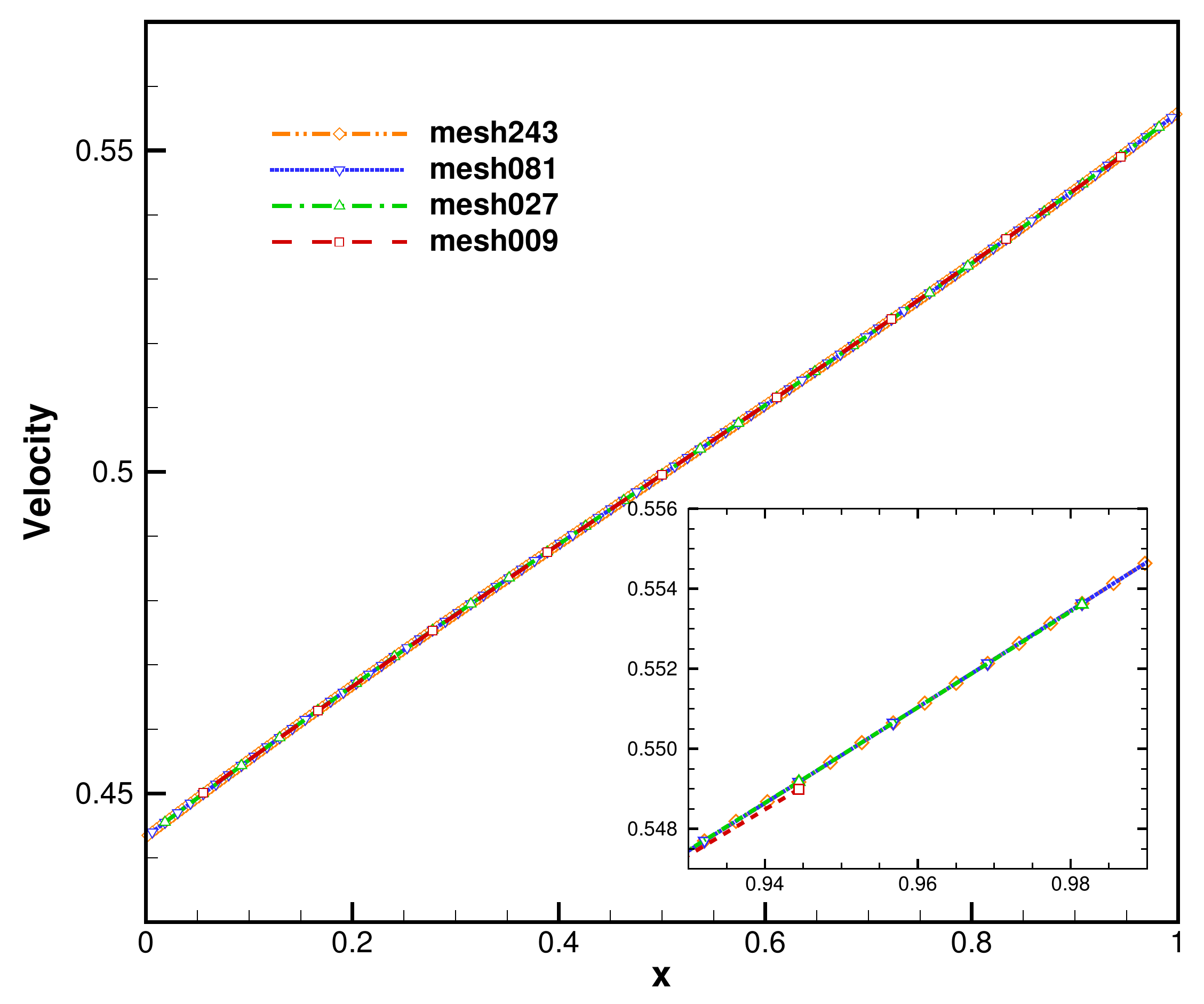}}
\subfigure[Temperature]{\includegraphics[width=0.46\textwidth]{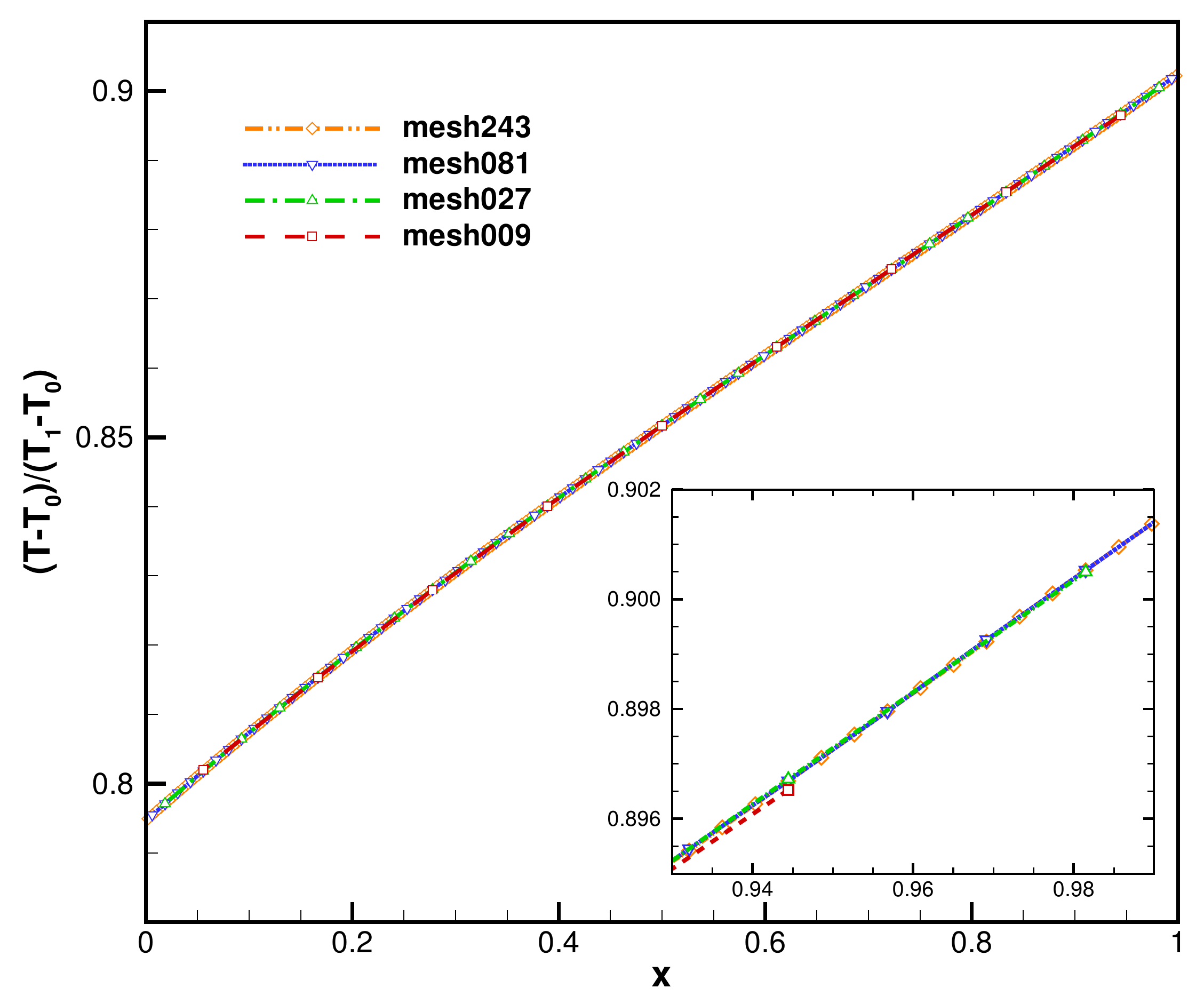}}
\caption{\label{fig:CouetteRarefied}The normalized velocity and temperature distribution obtained by IUGKS on different meshes  at $Kn = 10$.}
\end{figure}

In order to measure the spatial convergence of the IUGKS in rarefied regime, the same case at $Kn=10$ is computed on a group of meshes of $9$, $27$, $81$ and $243$ cells. The results are given in Fig.~\ref{fig:CouetteRarefied}. Since there is no analytic solution for the Couette flow in rarefied regime, a relative method to measure the error is adopted, i.e.,
\begin{equation}\label{eq:normB}
E_{L_2} = \dfrac{\|y_c (x) - y_f(x) \|_2}{\|y_f(x)\|_2},
\end{equation}
where $y_c(x)$ and $y_f(x)$ are the numerical solutions on the coarse and the fine meshes, respectively. The results given in Fig.~\ref{fig:CouetteAccuracyRarefied} show that a second-order accuracy in space can be obtained as well for the case in the collisionless limit.

\subsection{Advection of density perturbation}\label{sec:sineWaveCase}
In order to test the temporal accuracy of the IUGKS for unsteady flow simulations, the one-dimension case of advection of density perturbation \cite{pan2016efficient} is employed. The initial condition is set as follows
\begin{equation}
\rho (x) = 1+ 0.2 \sin (\pi x), \quad U(x)=1, \quad p(x) = 1, \quad x \in \left[0,2\right].
\end{equation}
The periodic boundary condition is implemented and it gives an analytic solution
\begin{equation}
	\rho_e(x, t) = 1 + 0.2 \sin \left(\pi(x-t)\right).
\end{equation}

Since the IUGKS is designed to be second-order accurate both in space and time, it is supposed to have an error on order of $\mathcal{O} (\Delta x^2, \Delta t^2)$. When $\mathcal{O}(\Delta t^2)$ is dominant, we can capture the convergence order of the IUGKS with respect to time step. Therefore, in this case we adopt a very fine uniform mesh with $10000$ cells and use very large time steps. A small mean collision time $\tau = 0.001 \Delta t_s$ is used here to drive the IUGKS in a continuum limit to the inviscid case.

\begin{figure}[H]
\centering
\includegraphics[width=0.5\textwidth]{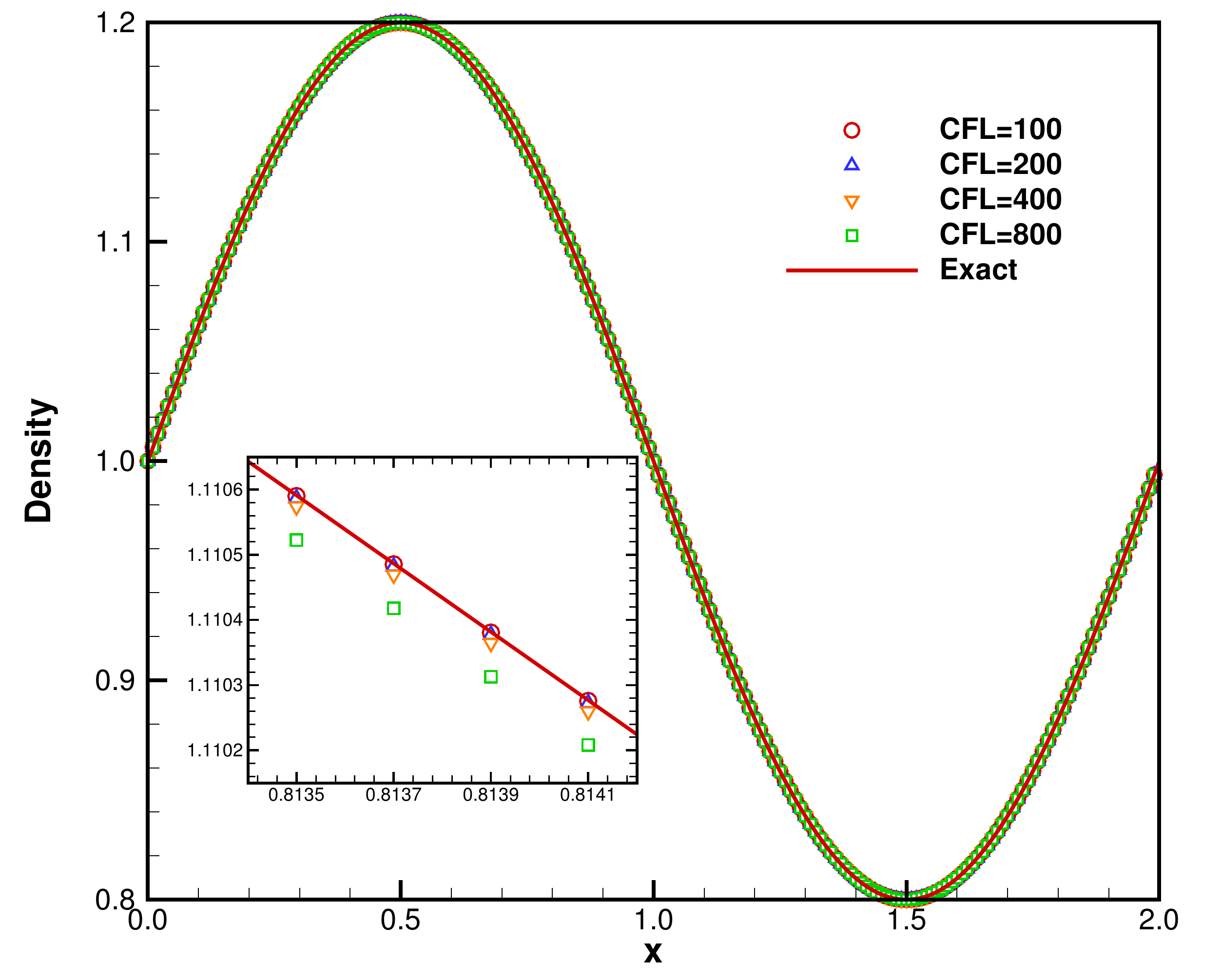}
\caption{\label{fig:sineDensity}The density distribution at time $t=2$ obtained by the IUGKS with different time steps.}
\end{figure}

\begin{figure}[H]
\centering
\subfigure[\label{fig:sineNormAb2}]{\includegraphics[width=0.48\textwidth]{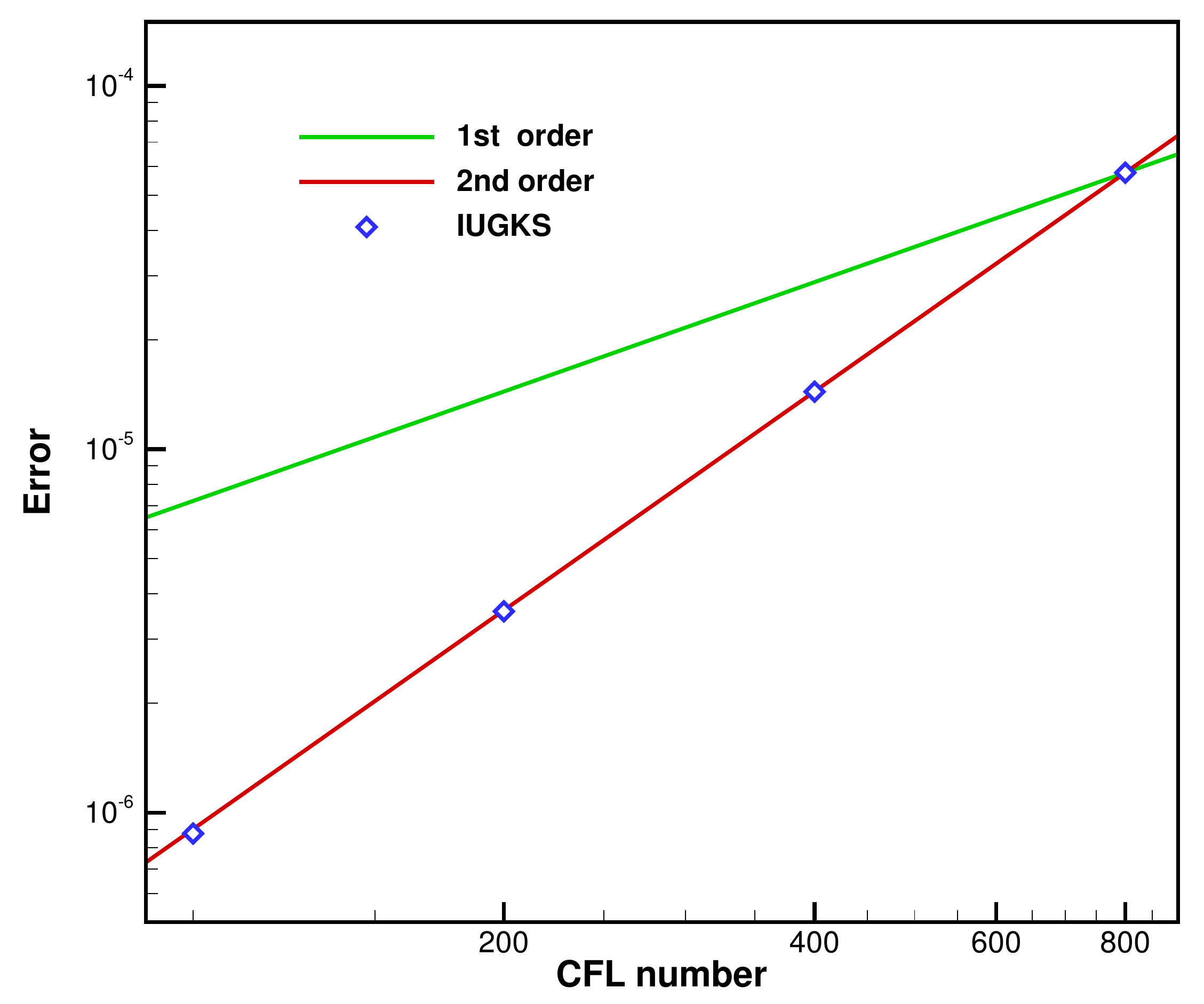}}
\subfigure[\label{fig:sineNormRe2}]{\includegraphics[width=0.48\textwidth]{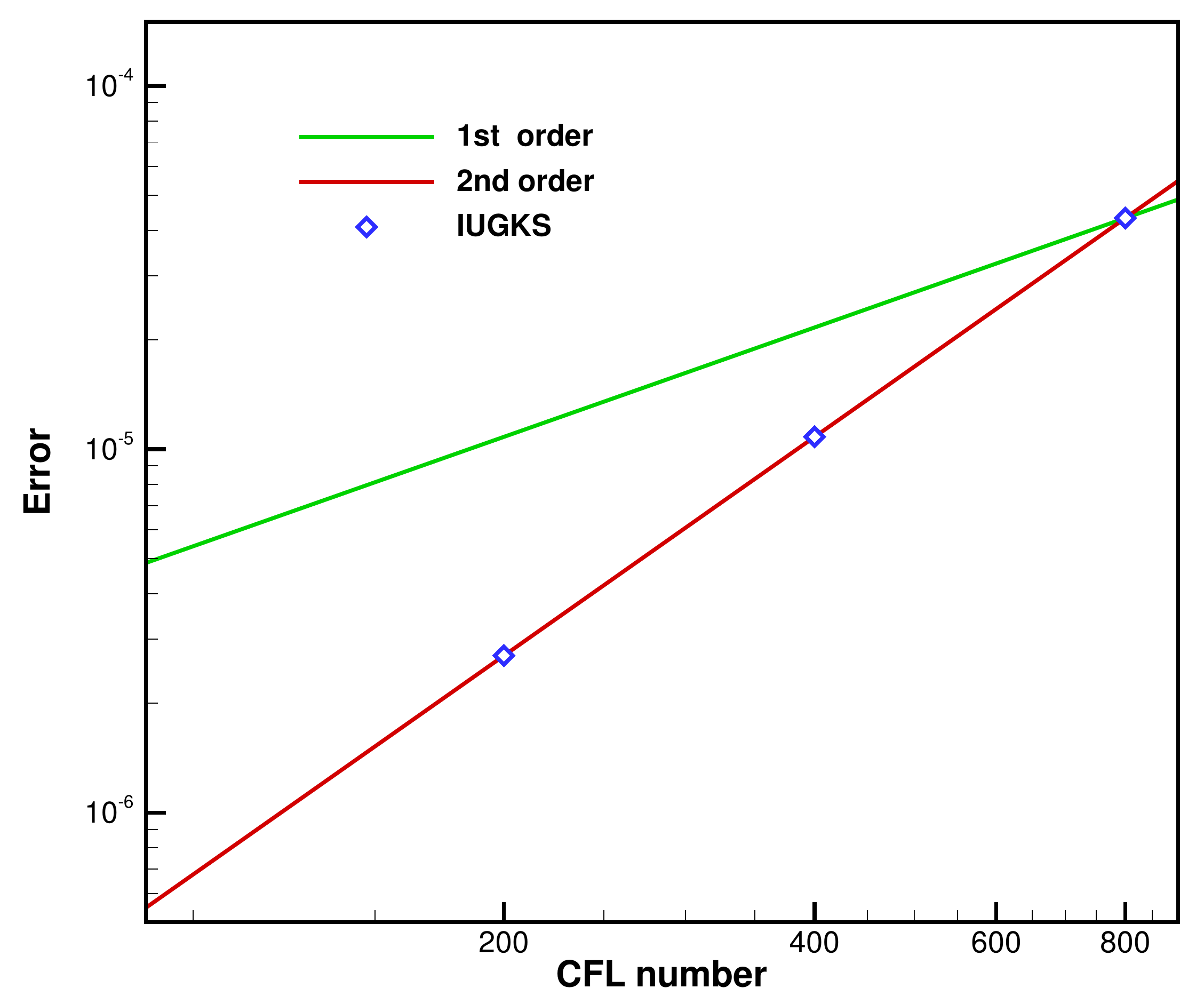}}
\caption{\label{fig:sineNormSecond}Temporal accuracy of the IUGKS with $\epsilon = 0.5$ measured by (a) Eq.~(\ref{eq:normA}) and (b) Eq.~(\ref{eq:normB}).}
\end{figure}

Fig.~\ref{fig:sineDensity} shows the density distribution at time $t=2$ that obtained by the IUGKS with different time steps. It can be predicted that the IUGKS with $\epsilon=0.5$ for the C-N temporal discretization has a  second-order accuracy in time because it has been proved that the IUGKS can automatically reduce into the second-order accurate explicit UGKS for small time step cases. In order to further validate this, the errors for different time step cases are calculated by both Eq.~(\ref{eq:normA}) and Eq.~(\ref{eq:normB}) to measure the temporal accuracy of the IUGKS. From Fig.~\ref{fig:sineNormSecond}, it shows that the C-N scheme with $\epsilon=0.5$ achieves the expected temporal accuracy.

\begin{figure}[H]
	\centering
	\subfigure[\label{fig:sineNormAb1}]{\includegraphics[width=0.48\textwidth]{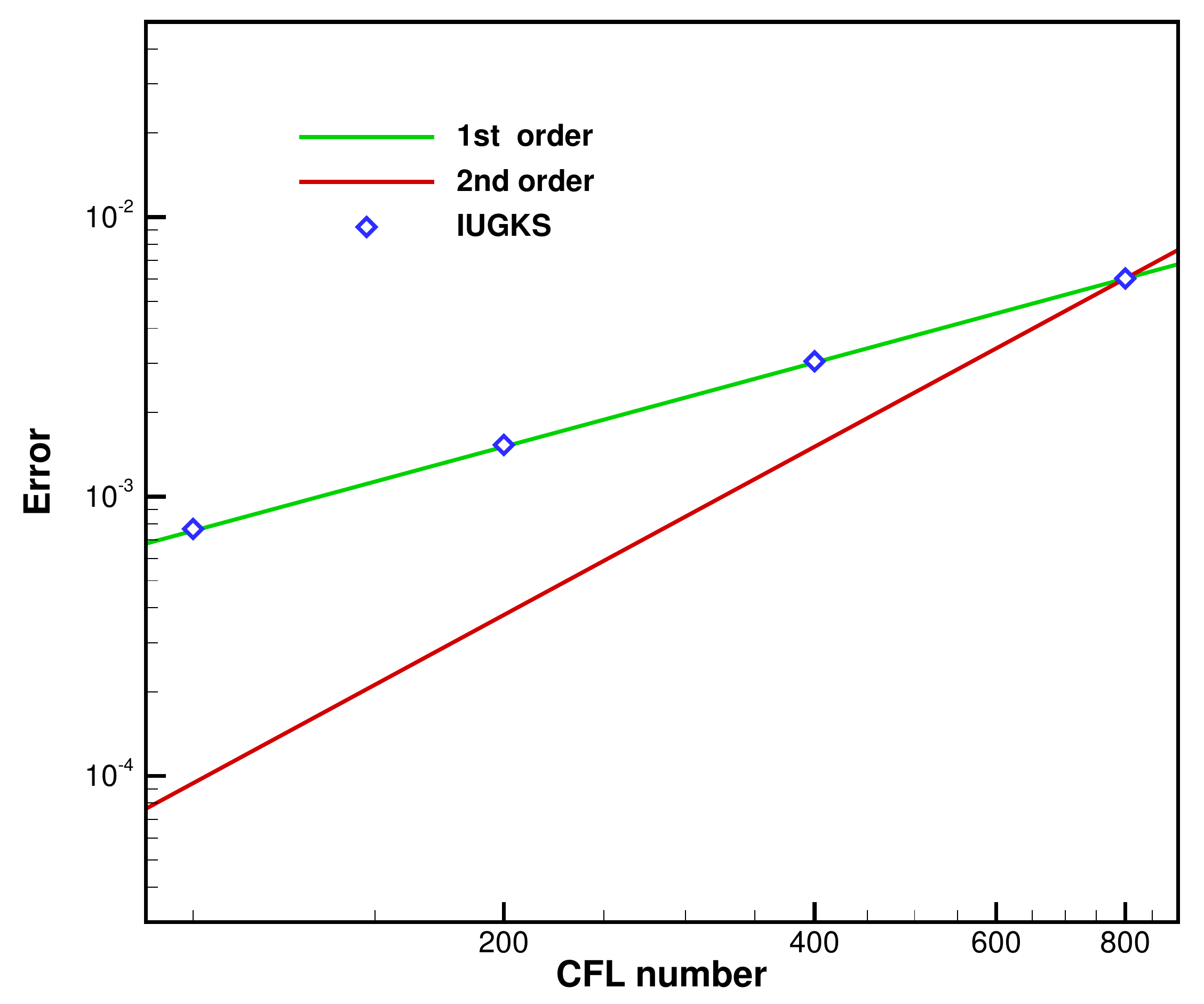}}
	\subfigure[\label{fig:sineNormRe1}]{\includegraphics[width=0.48\textwidth]{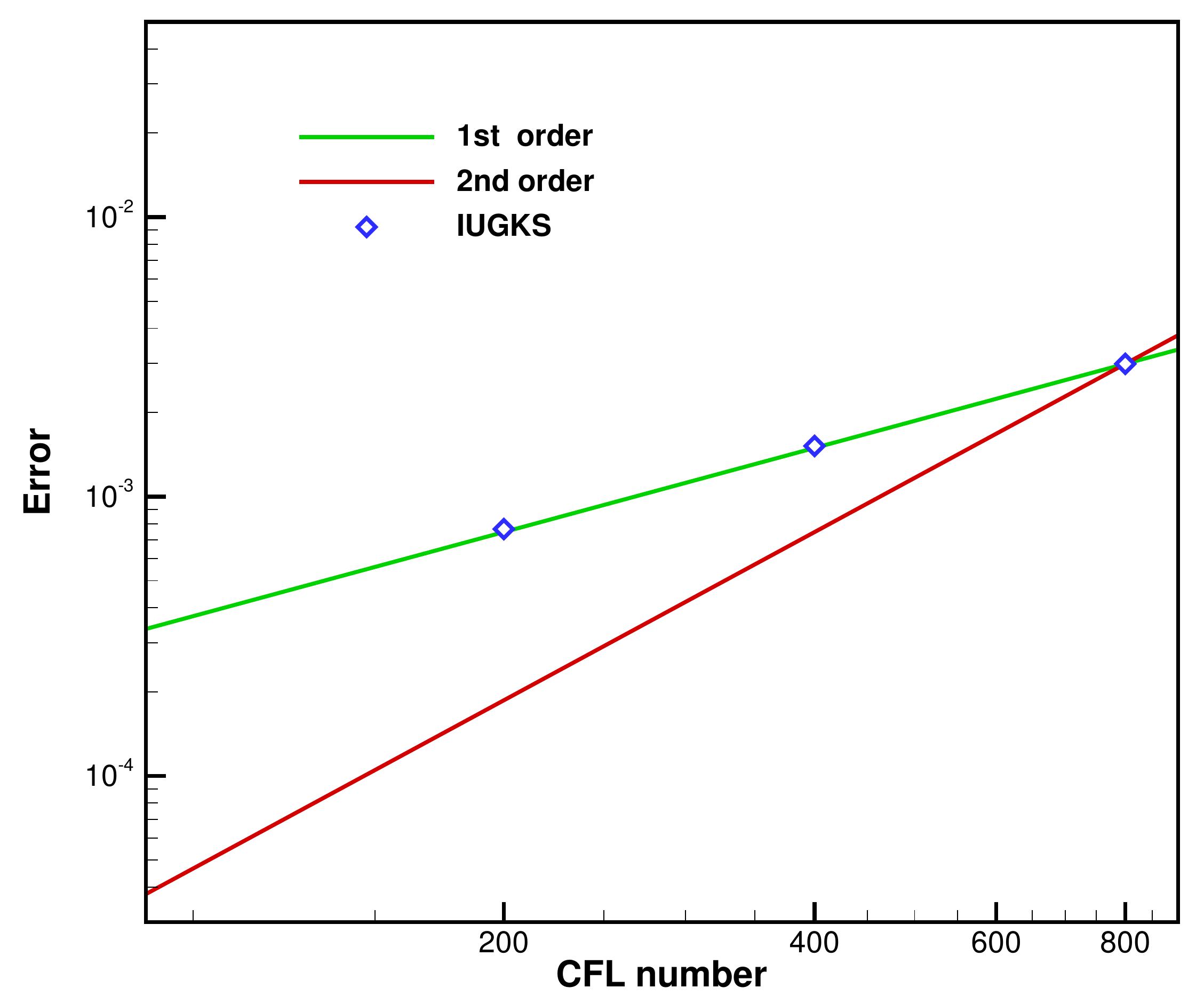}}
	\caption{\label{fig:sineNormFirst}Temporal accuracy of the IUGKS with $\epsilon = 0.75$ measured by (a) Eq.~(\ref{eq:normA}) and (b) Eq.~(\ref{eq:normB}).}
\end{figure}

The accuracy of IUGKS with a temporal discretization of $\epsilon =0.75$ has also been tested. The results obtained by both Eq.~(\ref{eq:normA}) and Eq.~(\ref{eq:normB}) are given in Fig.~\ref{fig:sineNormFirst}. As expected, the IUGKS for the case with $\epsilon > 0.5$ has a first-order accuracy in time.

\begin{figure}[H]
\centering
\subfigure[\label{fig:sinePeriod}]{\includegraphics[width=0.48\textwidth]{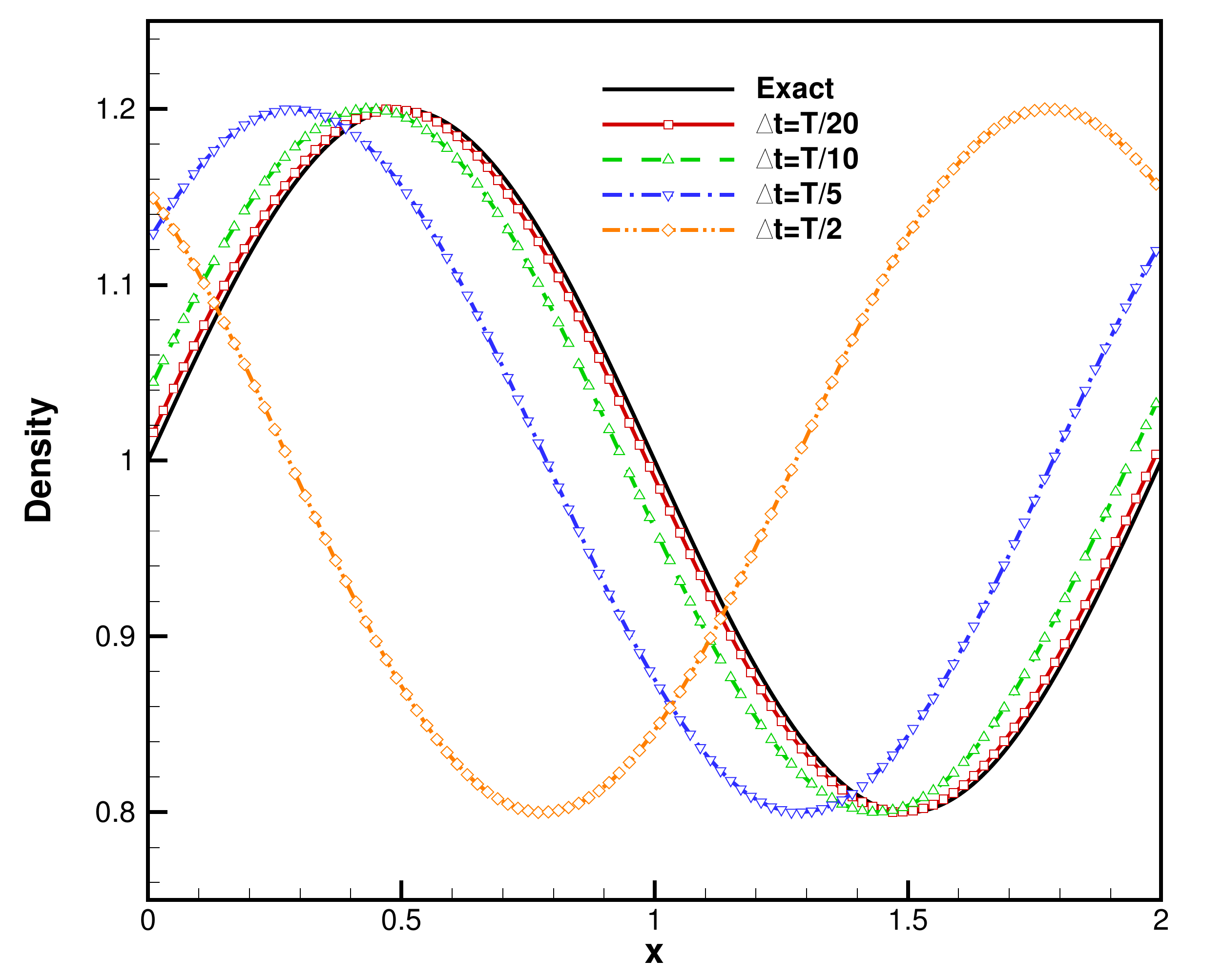}}
\subfigure[\label{fig:sineCorrection}]{\includegraphics[width=0.48\textwidth]{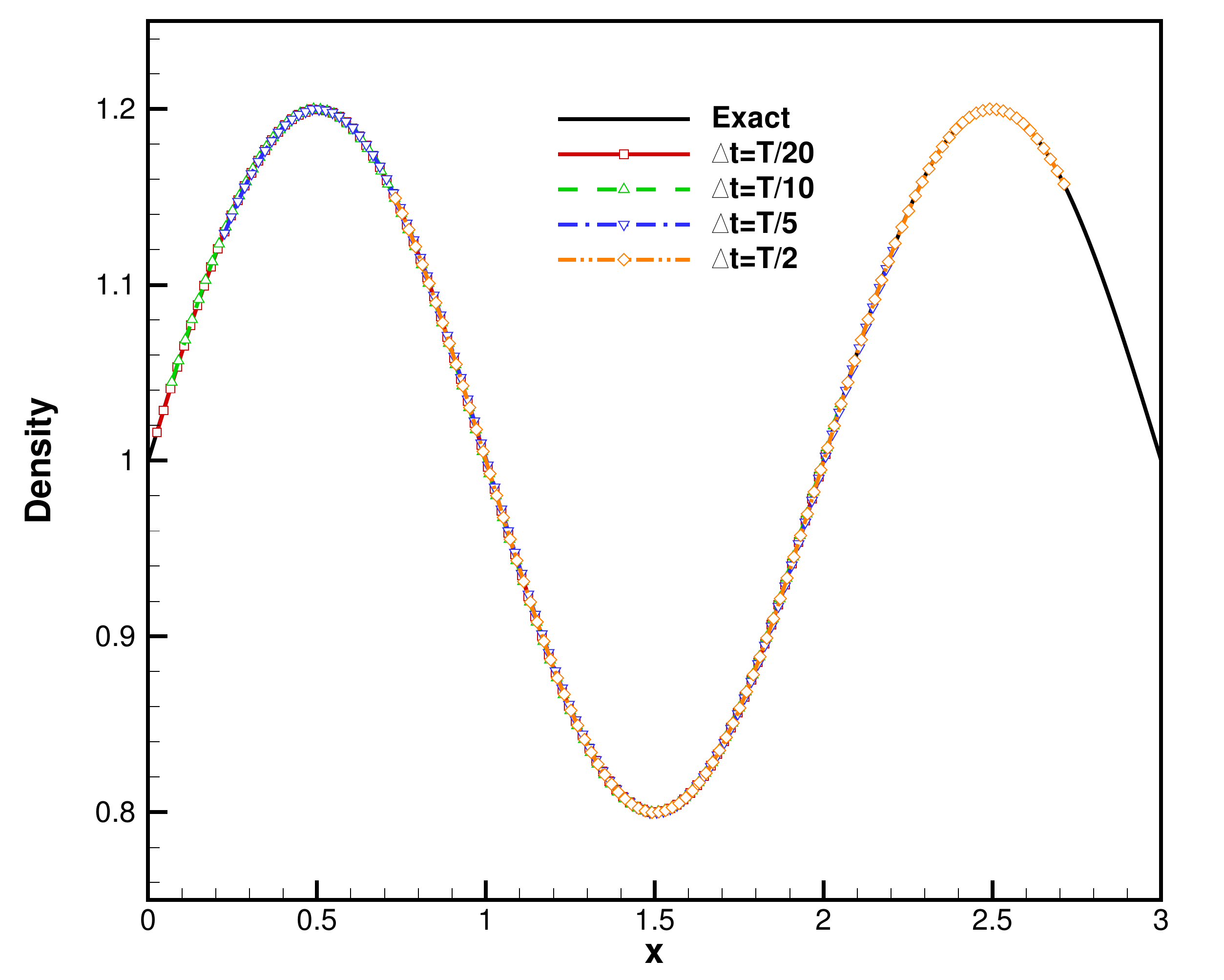}}
\caption{\label{fig:sineExplore}Density distribution obtained with $\Delta t = T/20$, $T/10$,  $T/5$ and $T/2$. (a) Phase error and (b) shifted results.}
\end{figure}

Generally, the stability of IUGKS can be achieved by adjusting the parameter $\epsilon$ so that the IUGKS is able to evolve in time using any large time step. However, large time steps may lead to unsatisfactory results due to the decrease of time accuracy. Even though large time steps are permissible, the time resolution of flow physics evolutions should be considered. In order to explore the applicability of the IUGKS, we utilize this periodic case on a uniform mesh with $100$ cells to further validate the scheme. In Fig.~\ref{fig:sinePeriod}, we give the results obtained with different time steps, i.e., $\Delta t = T/20$, $T/10$,  $T/5$ and $T/2$, where $T=2$ is the period of this sine wave. With respect to the wave length, the relative error of the location of extremum value is about $0.8\%$, $3\%$, $10.8$ and $36.1\%$. From Fig.~\ref{fig:sinePeriod}, it can be found that the shape of the sine wave is well preserved even for the case with $\Delta t=T/2$, and the solutions can coincide to the exact one after a shift, see Fig.~\ref{fig:sineCorrection}.  Based on this case, we can roughly draw a conclusion that for the requirement of time accuracy, in order to capture the flow variable fluctuations, the numerical time step chosen to evolve the unsteady flow field should not be larger than one-tenth of the period of fluctuations.

\subsection{Rayleigh flow}\label{sec:RayleighCase}
The Rayleigh flow is an unsteady gas flow around a vertical plate with infinite length. Initially, the argon gas with molecular mass $m_0 = 6.63 \times 10^{-26} {\rm kg}$ is stationary and has a temperature of $273{\rm K}$, and suddenly the plate obtains a constant vertical velocity of $10{\rm m/s}$ and gets a higher temperature of $373{\rm K}$. The computational domain is $1 {\rm m}$ long, which is the characteristic length to define the Knudsen number $Kn$ by the variable hard sphere (VHS) model. The dynamic viscosity is computed by $\mu = \mu_0 (T/T_0)^{\omega}$ with $\omega = 0.81$. Results at time $t=0.7{\rm ms}$ are discussed here.
\begin{figure}[H]
\centering
\subfigure[Kn=2.66]{\includegraphics[width=0.48\textwidth]{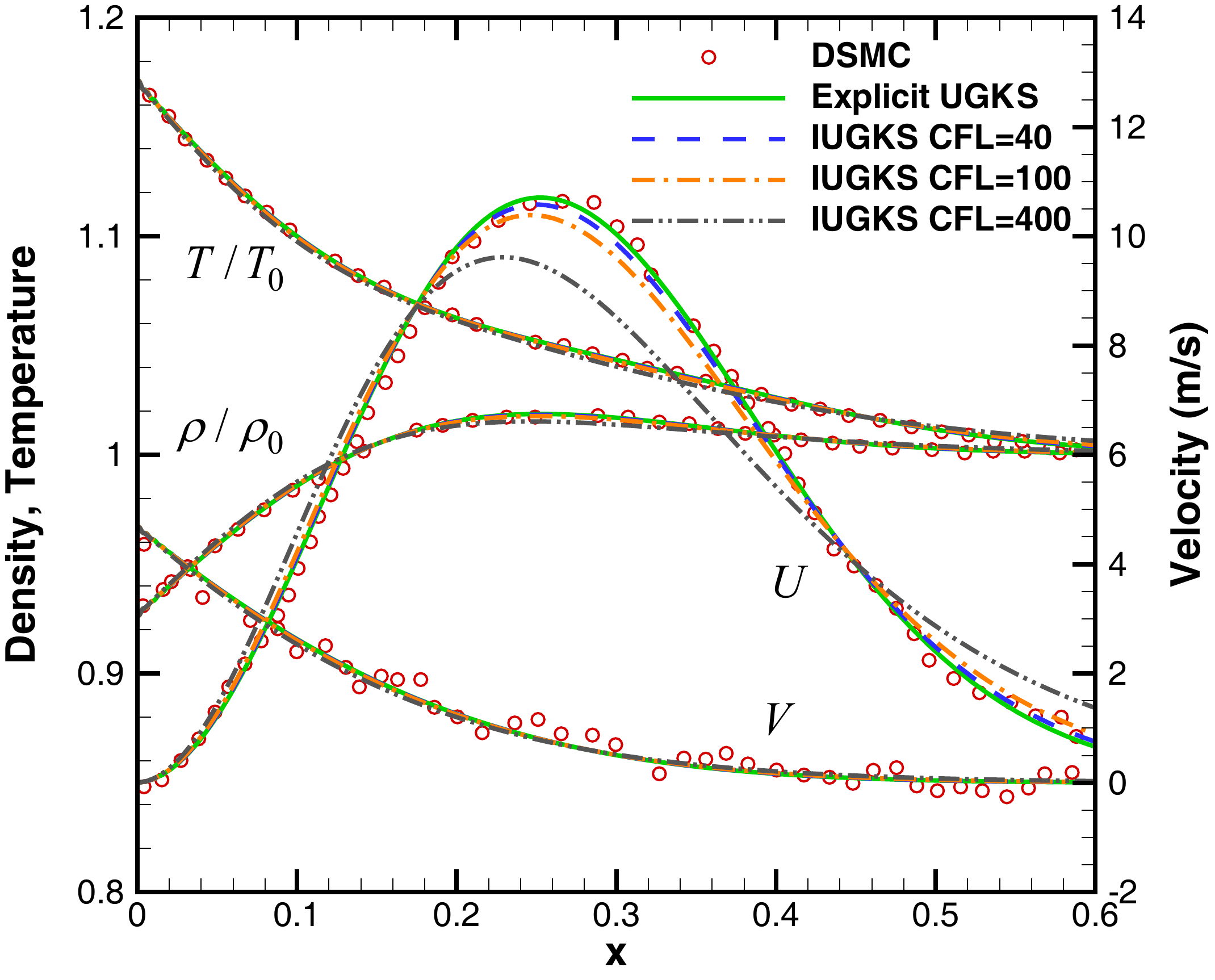}} \subfigure[Kn=0.266]{\includegraphics[width=0.48\textwidth]{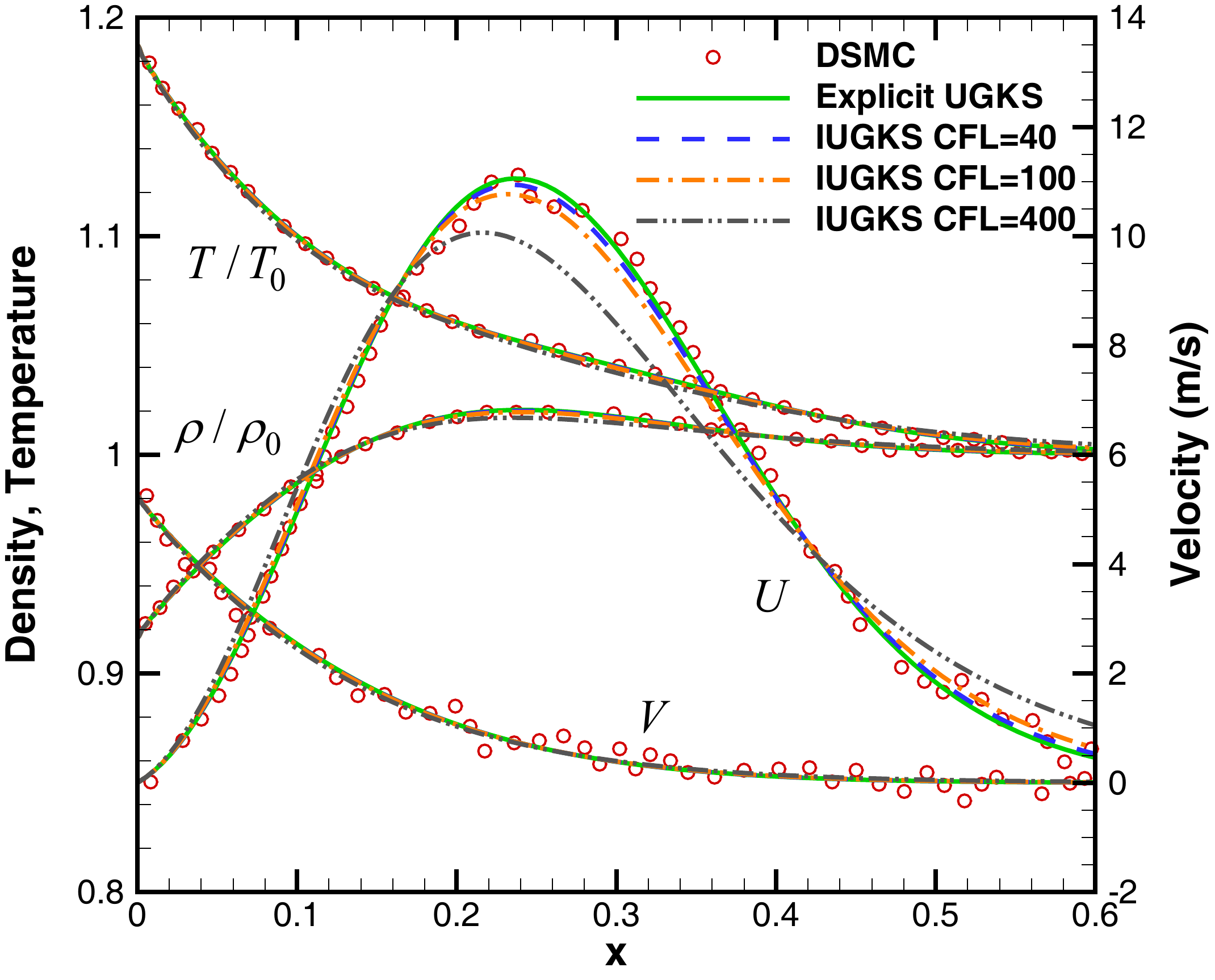}}\\
\subfigure[Kn=0.0266]{\includegraphics[width=0.48\textwidth]{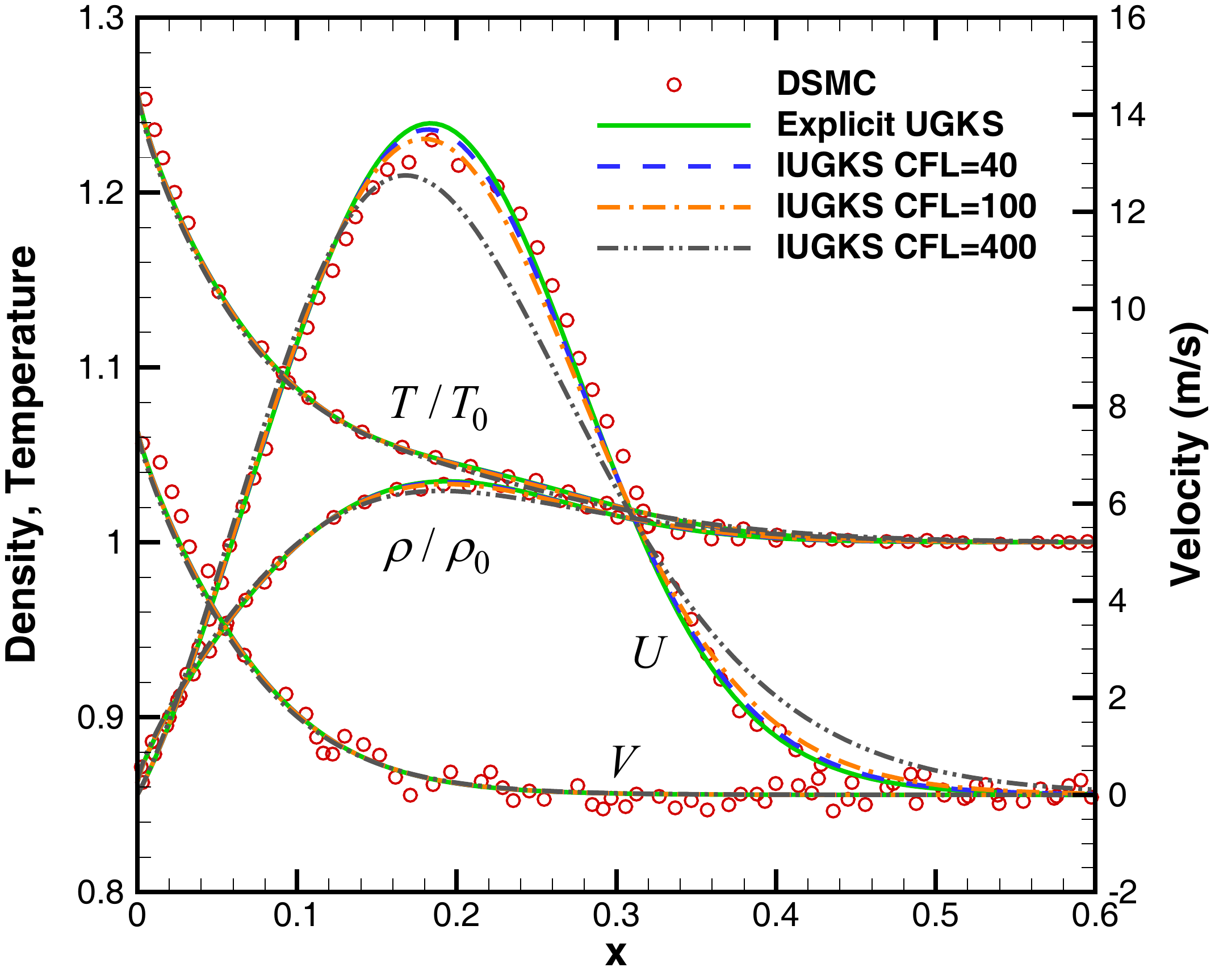}}
\subfigure[Kn=0.00266]{\includegraphics[width=0.48\textwidth]{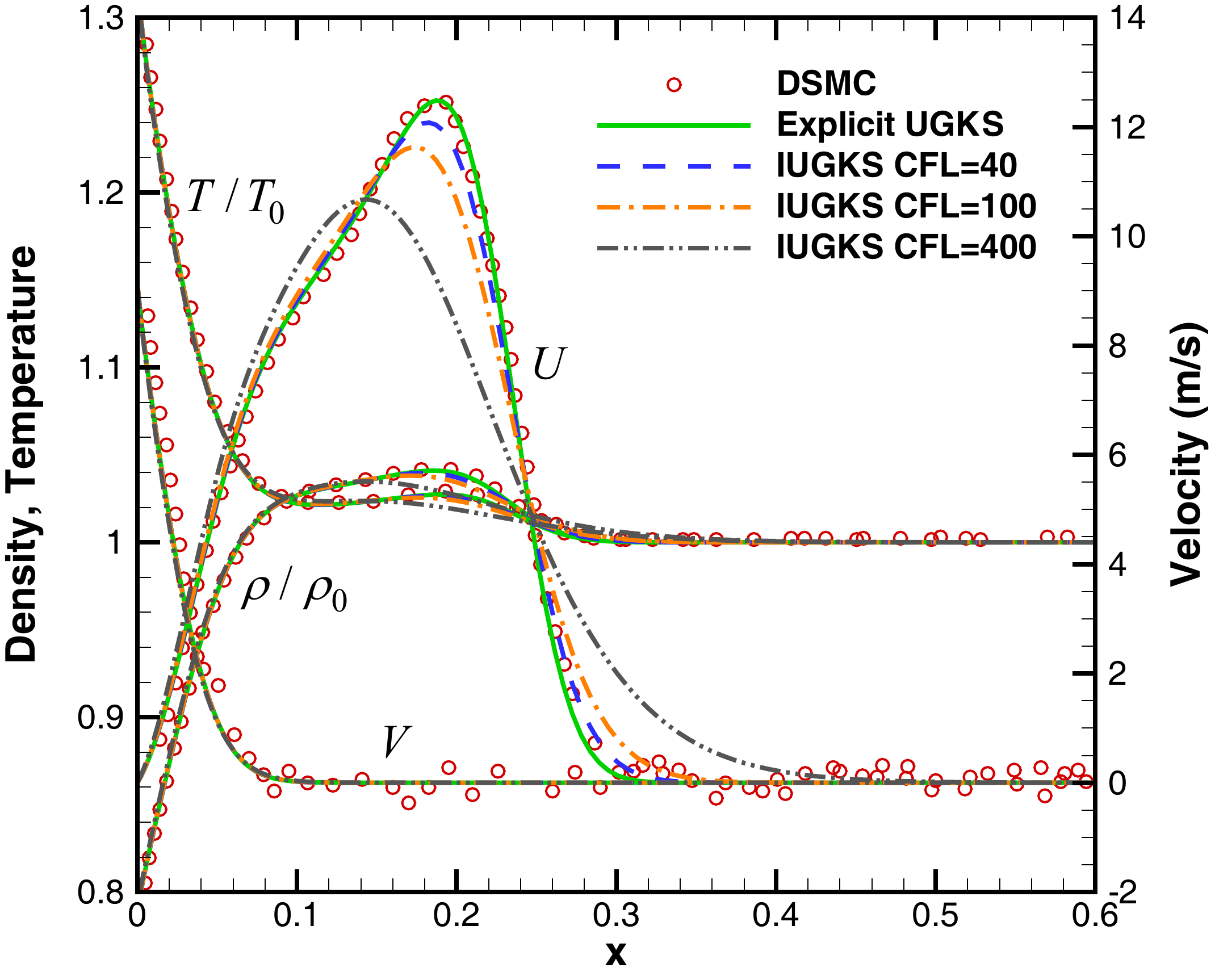}}
\caption{\label{fig:RayleighResults}Rayleigh problems at different Knudsen numbers which cover the free molecular, transition and continuum regimes. IUGKS here denotes the IUGKS-$\epsilon^{\prime}$ with $\epsilon=0.75$.}
\end{figure}

The target of the IUGKS is to release the time step restriction from small-size cells on non-uniform mesh and thus to accelerate overall computational efficiency. For this test,  we employ non-uniform mesh with the minimum cell size of $0.0005{\rm m}$ near the plate. The results at different Knudsen numbers are plotted in Fig.~\ref{fig:RayleighResults}, where the density and temperature are normalized by $\rho_0$ and $T_0$. In comparison with the DSMC results obtained from the reference paper \cite{huang2013unified} and those from the explicit UGKS simulation, IUGKS-$\epsilon^{\prime}$ with $\epsilon=0.75$ can give satisfactory results with a relative large time step. With the continuous increment of time step, the solutions gradually deviate from the reference ones due to the increase of temporal discretization error.
For the collisionless limit which requires the interval of discrete velocities satisfy $\Delta u \sim \Delta x / t$,
in order to get smooth solutions $600$ velocity points uniformly in $u$-direction and 100 points in $v$-direction to cover a range from $-2023{\rm m/s}$ to $2023{\rm m/s}$ are used in both directions. Extra more discrete points are employed for discretization of the physical and velocity space in this case, so that it gives more reliable results in the efficiency testing due to long time program running. We present the computational cost in Table.~\ref{tab:RayleighEfficiency} at different Knudsen numbers with various time steps. Generally more inner iterations are required in the small Knudsen number case. The increase of computational efficiency for near continuum flows would not be as much as that in rarefied cases, but it still can be about ten times faster than that of the explicit scheme with ${\rm CFL}=40$.
Since the IUGKS is stable enough, we also give the computational cost for case with ${\rm CFL}=1200$ as a reference even though the solutions under such a condition may not make any sense for a satisfactory solution.

\begin{table}[H]
\centering
\begin{threeparttable}[c]
\caption{\label{tab:RayleighEfficiency}Computational cost for Rayleigh Problem with different time steps.}
\begin{tabular*}{0.9\textwidth}{@{\extracolsep{\fill}}llcrrrr}
\toprule
& \multirow{2}{*}{Kn} & Explicit UGKS & \multicolumn{4}{c}{IUGKS-$\epsilon^{\prime}$ with $\epsilon = 0.75$}\\\cline{4-7}
& ~    & CFL=0.5           & 40    & 100   & 400   & 1200\\
\midrule
\multirow{4}{*}{\begin{tabular}{c} CPU time \\(min)\end{tabular}}
 &2.66      & 437.8  & 22.5 & 10.3  & 4.0  & 1.9\\
~&0.266     & 435.0  & 25.9 & 11.5  & 4.8  & 1.8\\
~&0.0266    & 436.7  & 31.0 & 14.3  & 7.1  & 3.3\\
~&0.00266   & 439.8  & 48.2 & 28.2  & 13.7  & 11.1\\
\midrule
\multirow{4}{*}{Speedup}
 &2.66      & 1.0     & 19.5  & 42.3 & 108.7 & 236.6\\
~&0.266     & 1.0     & 16.8  & 37.7 & 91.0 & 236.8\\
~&0.0266    & 1.0     & 14.1  & 30.6 & 61.1 & 132.4\\
~&0.00266   & 1.0     & 9.1  & 15.6 & 32.1 & 39.5\\
\bottomrule
\end{tabular*}
\end{threeparttable}
\end{table}

\begin{figure}[H]
\centering
\includegraphics[width=0.8\textwidth]{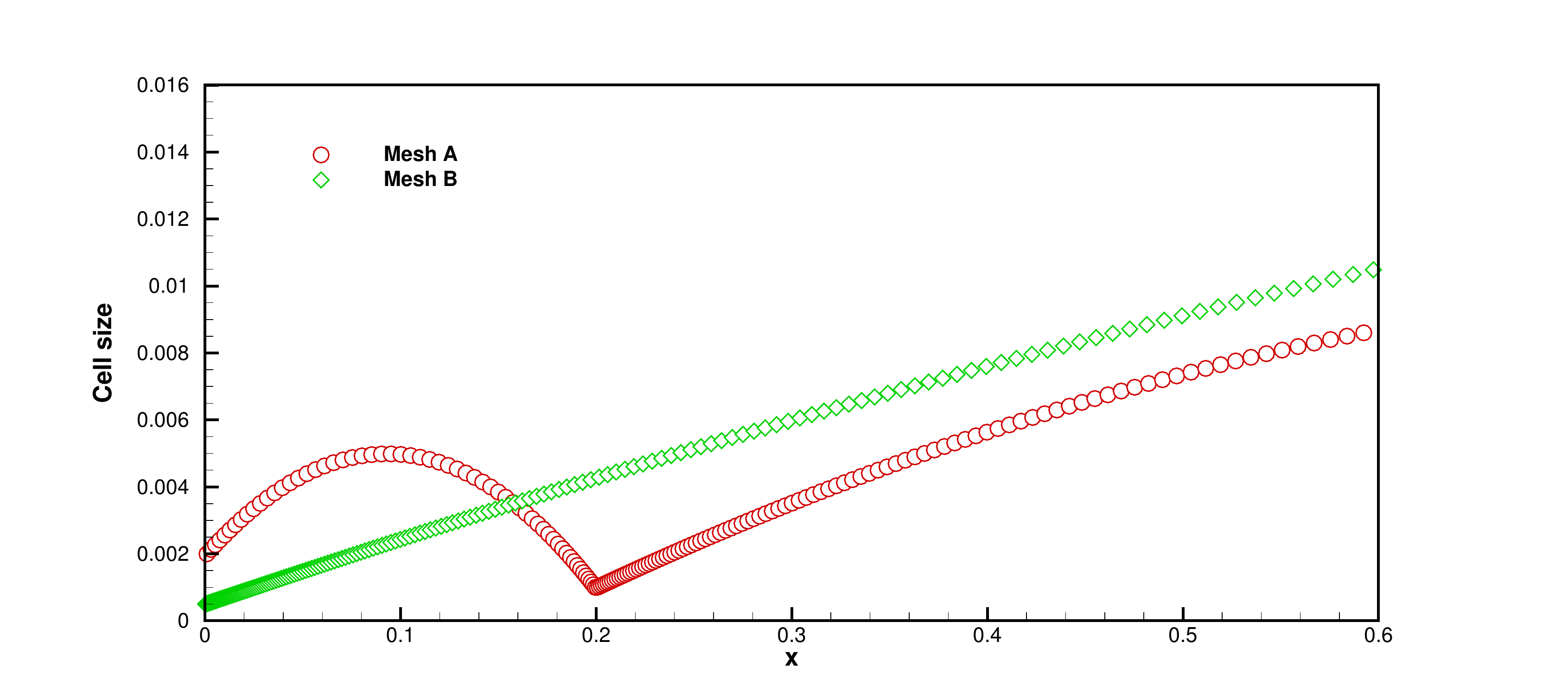}
\caption{\label{fig:meshShow}The distribution of the cell size used in simulation of Rayleigh problems.}
\end{figure}

\begin{figure}[H]
\centering
\subfigure[Kn=2.66]{\includegraphics[width=0.48\textwidth]{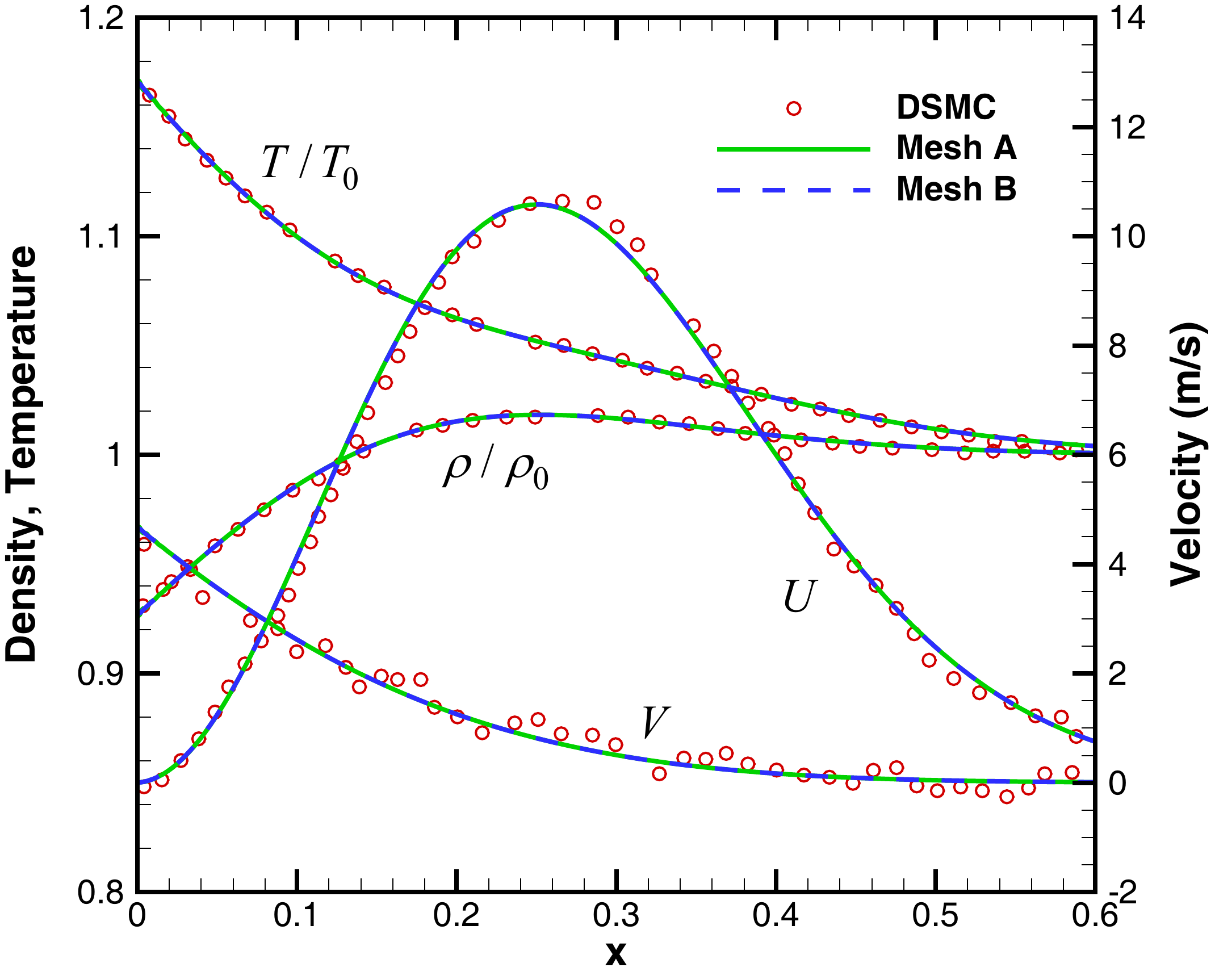}}
\subfigure[Kn=0.266]{\includegraphics[width=0.48\textwidth]{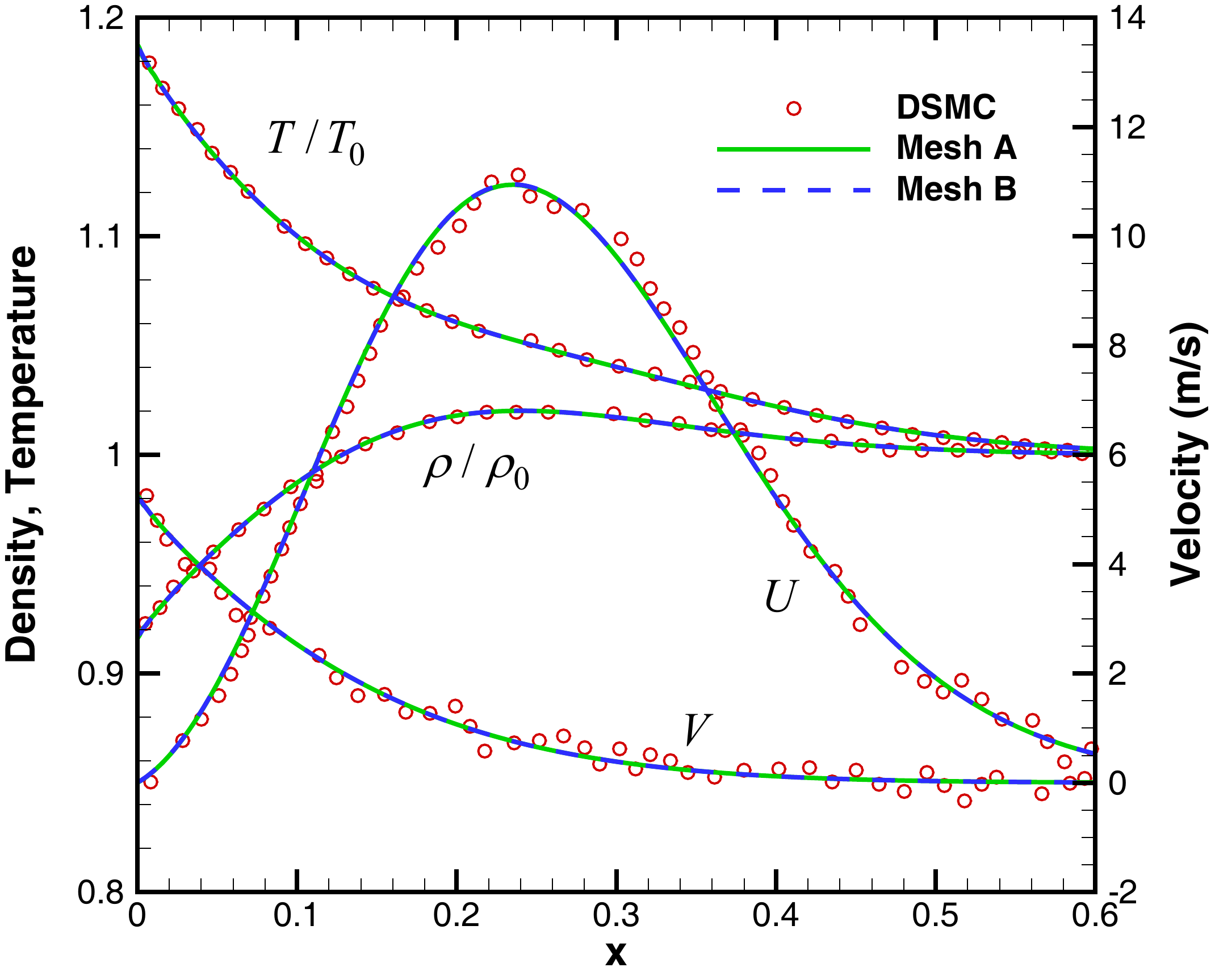}}\\
\subfigure[Kn=0.0266]{\includegraphics[width=0.48\textwidth]{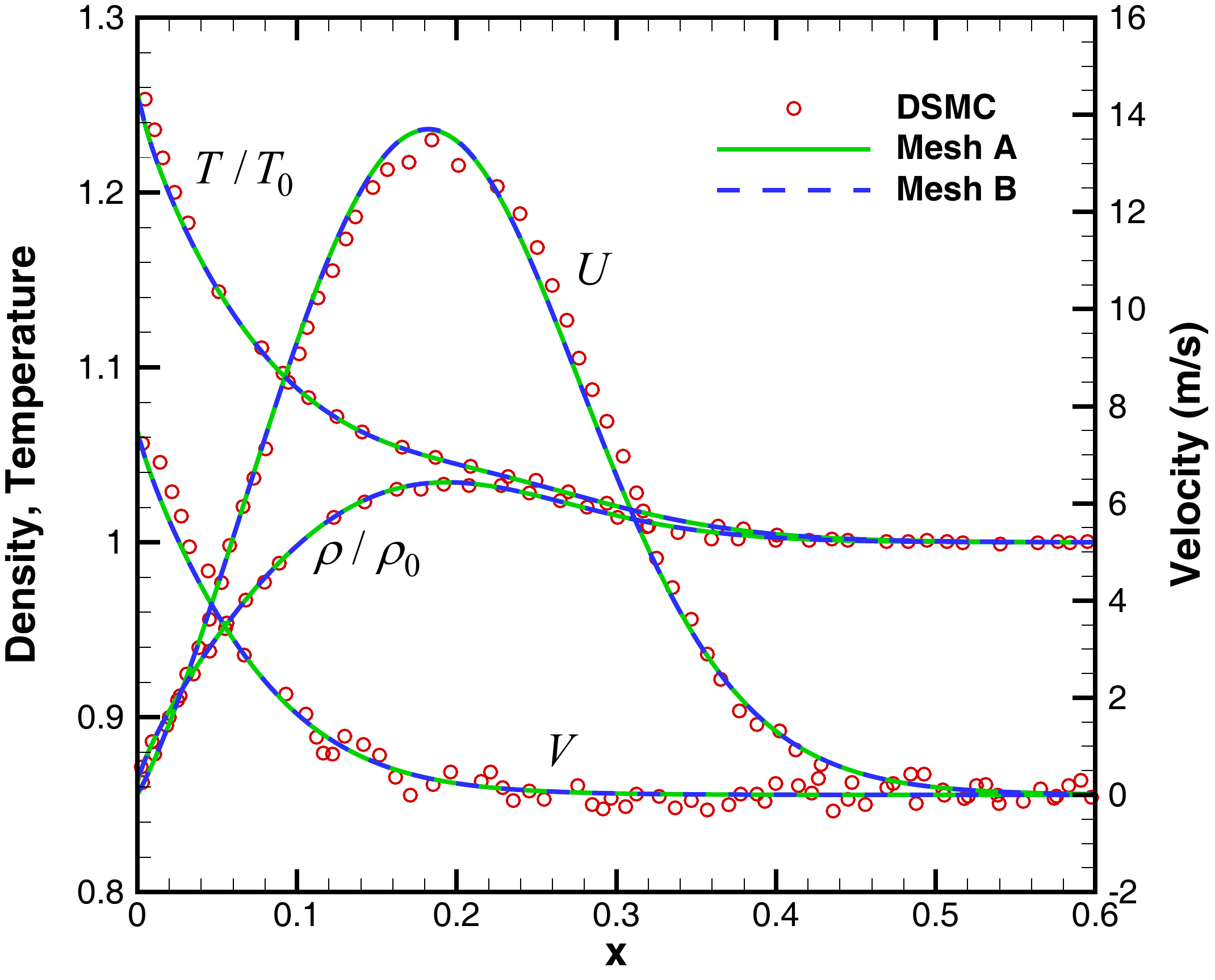}}
\subfigure[Kn=0.00266]{\includegraphics[width=0.48\textwidth]{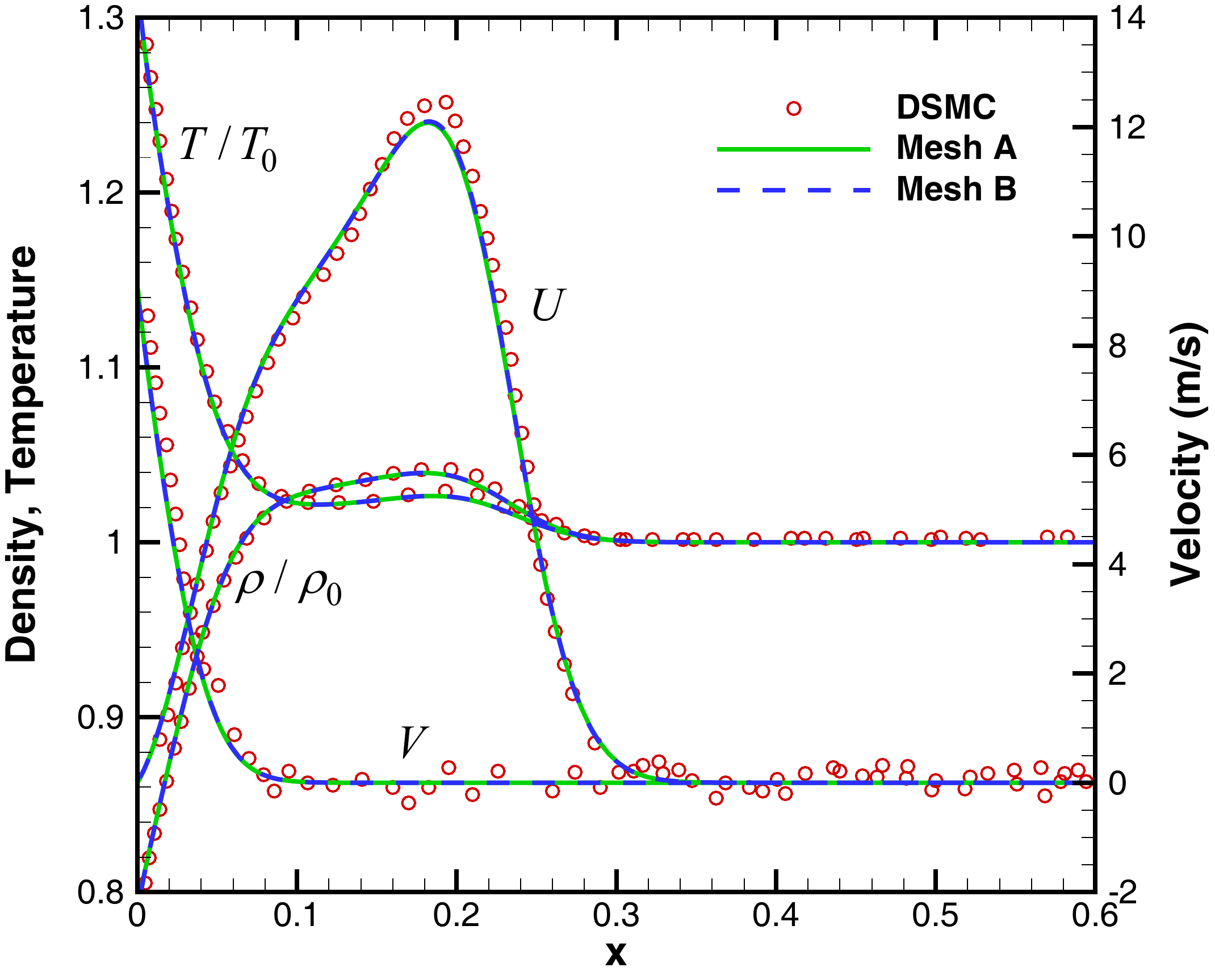}}
\caption{\label{fig:RayleighComparison}Comparison of the results for Rayleigh problems computed on different meshes with a fixed time step $\Delta t = 8.9{\rm \mu s}$.}
\end{figure}

In order to check the sensitivity of the IUGKS to the mesh size, the Rayleigh problems are computed on two kinds of meshes, whose cell size distributions are given in Fig.~\ref{fig:meshShow}. The mesh denoted by the green diamonds is used in previous calculations, and the one denoted by red circles has a cell size shrinking at $x=0.2{\rm m}$. Comparison between results obtained from these two meshes are given in Fig.~\ref{fig:RayleighComparison}, from which it can be observed that flow can efficiently spread through the small cell size region, and identical solutions to the previous ones can be obtained. A fixed time step $\Delta t = 8.9{\rm \mu s}$ used corresponds to the previous case at ${\rm CFL}=40$, which is shown in Fig.~\ref{fig:RayleighResults}.

\subsection{Wall bounded Rayleigh flow}\label{sec:wallRayleighCase}
\begin{figure}[H]
\centering
\includegraphics[width=0.85\textwidth]{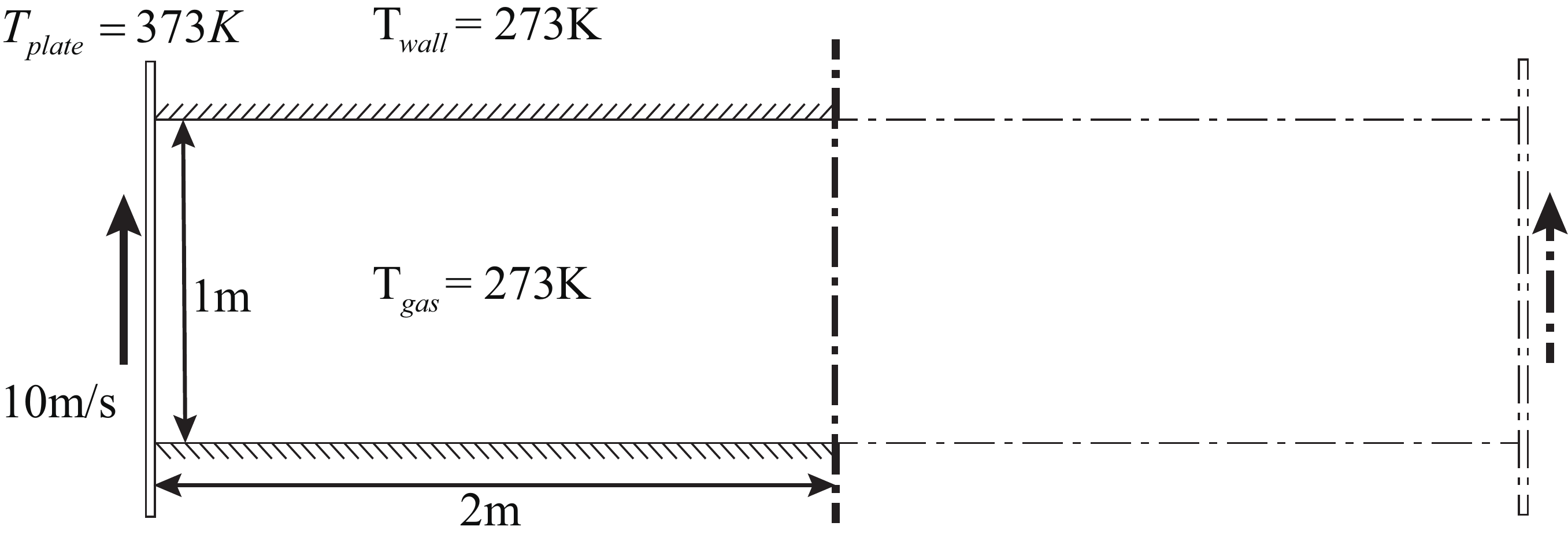}
\caption{\label{fig:schematic}The schematic for wall bounded Rayleigh flow.}
\end{figure}

The Rayleigh problems tested in Section \ref{sec:RayleighCase} are one-dimensional, where periodic boundary conditions are imposed in the $y$-direction. Here we added two parallel solid walls along x-direction at two locations separated by $1{\rm m}$ in the $y$-axis,
which restrict the movement of the flow in the vertical direction.
The schematic for the setting is shown in Fig.~\ref{fig:schematic}.
The lower and upper solid walls have a length of $4{\rm m}$ and a constant temperature of $T_{wall} = 273{\rm K}$.
The argon gas has molecular mass $6.63\times 10^{-26} {\rm kg}$, which corresponds to a specific gas constant of $R=208.13 {\rm J/(kg \cdot K)}$.  The initial gas temperature is $T_{0} = 273{\rm K}$. The side plates move vertically at $10{\rm m/s}$ with a higher temperature of $373{\rm K}$.  The Maxwellian diffusive boundary condition is imposed on the solid isothermal walls. In addition, symmetric boundary condition is imposed at $x=2{\rm m}$, so that half domain can be used in computation. The case of $Kn=0.05$ is considered here.

\begin{figure}[H]
\centering
\subfigure[]{\includegraphics[width=0.48\textwidth]{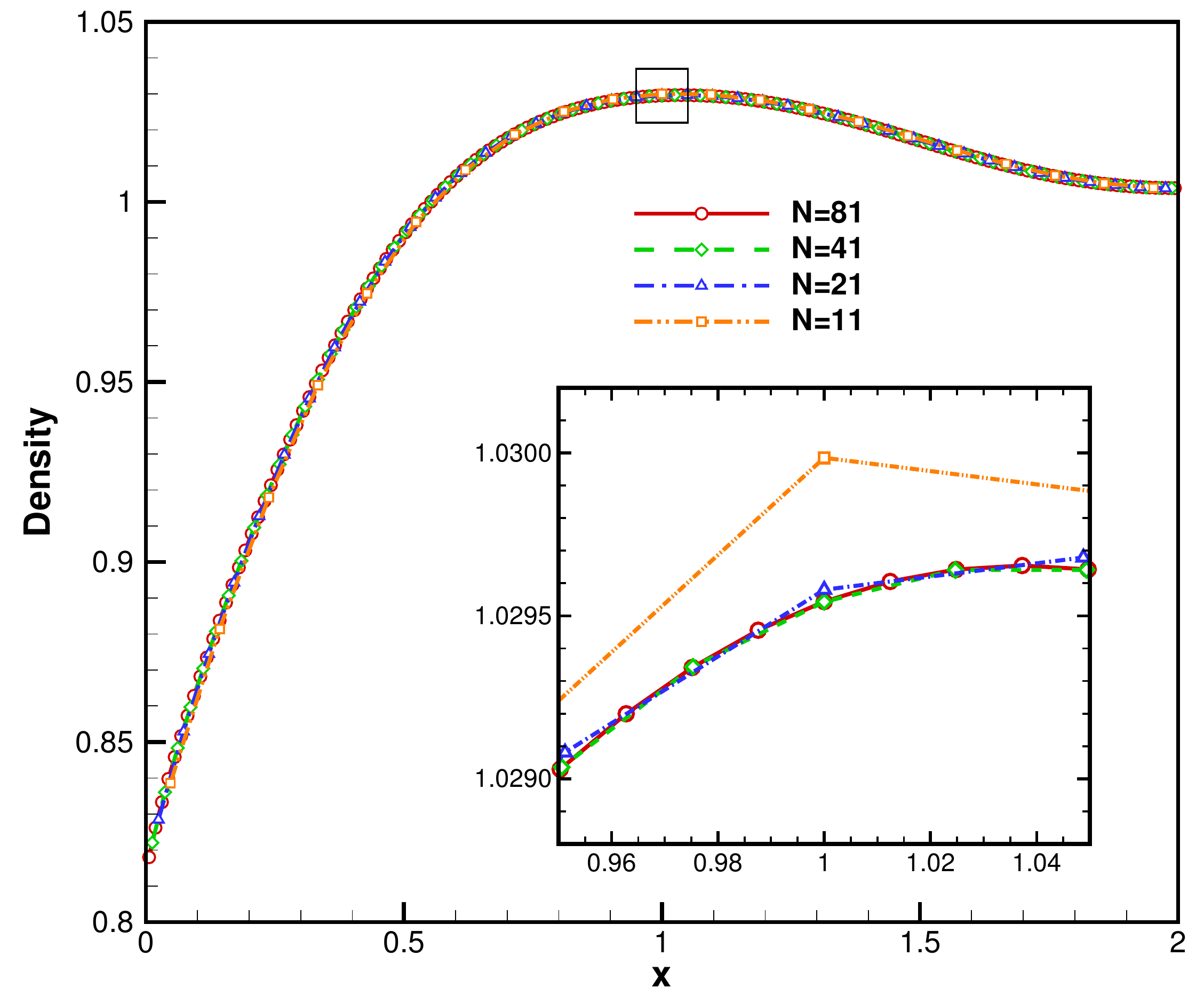}}	
\subfigure[]{\includegraphics[width=0.48\textwidth]{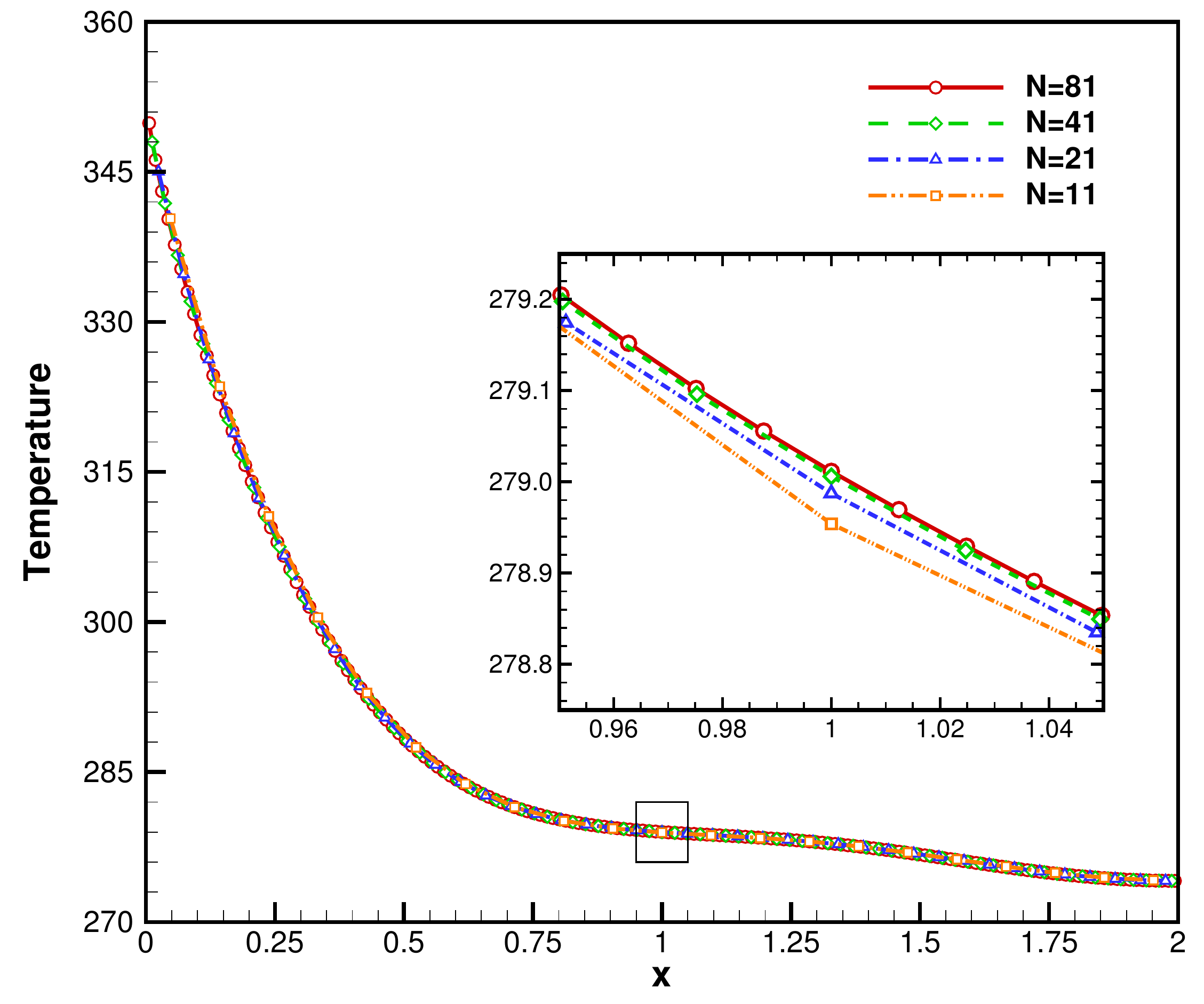}}\\
\subfigure[]{\includegraphics[width=0.48\textwidth]{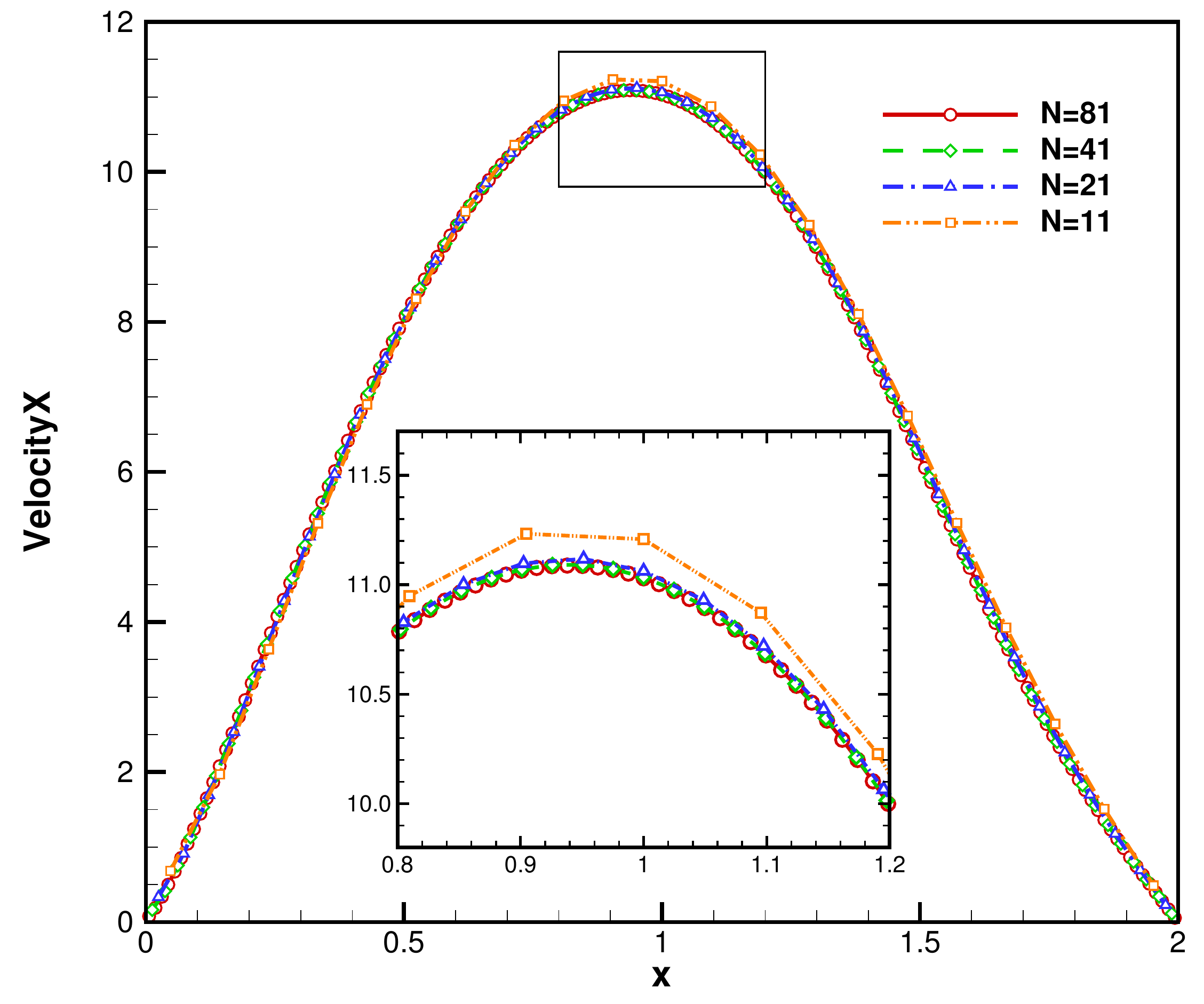}}
\subfigure[]{\includegraphics[width=0.48\textwidth]{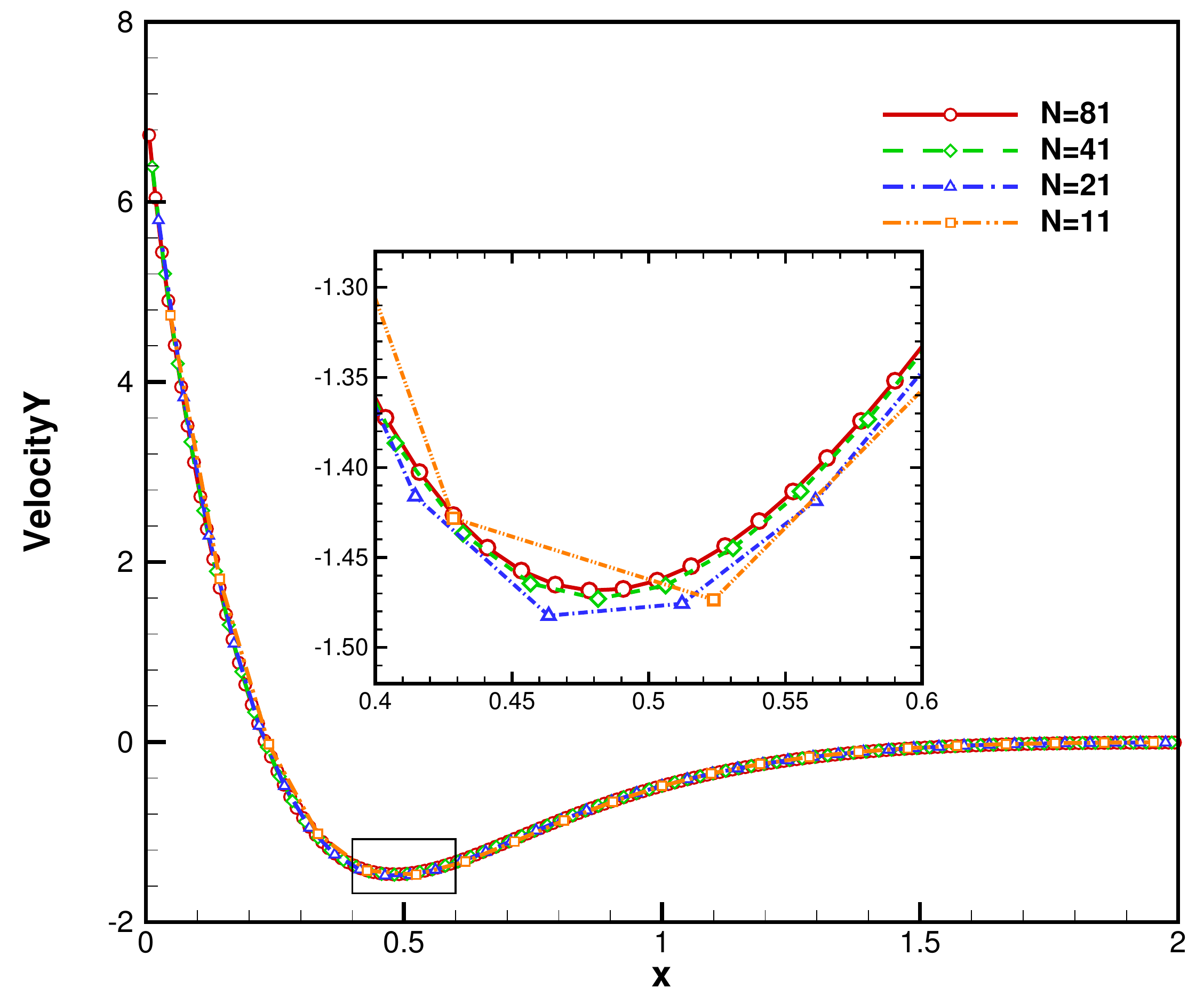}}
\caption{\label{fig:meshConvergeX}Flow variables along the central horizontal line obtained on different meshes. $N$ denotes the discrete cell number along the vertical plate.}
\end{figure}

\begin{figure}[H]
	\centering
	\subfigure[]{\includegraphics[width=0.48\textwidth]{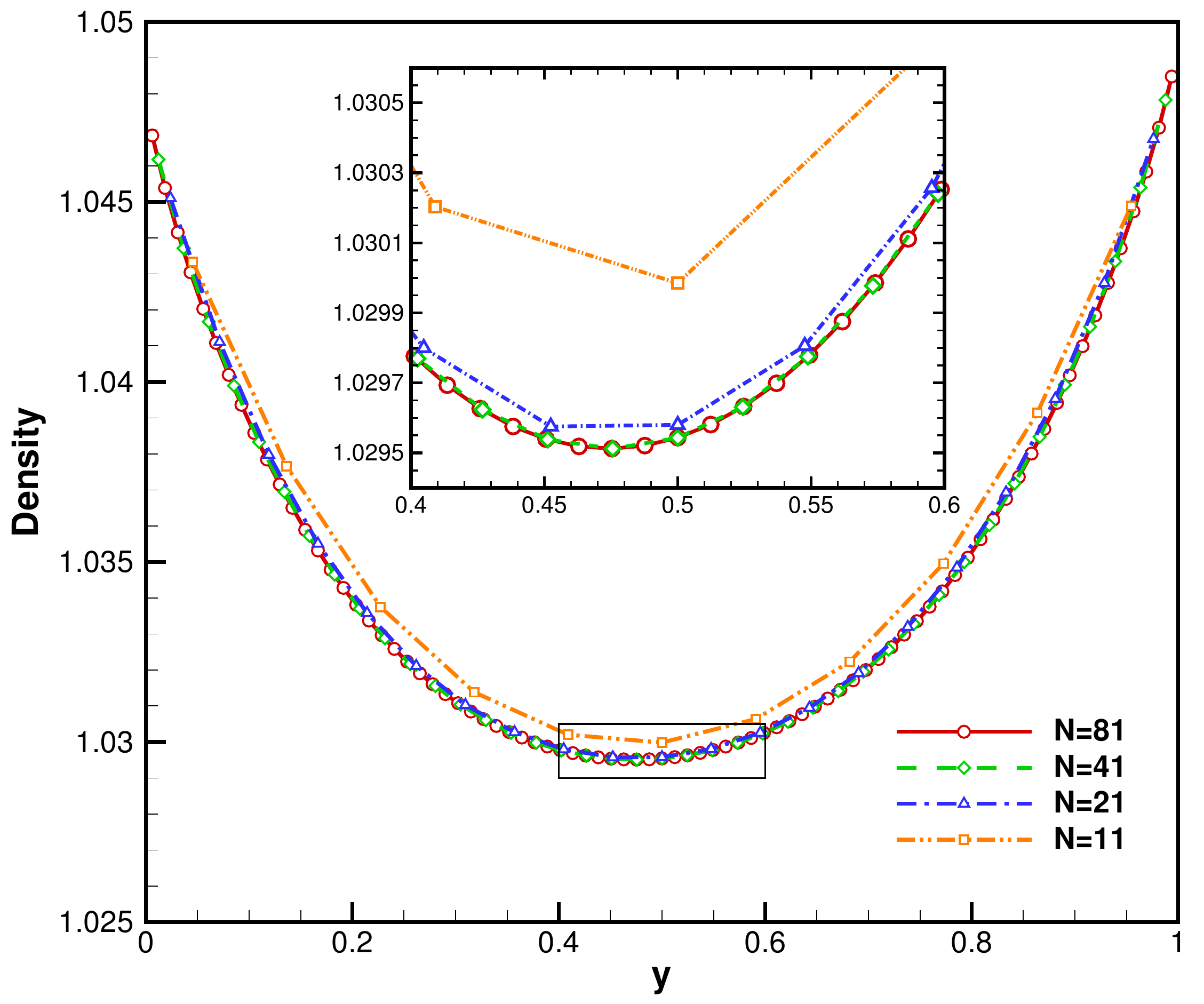}}	
	\subfigure[]{\includegraphics[width=0.48\textwidth]{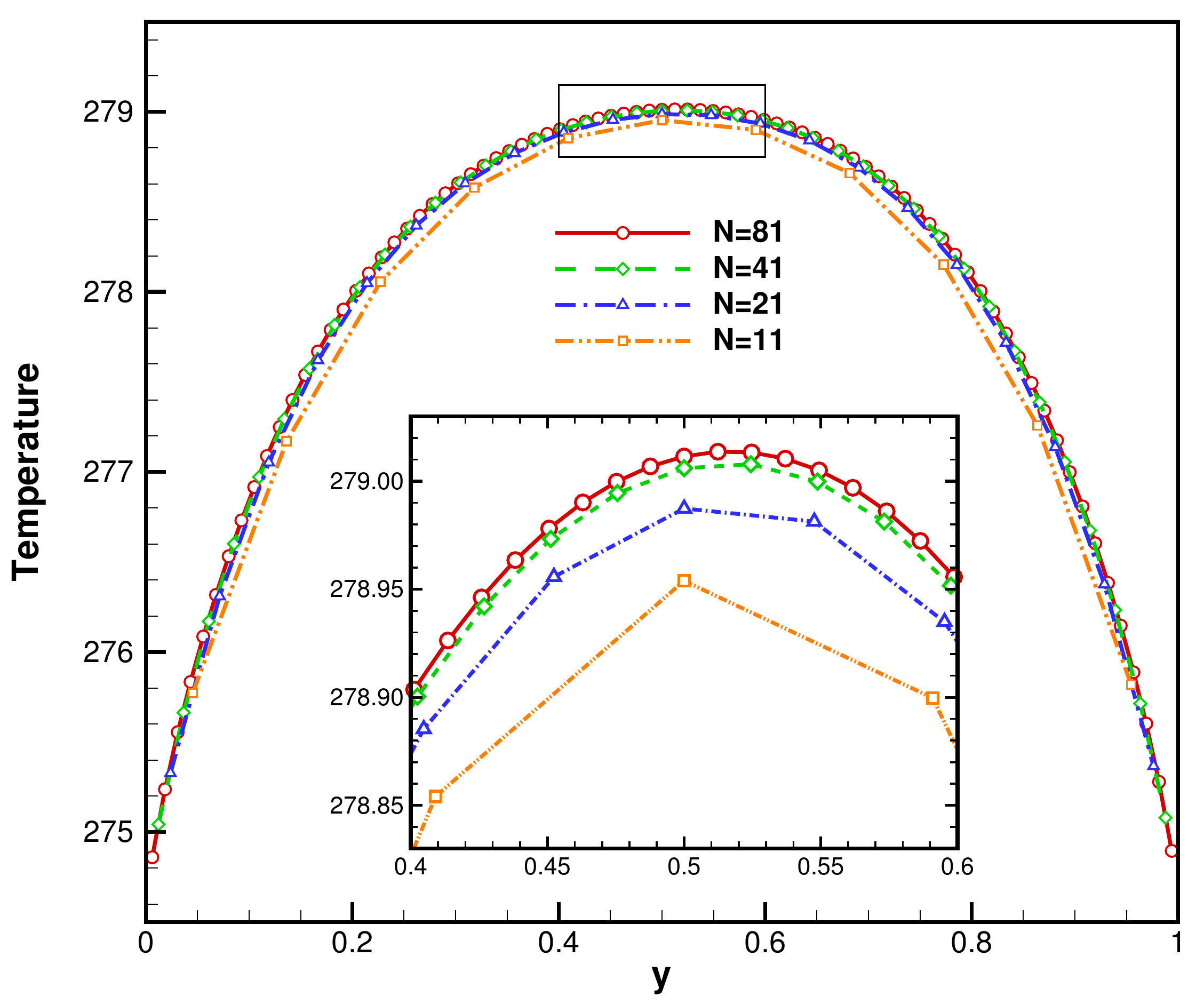}}\\
	\subfigure[]{\includegraphics[width=0.48\textwidth]{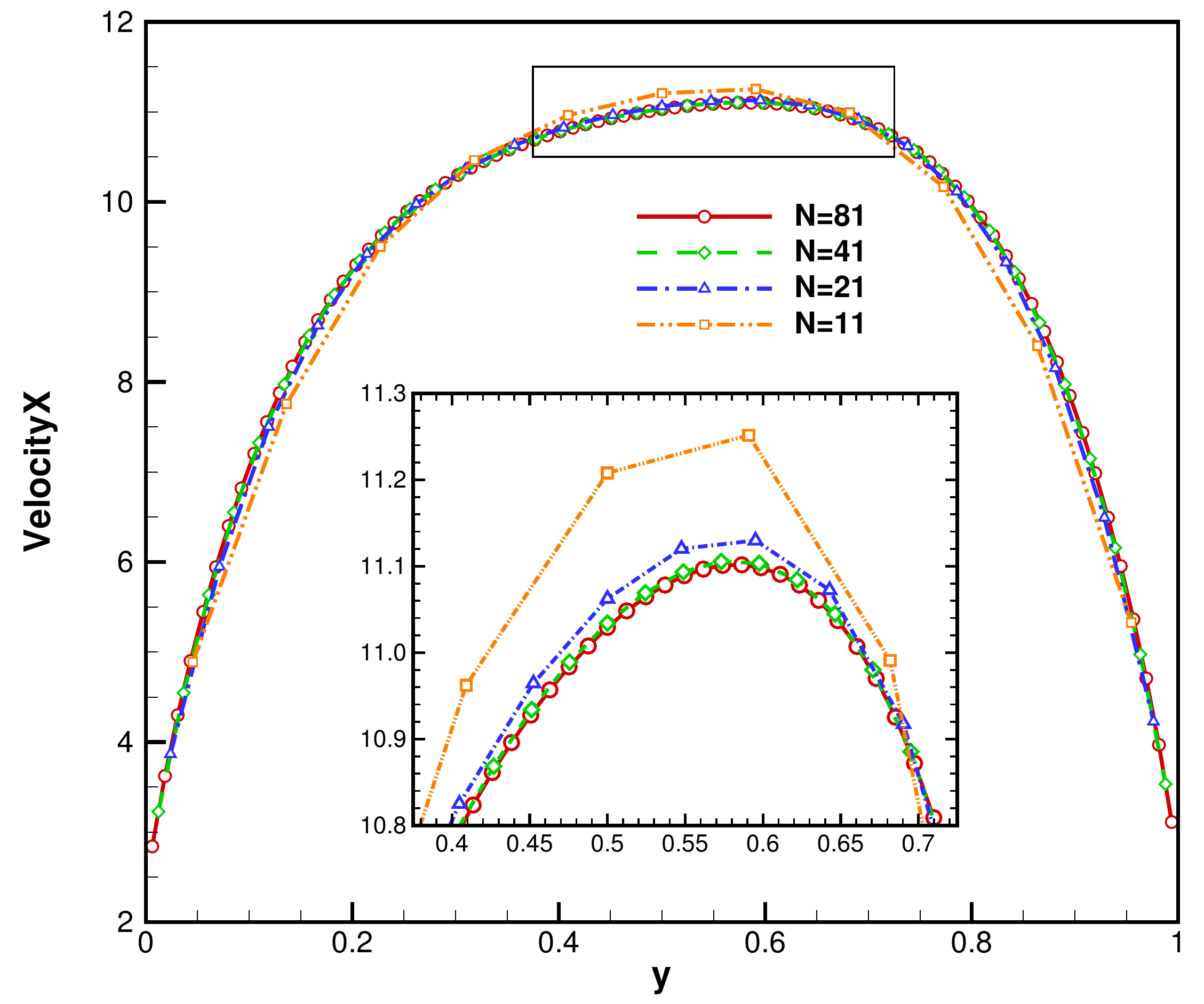}}
	\subfigure[]{\includegraphics[width=0.48\textwidth]{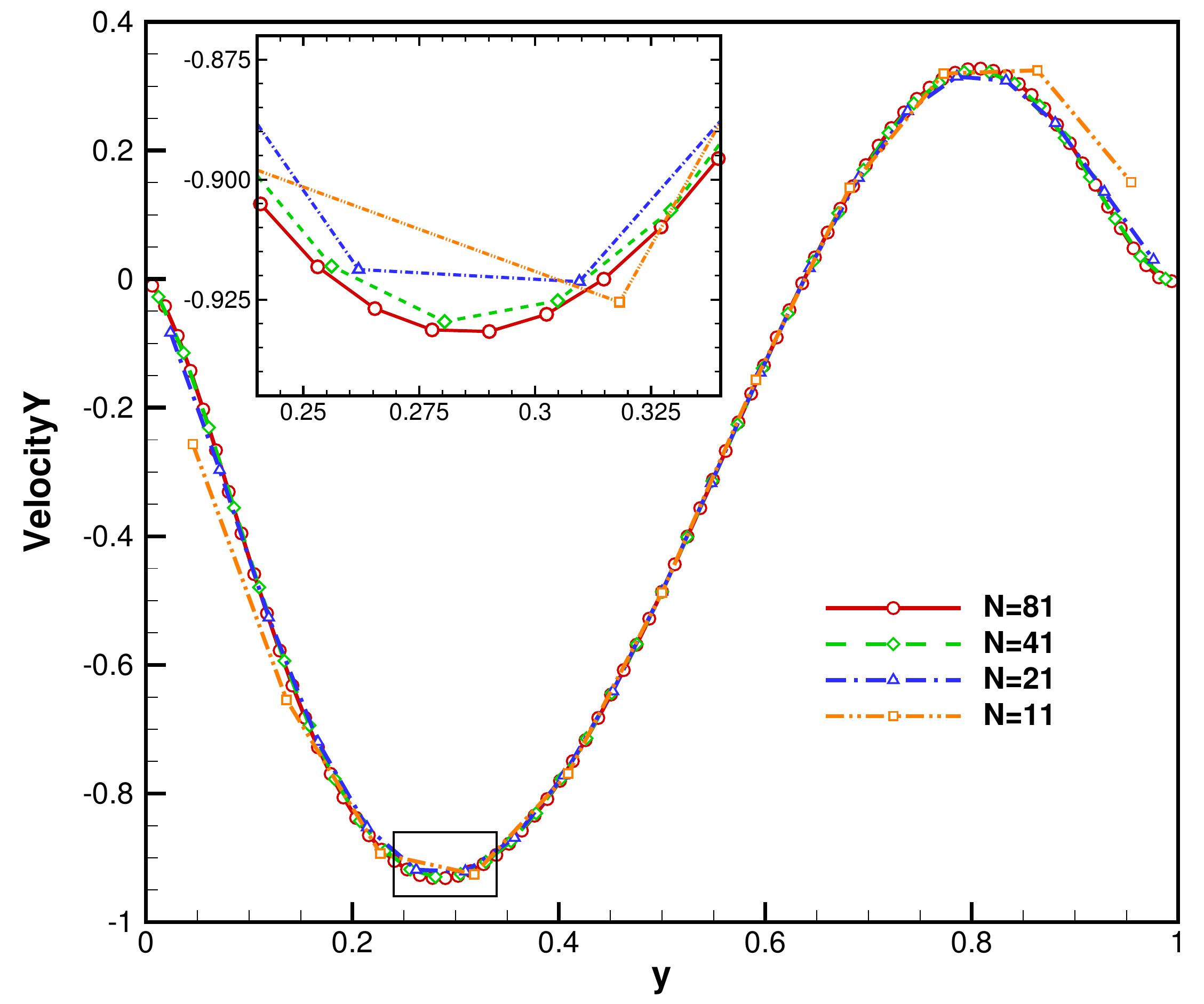}}
	\caption{\label{fig:meshConvergeY}Flow variables along the central vertical line obtained on different meshes. $N$ denotes the discrete cell number along the vertical plate.}
\end{figure}

\begin{figure}[H]
	\centering
	\subfigure[]{\includegraphics[width=0.48\textwidth]{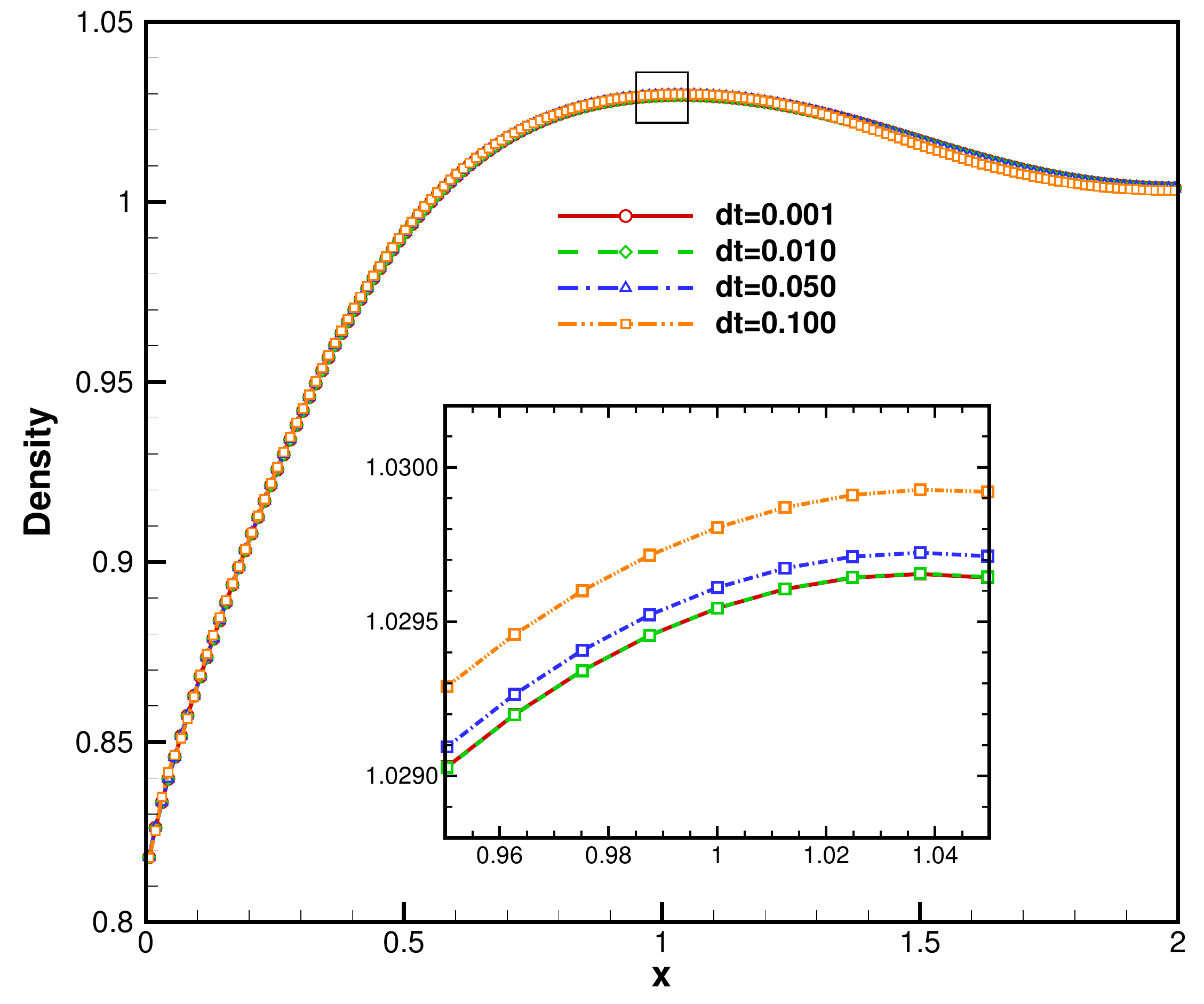}}	
	\subfigure[]{\includegraphics[width=0.48\textwidth]{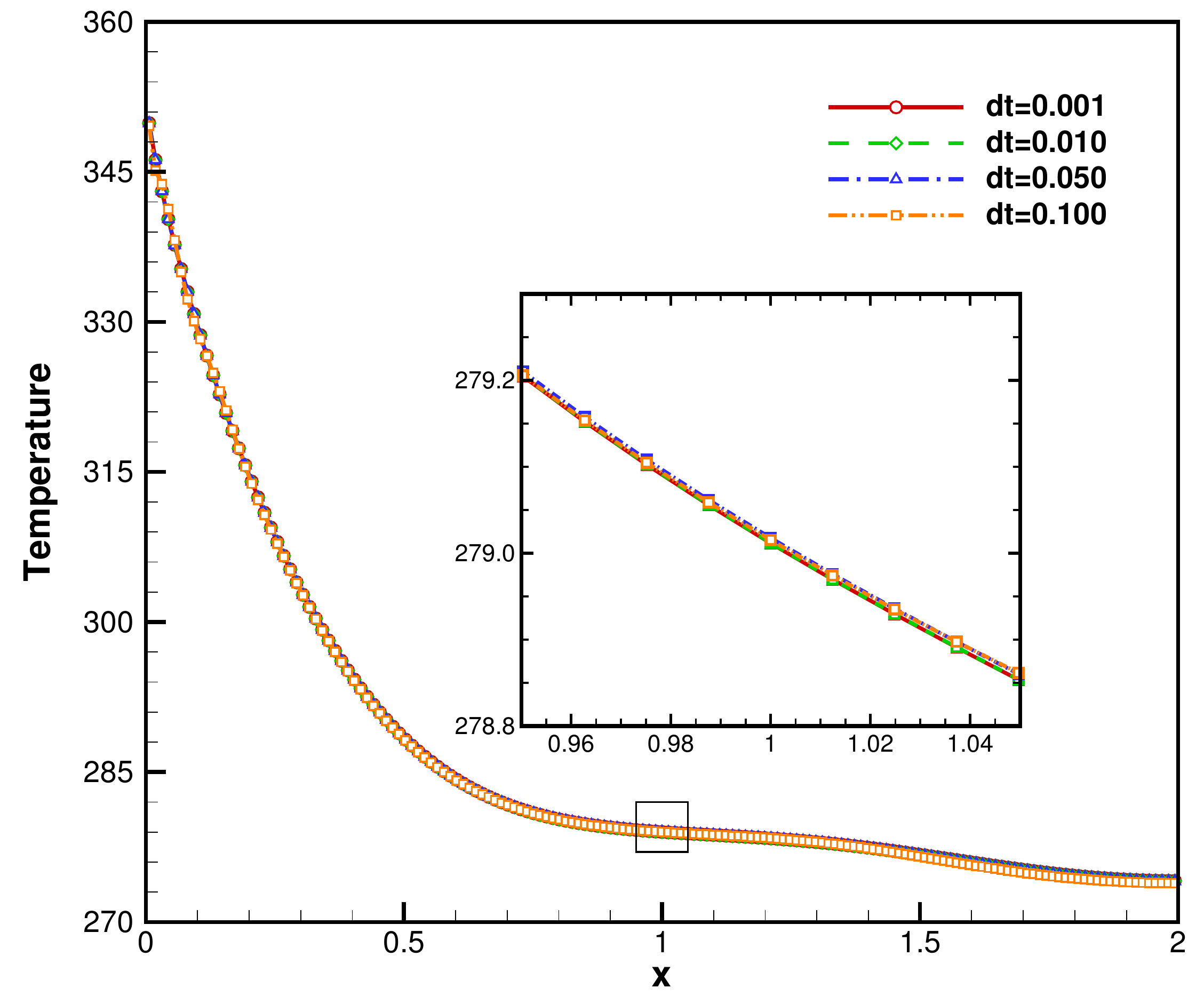}}\\
	\subfigure[]{\includegraphics[width=0.48\textwidth]{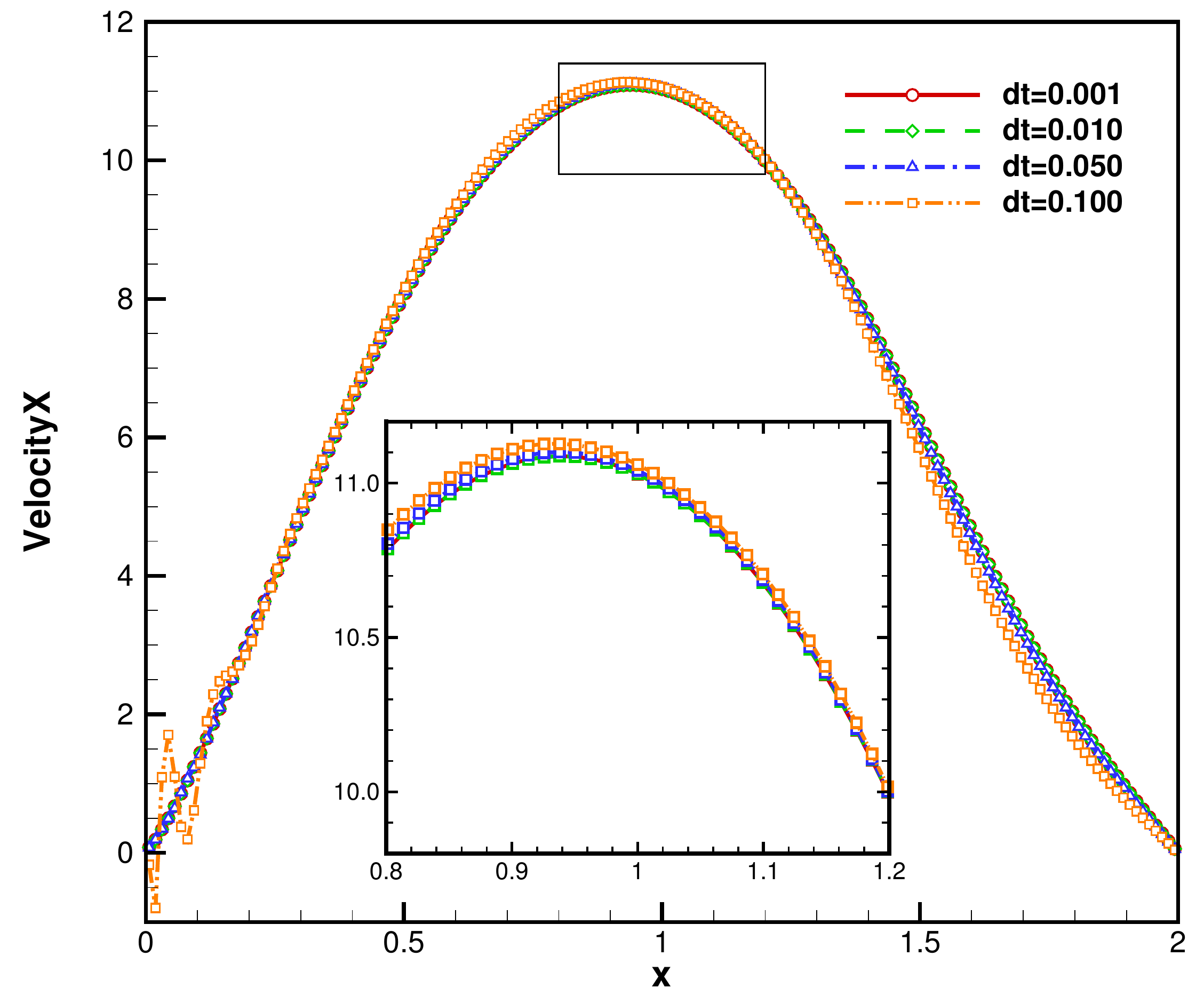}}
	\subfigure[]{\includegraphics[width=0.48\textwidth]{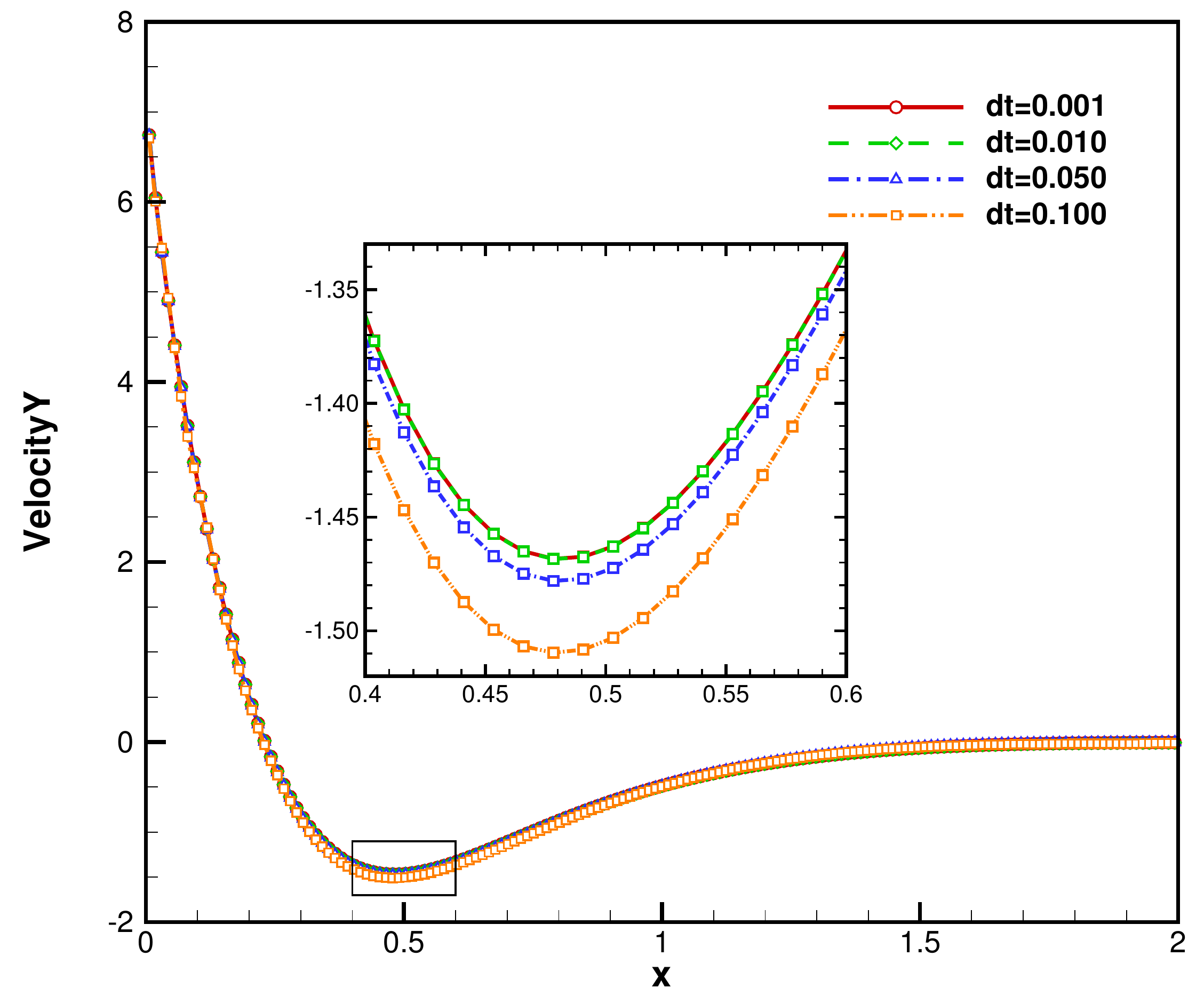}}
	\caption{\label{fig:timeConvergeX}Flow variables along the central horizontal line obtained with different time steps.}
\end{figure}

\begin{figure}[H]
	\centering
	\subfigure[]{\includegraphics[width=0.48\textwidth]{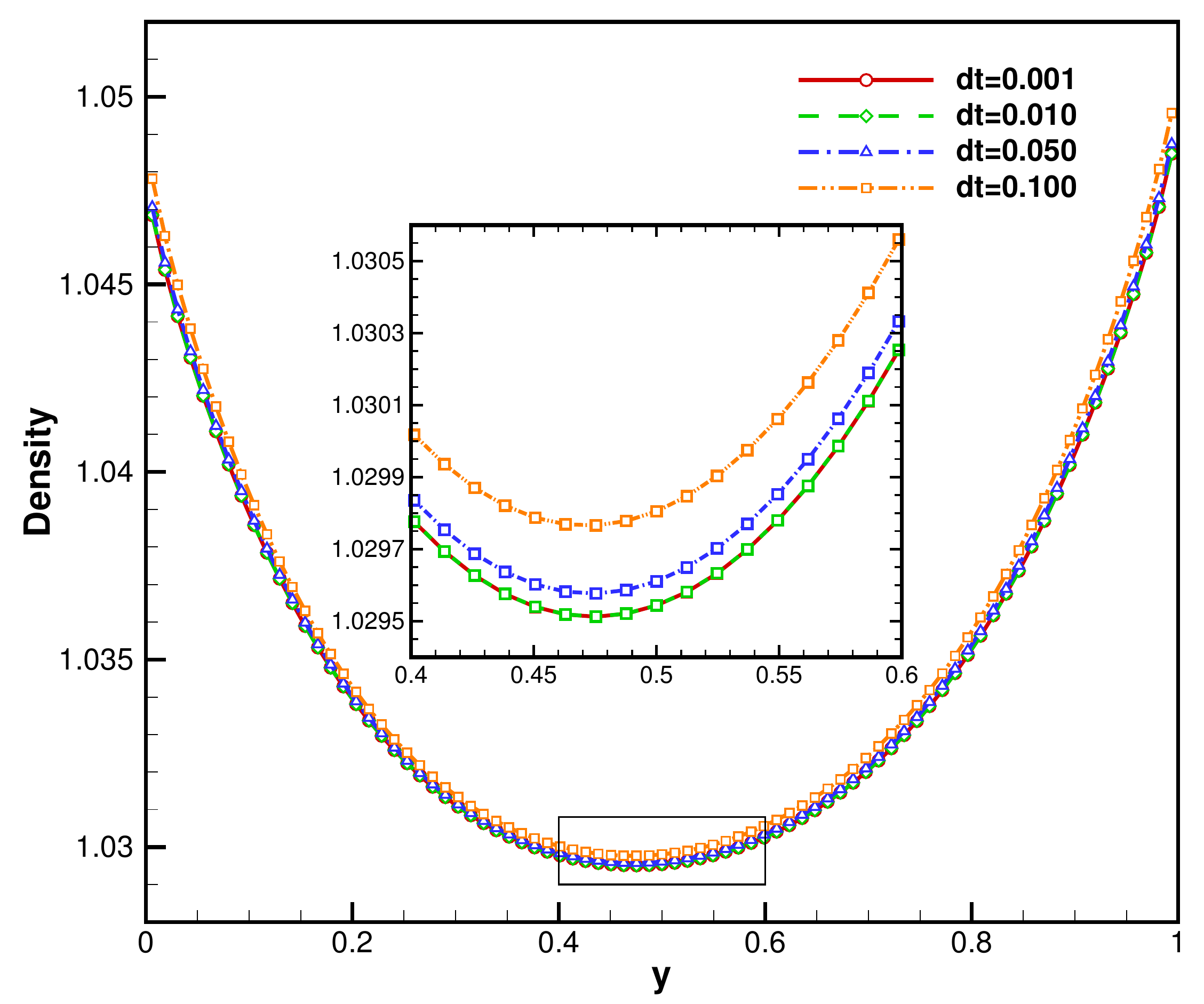}}	
	\subfigure[]{\includegraphics[width=0.48\textwidth]{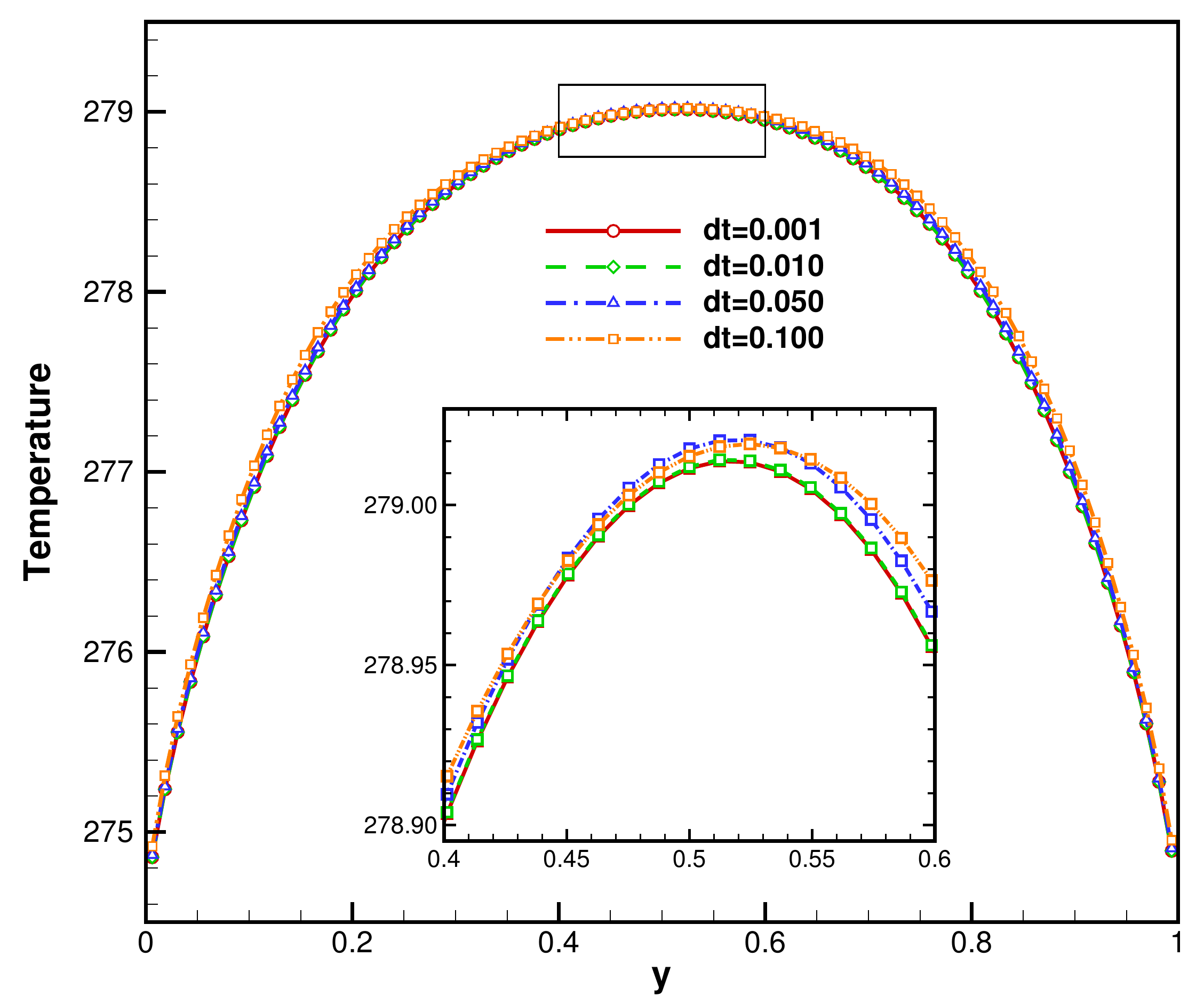}}\\
	\subfigure[]{\includegraphics[width=0.48\textwidth]{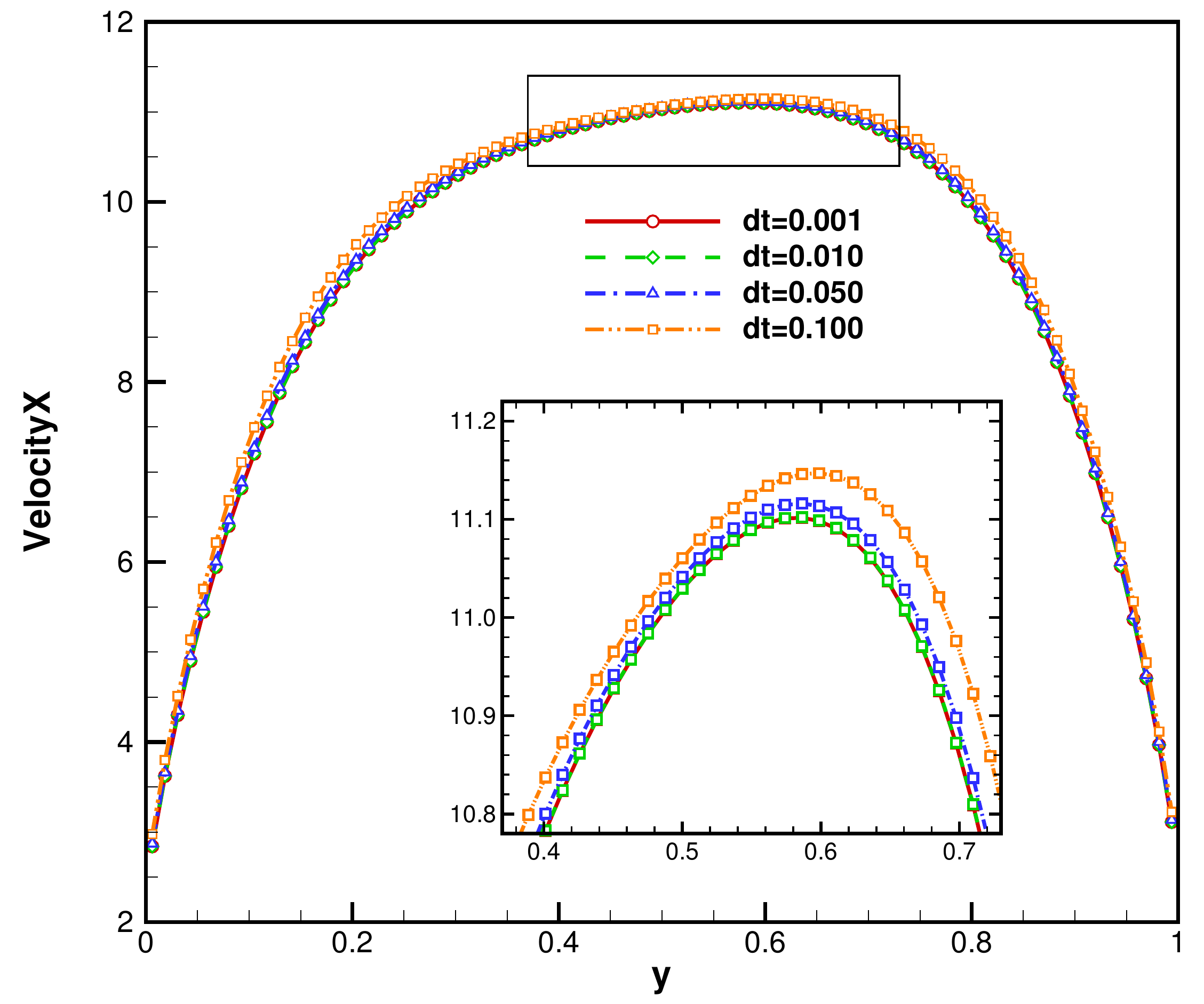}}
	\subfigure[]{\includegraphics[width=0.48\textwidth]{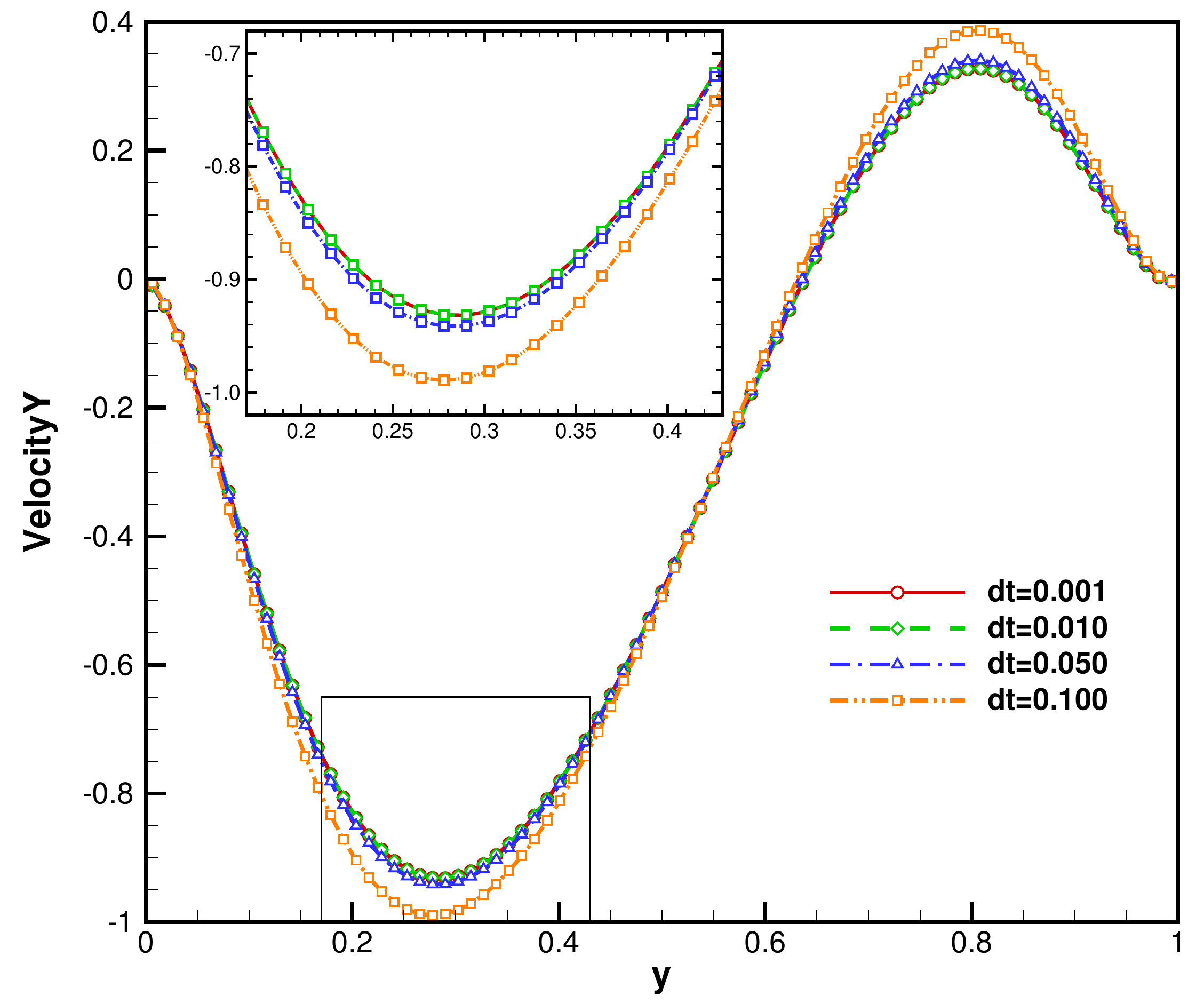}}
	\caption{\label{fig:timeConvergeY}Flow variables along the central vertical line obtained with different time steps.}
\end{figure}

In order to check the convergence of solutions with respect to the cell size and the time step, numerical simulations are carried out on different meshes using different time steps. The spatial discretizations used in the test case are $11\times21$, $21 \times 41$, $41 \times 81$ and $81 \times 161$. In the following, these meshes will be denoted by $N=11$, $N=21$, $N=41$ and $N=81$ for simplicity, where $N$ means the number of discrete cells along the vertical plate. In this case, the non-dimensional time and time step are used, and the normalization is based on a reference time scale $t_0 = L_0/U_0$, where $L_0 = 1{\rm m}$ and $U_0 = \sqrt{2 R T_0}$. The time steps used in these cases for the IUGKS are $\Delta t = 0.01$, $0.05$ and $0.1$. As a reference, the cases with $\Delta t = 0.001$ are computed as well using the explicit UGKS.
Flow variables along the central vertical and horizontal lines ($x=1{\rm m}$ and $y=0.5{\rm m}$) at $t=1.5$ are monitored to illustrate the mesh convergence solutions of the unsteady flow, where the results are plotted in Fig.~\ref{fig:meshConvergeX} and Fig.~\ref{fig:meshConvergeY}.
It shows that the results on the mesh of $N=41$ have little difference from those on the mesh of $N=81$,
which can be regarded as a mesh convergent solution. Flow variables along central lines obtained using different time steps are given in Fig.~\ref{fig:timeConvergeX} and Fig.~\ref{fig:timeConvergeY} which show the convergence of the evolving solutions using different numerical time steps.

\begin{figure}[H]
	\centering
	\subfigure[$t=0.5$]{\includegraphics[width=0.48\textwidth]{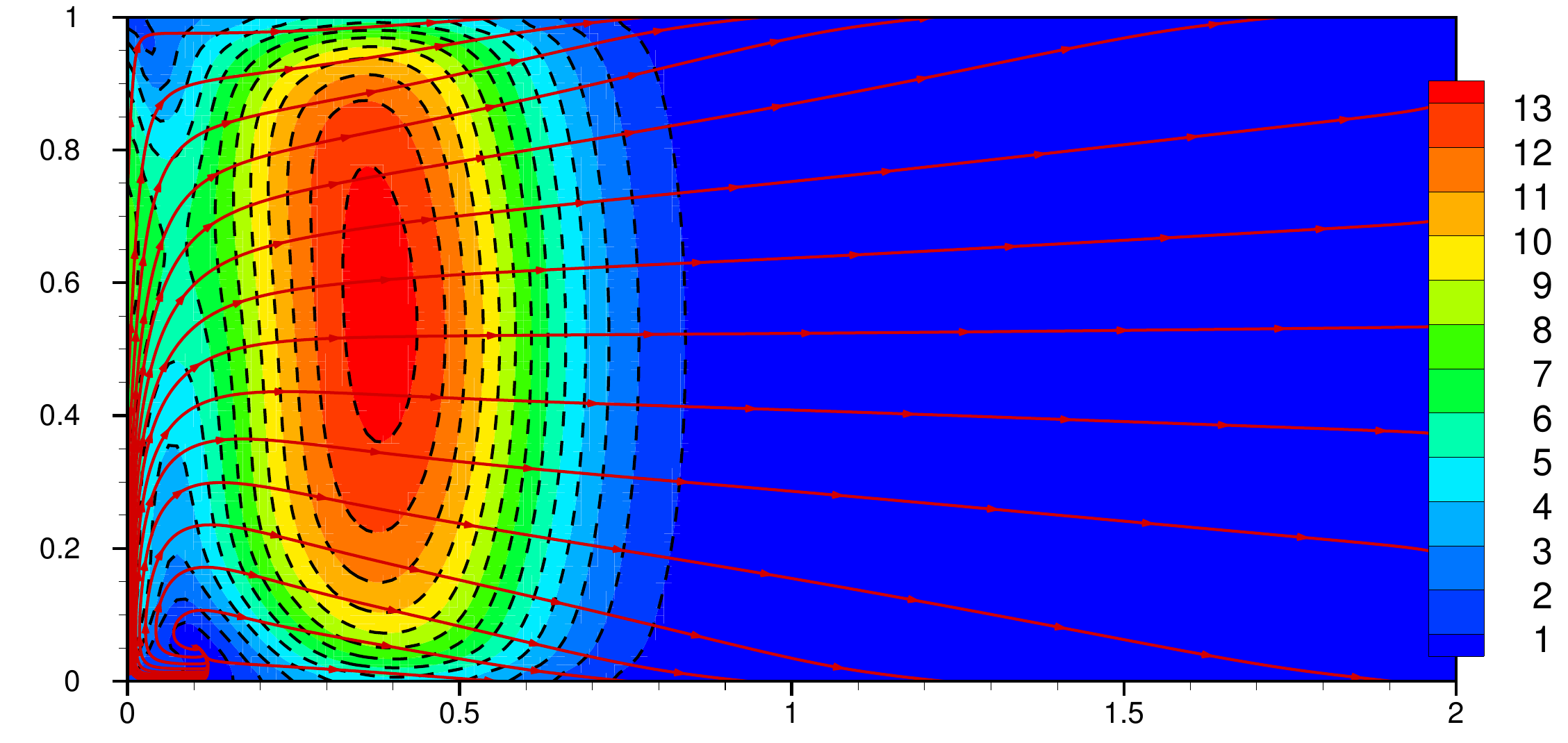}}
	\subfigure[$t=1.0$]{\includegraphics[width=0.48\textwidth]{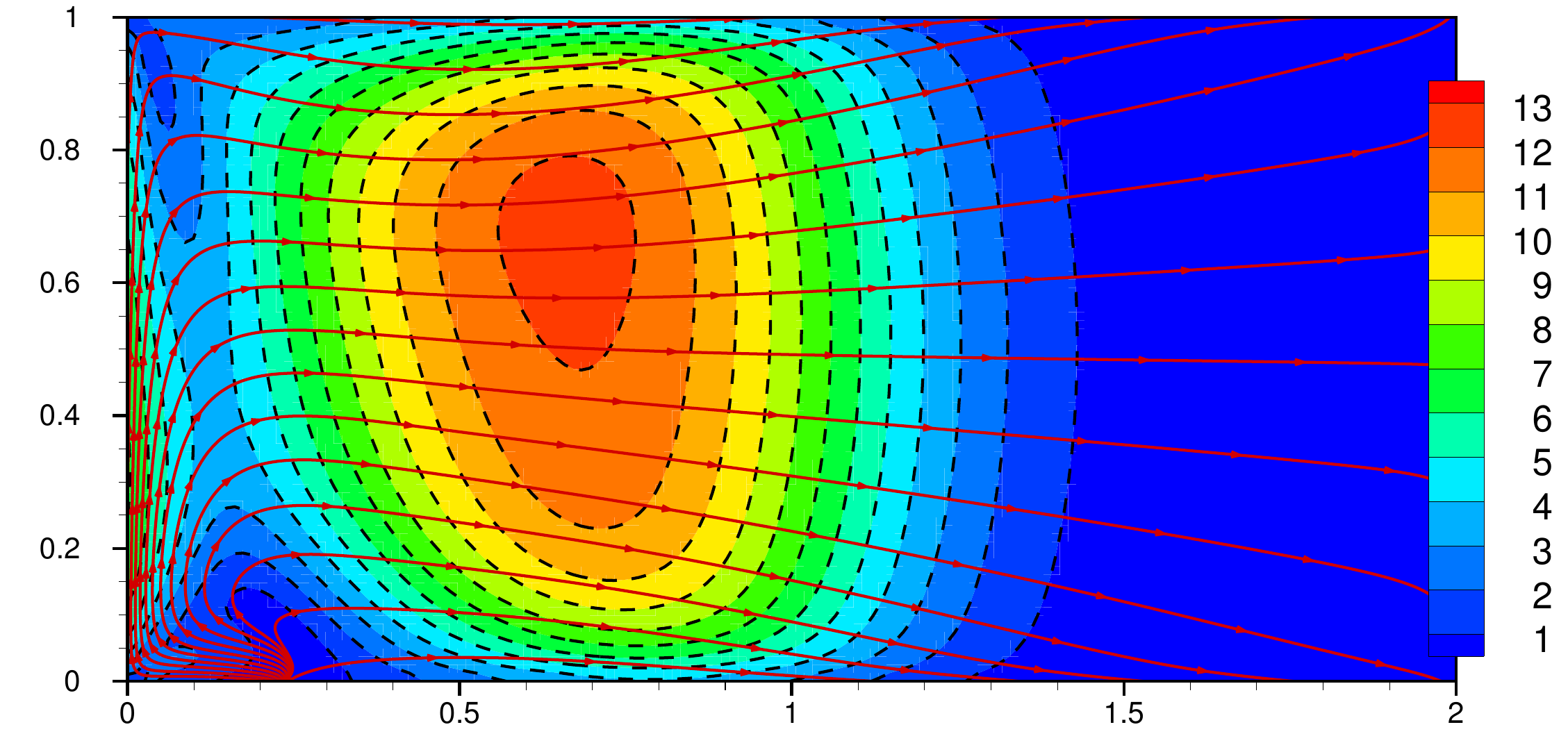}}\\
	\subfigure[$t=1.5$]{\includegraphics[width=0.48\textwidth]{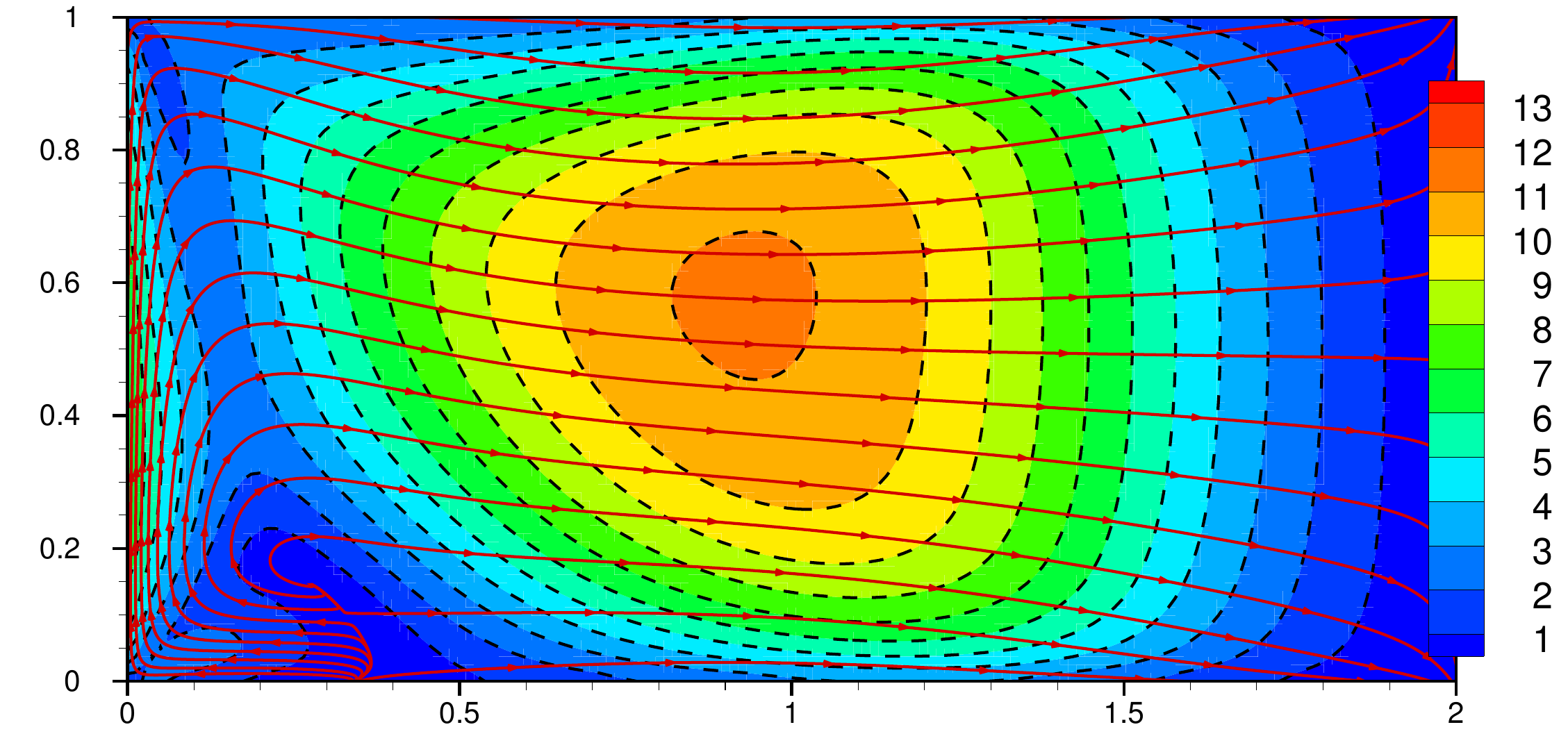}}
	\subfigure[$t=2.0$]{\includegraphics[width=0.48\textwidth]{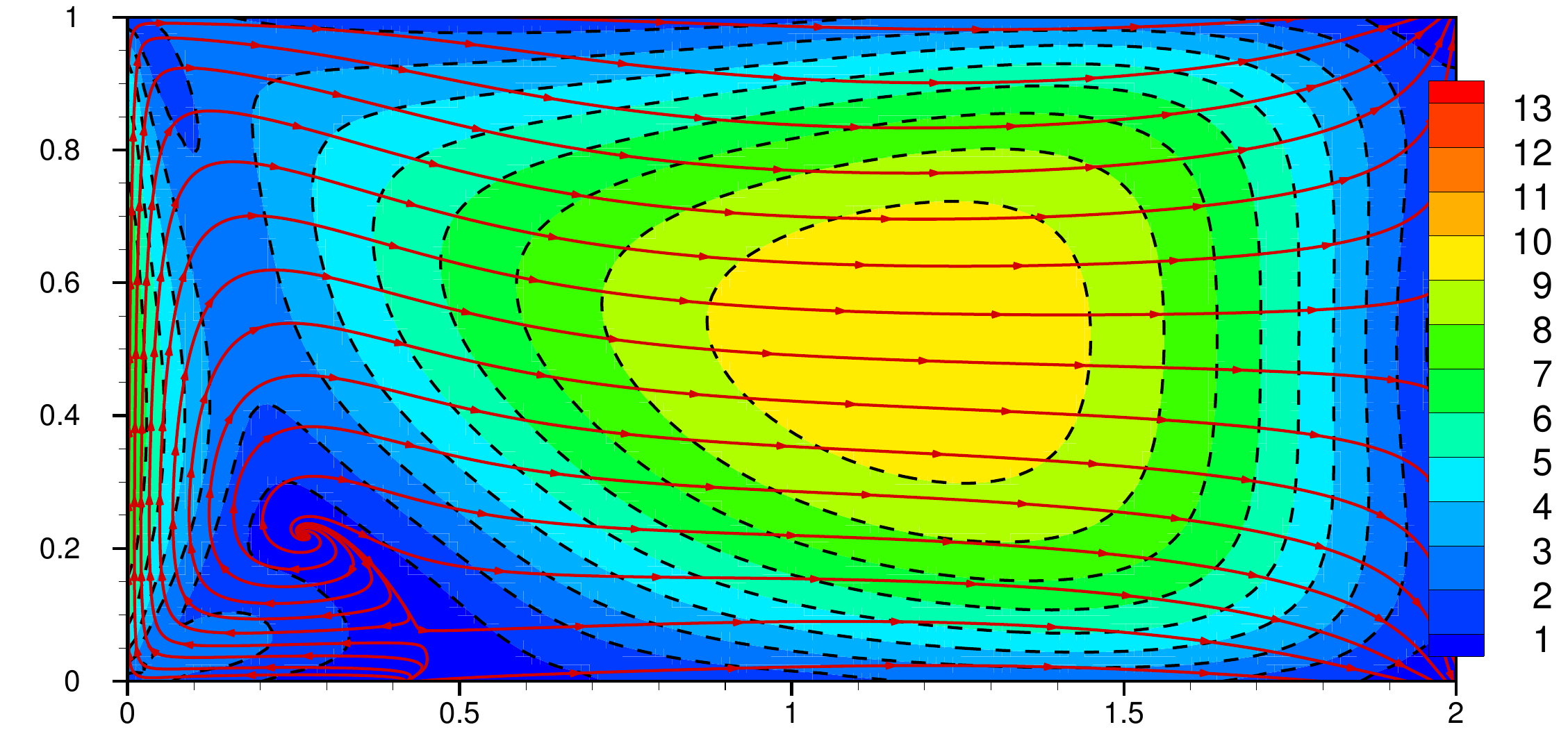}}
	\caption{\label{fig:RayleighWallSpeed}Magnitude of velocity distribution for wall bounded Rayleigh flows at different times. Background: explicit UGKS with $\Delta t = 0.001$; dotted line: IUGKS with time step $\Delta t = 0.01$; red solid line: streamlines.}
\end{figure}

\begin{figure}[H]
	\centering
	\subfigure[$t=0.5$]{\includegraphics[width=0.48\textwidth]{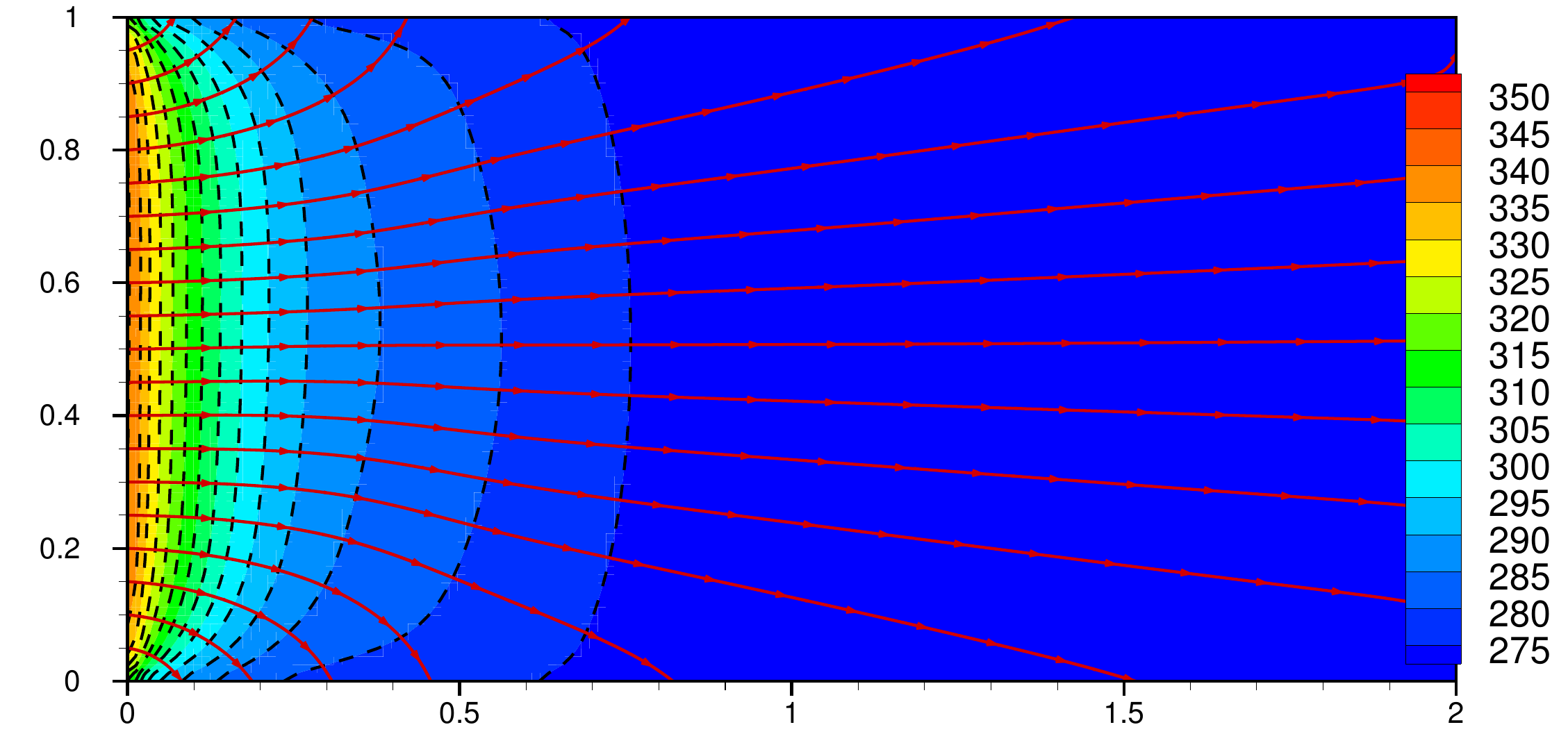}}
	\subfigure[$t=1.0$]{\includegraphics[width=0.48\textwidth]{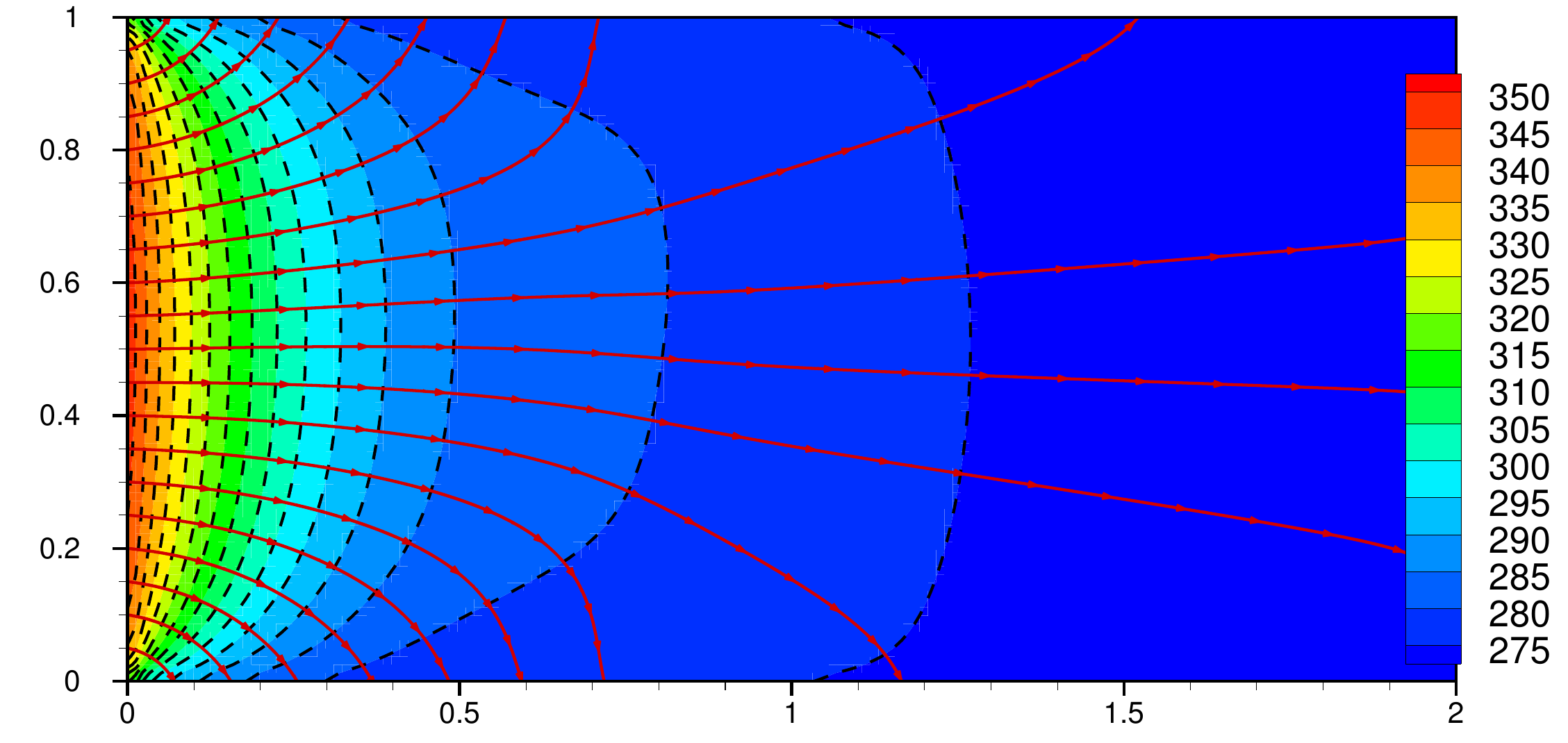}}\\
	\subfigure[$t=1.5$]{\includegraphics[width=0.48\textwidth]{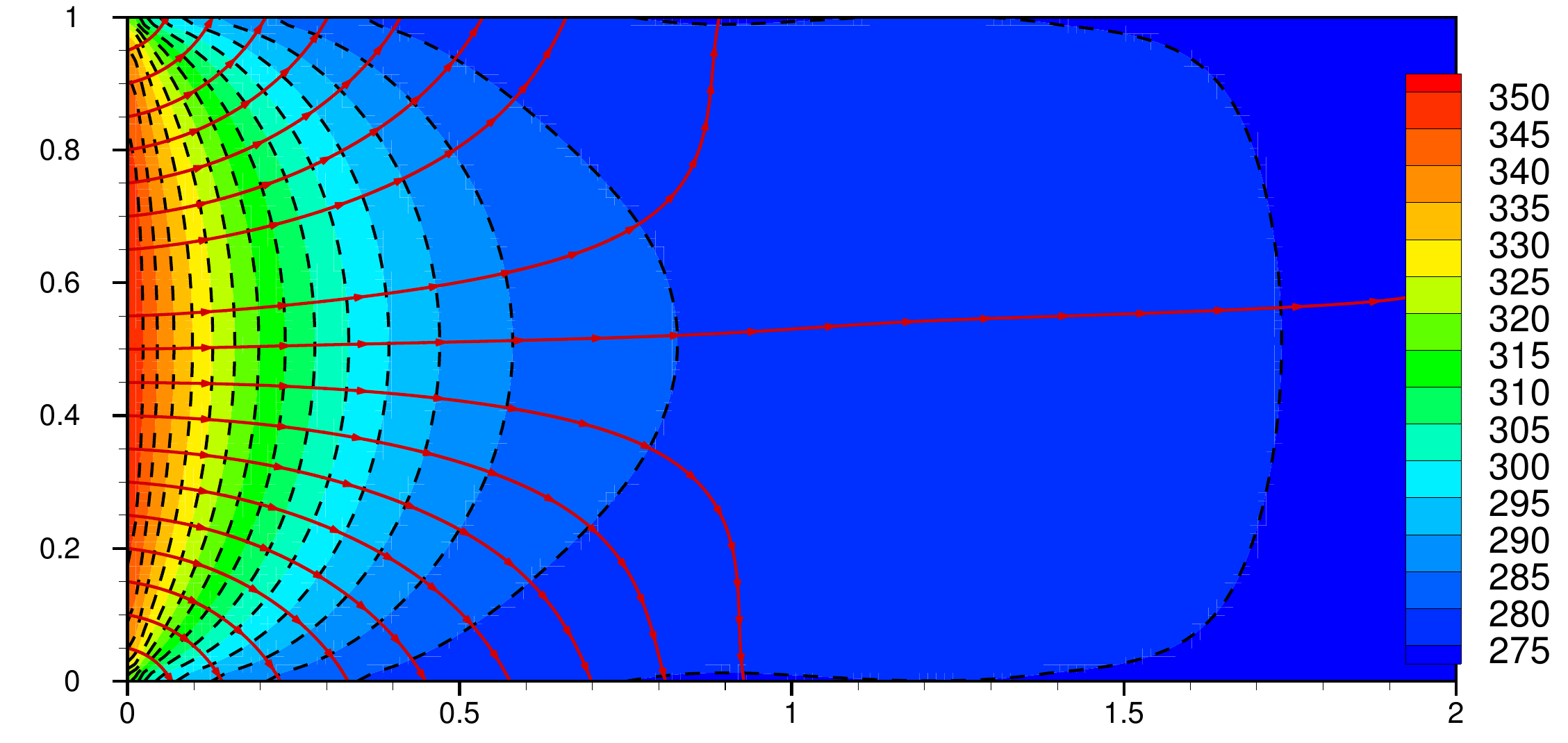}}
	\subfigure[$t=2.0$]{\includegraphics[width=0.48\textwidth]{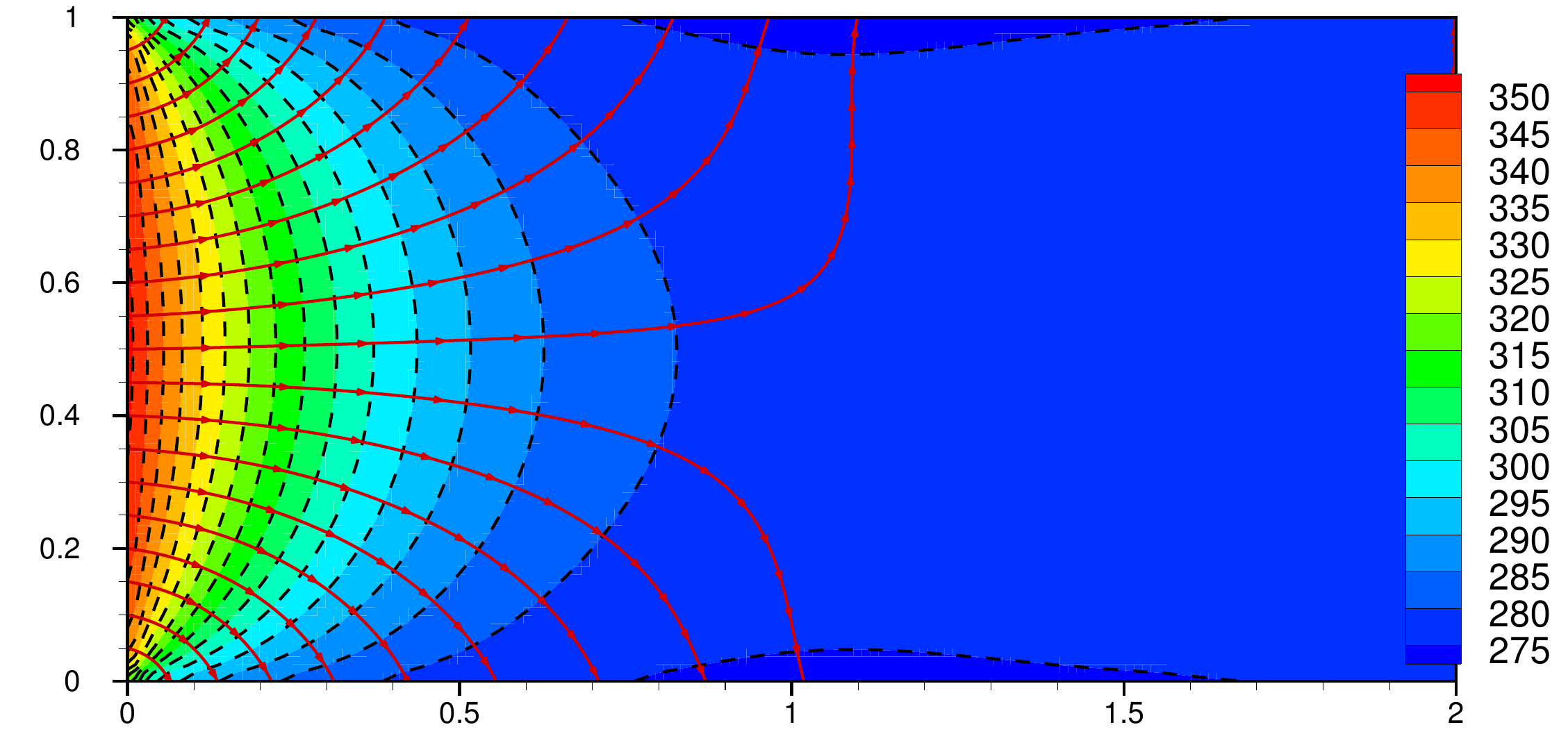}}
	\caption{\label{fig:RayleighWallTemperature}Temperature distribution for wall bounded Rayleigh flows at different times. Background: explicit UGKS with $\Delta t = 0.001$; solid line: IUGKS with time step $\Delta t = 0.01$; red solid line: heat flux.}
\end{figure}

The flow variables such as magnitude of velocity and temperature at different times are plotted in Fig.~\ref{fig:RayleighWallSpeed} and Fig.~\ref{fig:RayleighWallTemperature}, which clearly present the flow evolution.
The mechanism for the flow evolution comes from plate's shearing and thermal heating effect.
Based on the data on the central lines, the  rarefied gas effect appears, such as the velocity slip and temperature jump near the solid walls.


\begin{figure}[H]
\centering
\subfigure[Pressure]{\includegraphics[width=0.32\textwidth]{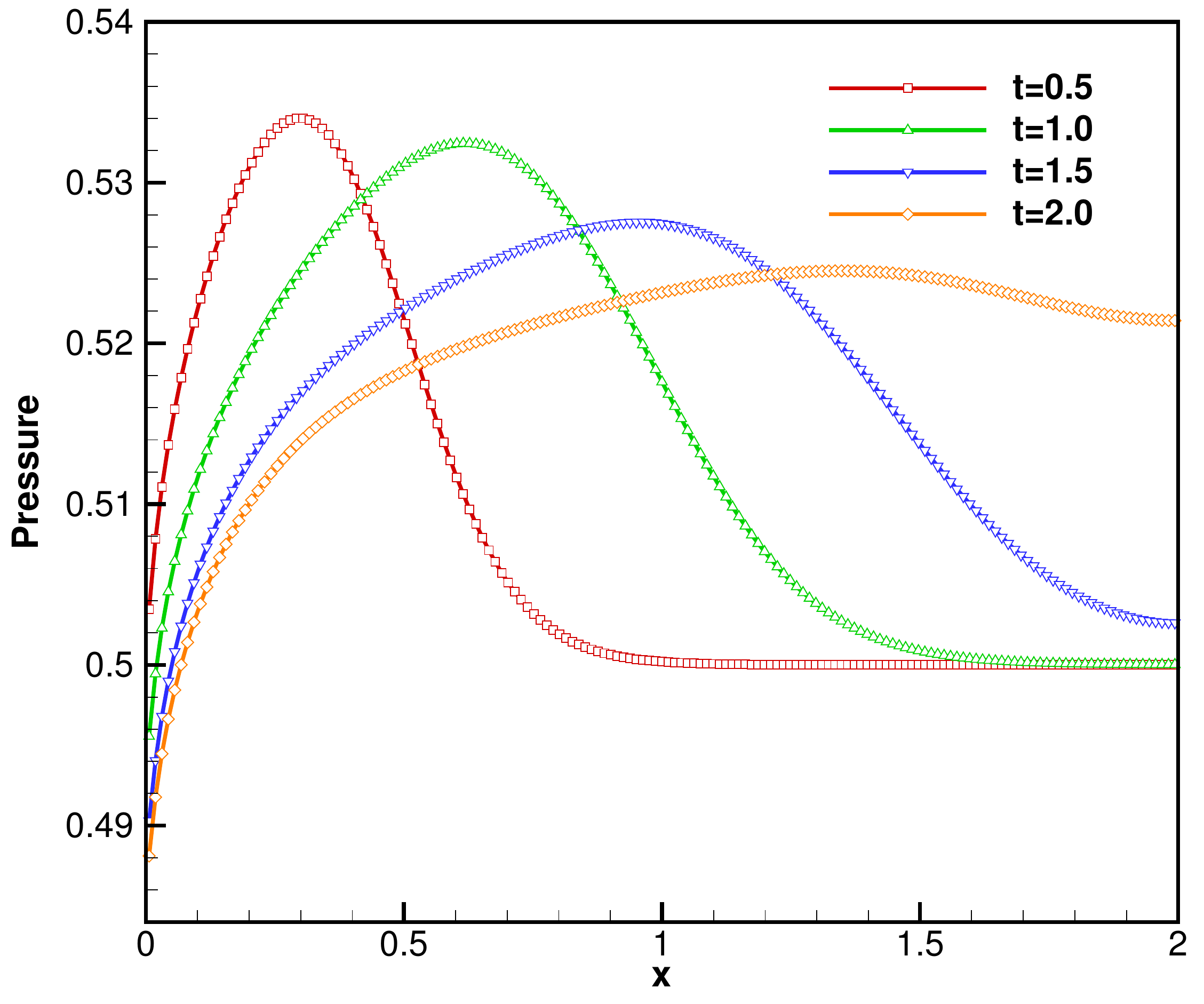}}
\subfigure[Shear stress]{\includegraphics[width=0.32\textwidth]{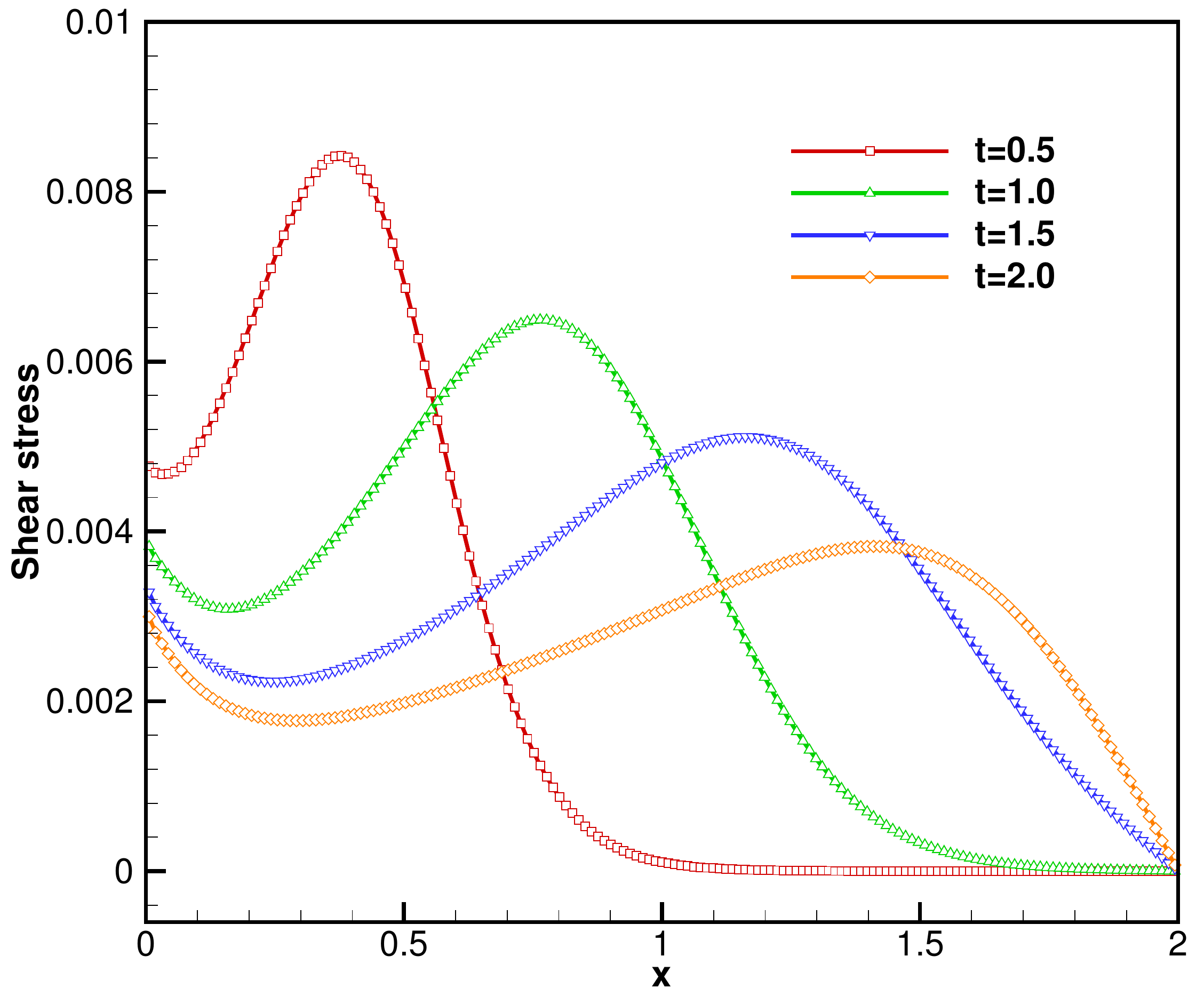}}
\subfigure[Heat flux]{\includegraphics[width=0.32\textwidth]{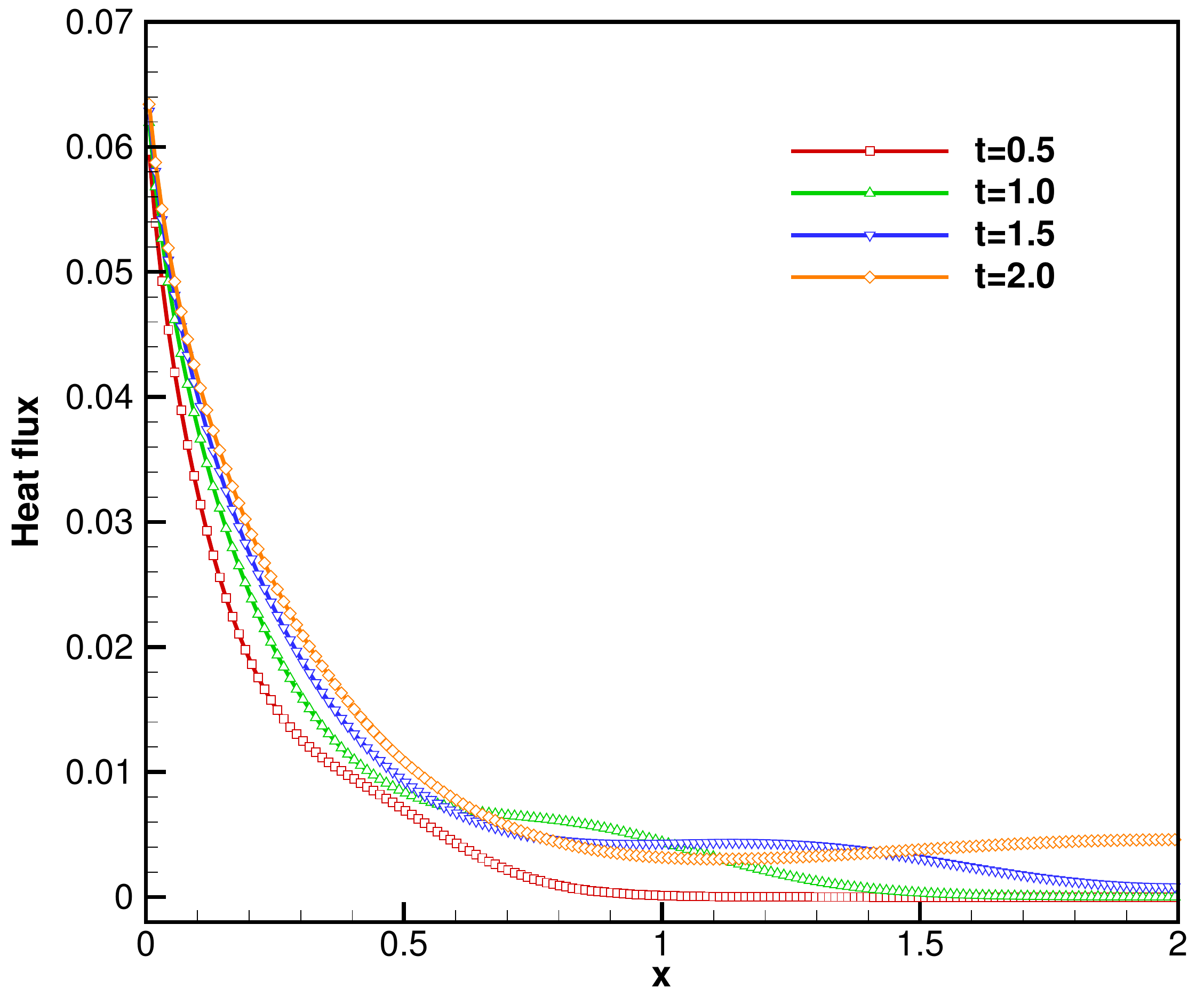}}
\caption{\label{fig:RayleighSurfaceLower}Surface quantities along the lower wall changing with time. The pressure and shear stress are normalized by $\rho_{0} C_{0}^2$ where $C_{0} = \sqrt{2 R T_{0}}$, and the heat flux is normalized by $\rho_{0} C_{0}^3$.}
\end{figure}

\begin{figure}[H]
\centering
\subfigure[Pressure]{\includegraphics[width=0.32\textwidth]{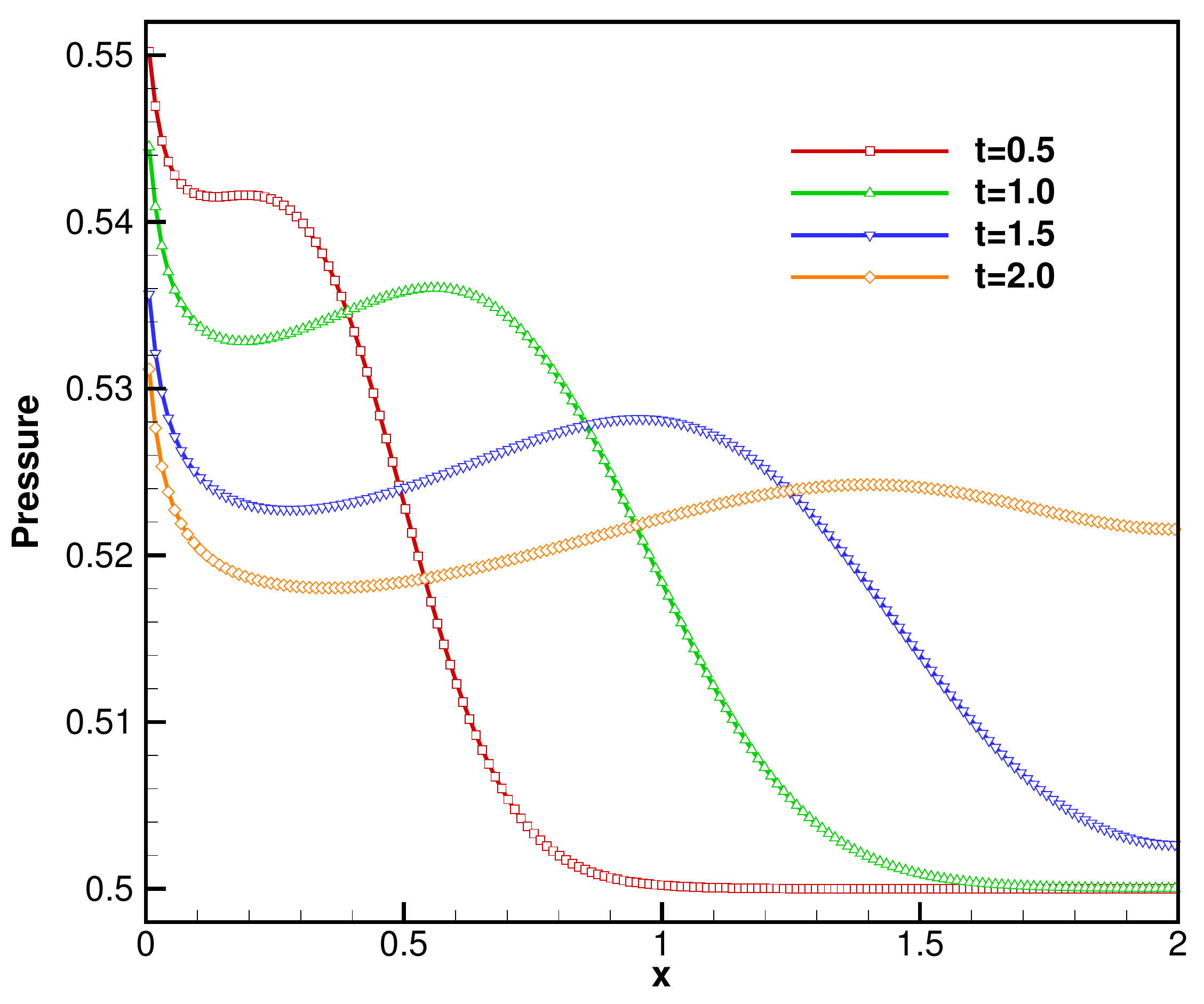}}
\subfigure[Shear stress]{\includegraphics[width=0.32\textwidth]{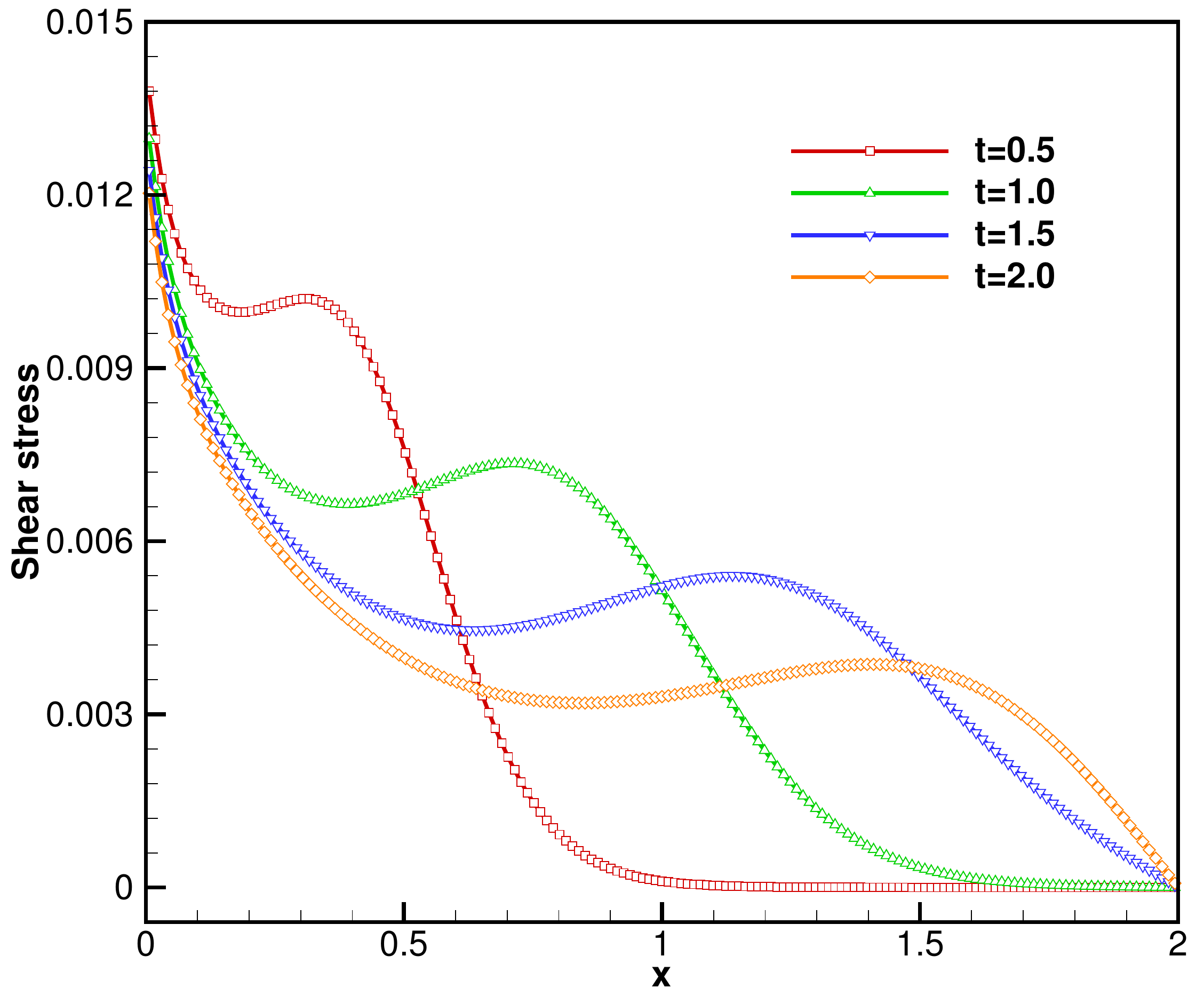}}
\subfigure[Heat flux]{\includegraphics[width=0.32\textwidth]{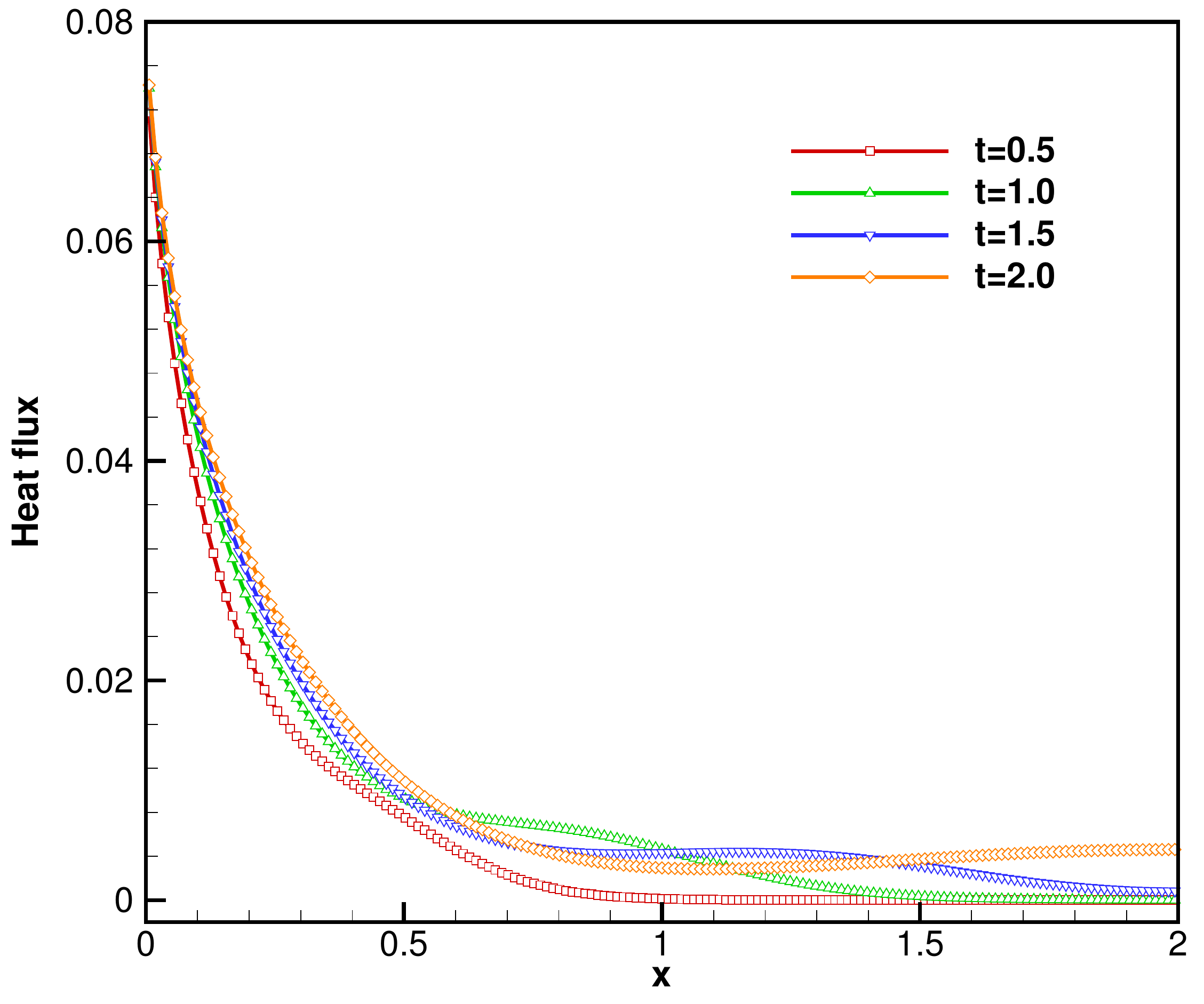}}
\caption{\label{fig:RayleighSurfaceUpper}Surface quantities along the upper wall changing with time. The pressure and shear stress are normalized by $\rho_{0} C_{0}^2$ where $C_{0} = \sqrt{2 R T_{0}}$, and the heat flux is normalized by $\rho_{0} C_{0}^3$.}
\end{figure}

The surface quantities, such as pressure, shear stress, and heat flux at four different instants are given in Fig.~\ref{fig:RayleighSurfaceLower}, Fig.~\ref{fig:RayleighSurfaceUpper}, and Fig.~\ref{fig:RayleighSurfaceSide}. Detailed data for this case have been listed in the Appendix
for future references.
\begin{figure}[H]
\centering
\subfigure[Pressure]{\includegraphics[width=0.32\textwidth]{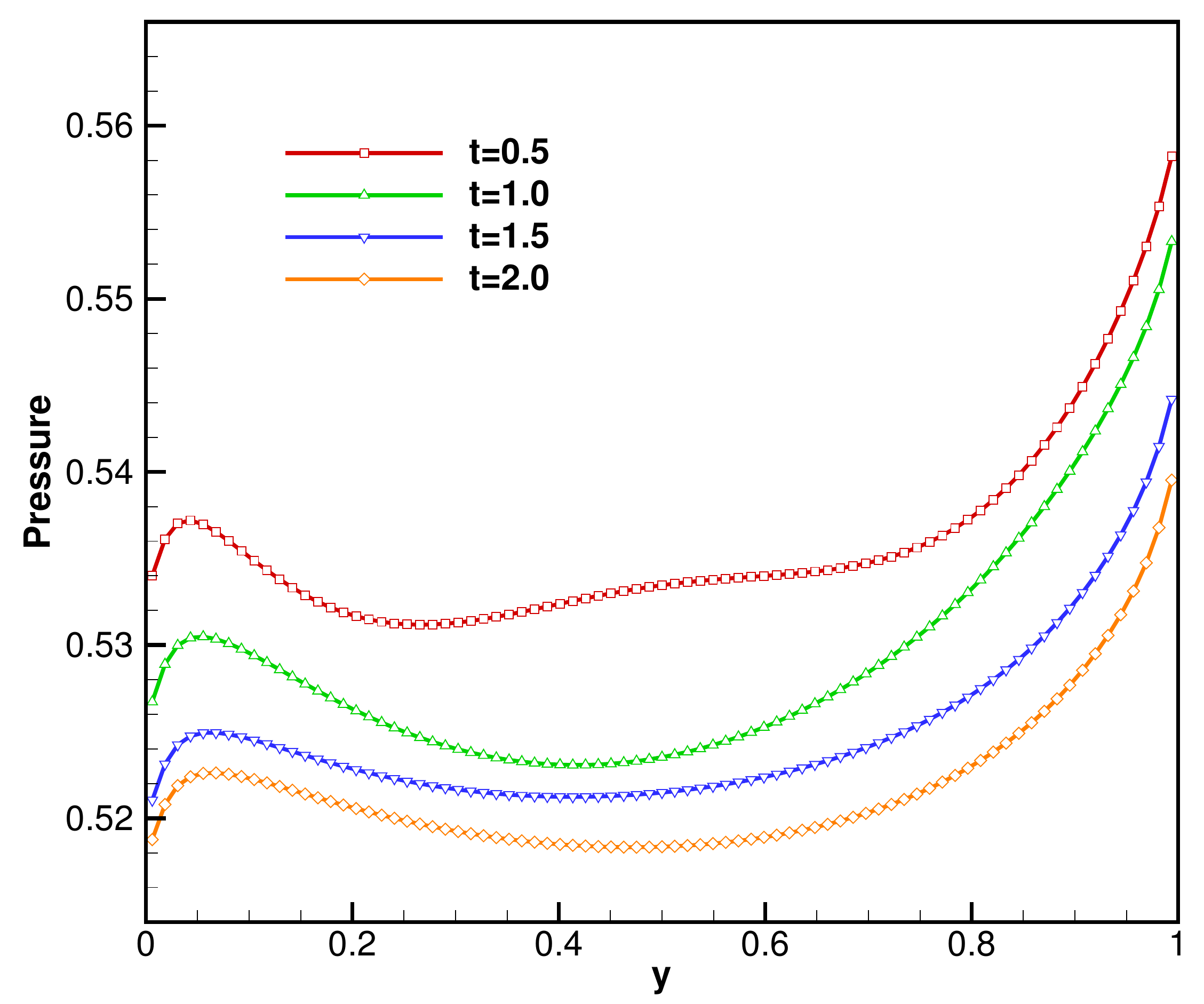}}
\subfigure[Shear stress]{\includegraphics[width=0.32\textwidth]{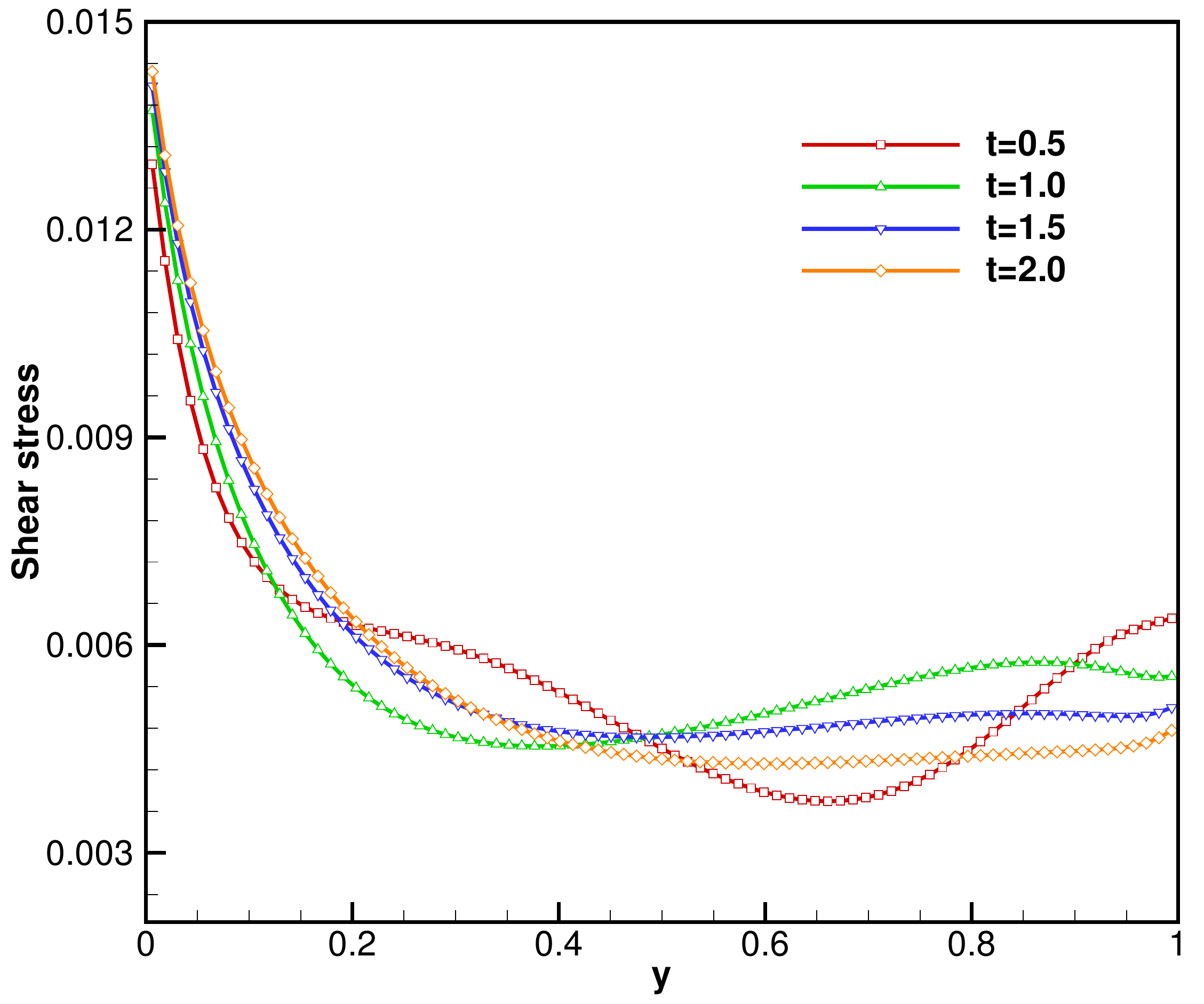}}
\subfigure[Heat flux]{\includegraphics[width=0.32\textwidth]{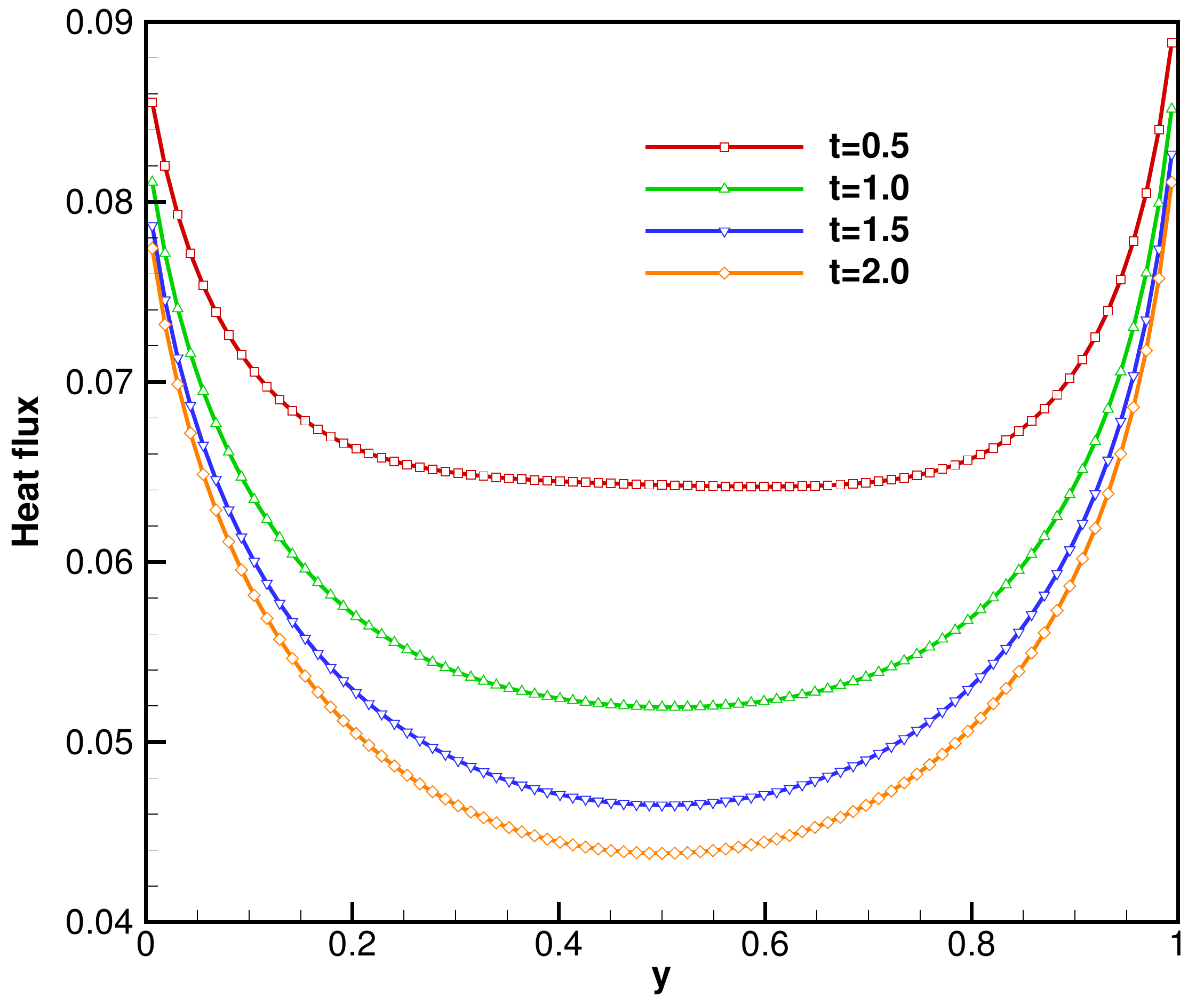}}
\caption{\label{fig:RayleighSurfaceSide}Surface quantities along the side plate changing with time. The pressure and shear stress are normalized by $\rho_{0} C_{0}^2$ where $C_{0} = \sqrt{2 R T_{0}}$, and the heat flux is normalized by $\rho_{0} C_{0}^3$.}
\end{figure}

\subsection{Hypersonic flow past a square cylinder}\label{sec:square}
This high-speed rarefied flow past a square cylinder is computed to validate the efficiency and stability of the current implicit scheme. The freestream is hypersonic flow of the argon gas at a Mach number $5$ with a temperature of $T_{\infty} = 273{\rm K}$. The computational domain is $\left[-0.06{\rm m},0.08{\rm m}\right] \times \left[-0.06{\rm m}, 0.06{\rm m}\right]$, and the square center locates at the origin. The spatial discretization is the same as that described in \cite{chen2017memory}. The density in the freestream is $8.58 \times 10^{-5} {\rm kg/m^3}$, resulting in a global Knudsen number $0.1$ which is defined with respect to the square side length $0.01{\rm m}$. The dynamic viscosity is calculated by $\mu = \mu_0 (T/T_0)^\omega$ where $\omega=0.81$.

\begin{figure}[H]
\centering
\subfigure[Normalized pressure]{\includegraphics[width=0.48\textwidth]{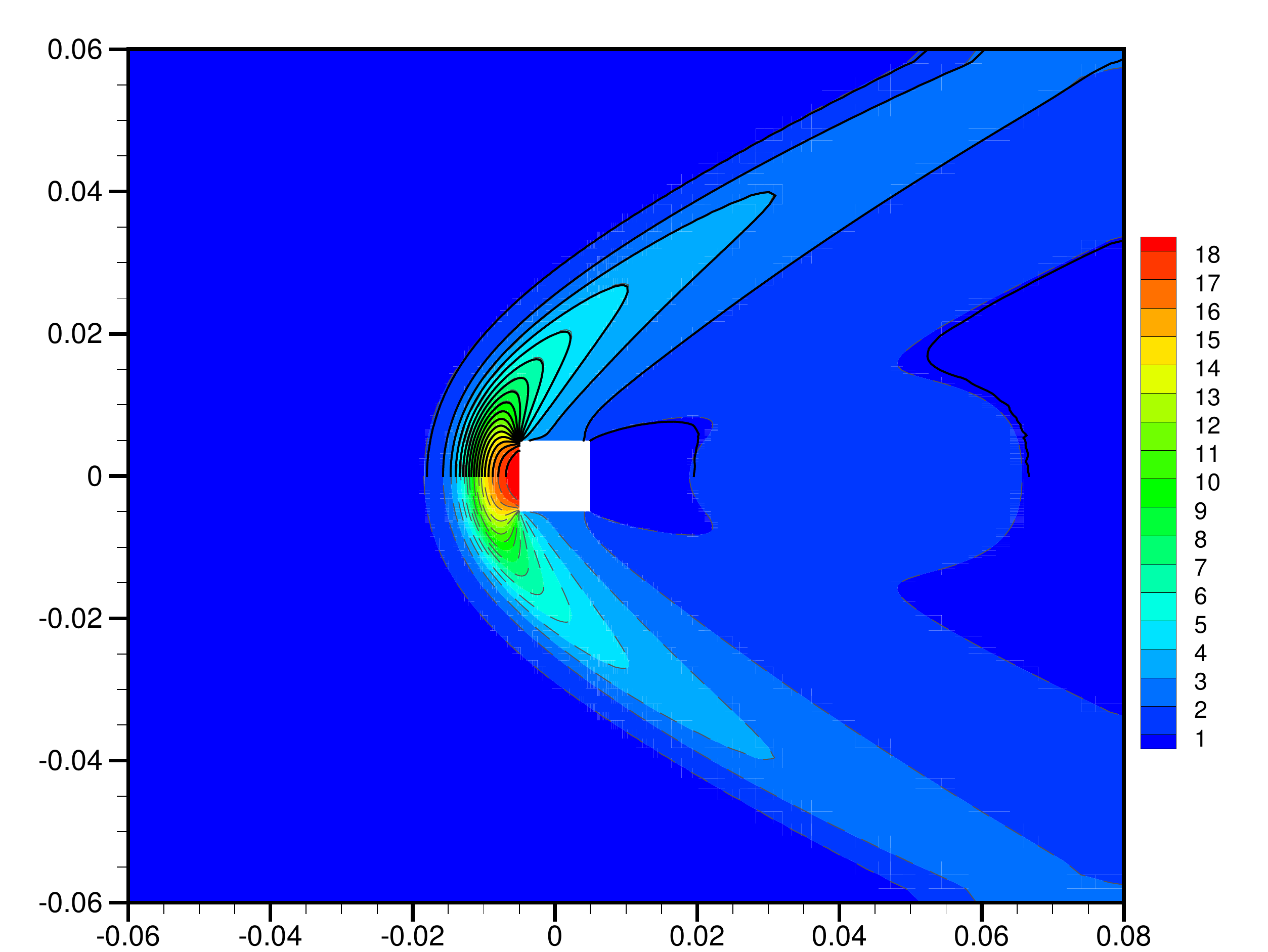}}
\subfigure[Temperature (K)]{\includegraphics[width=0.48\textwidth]{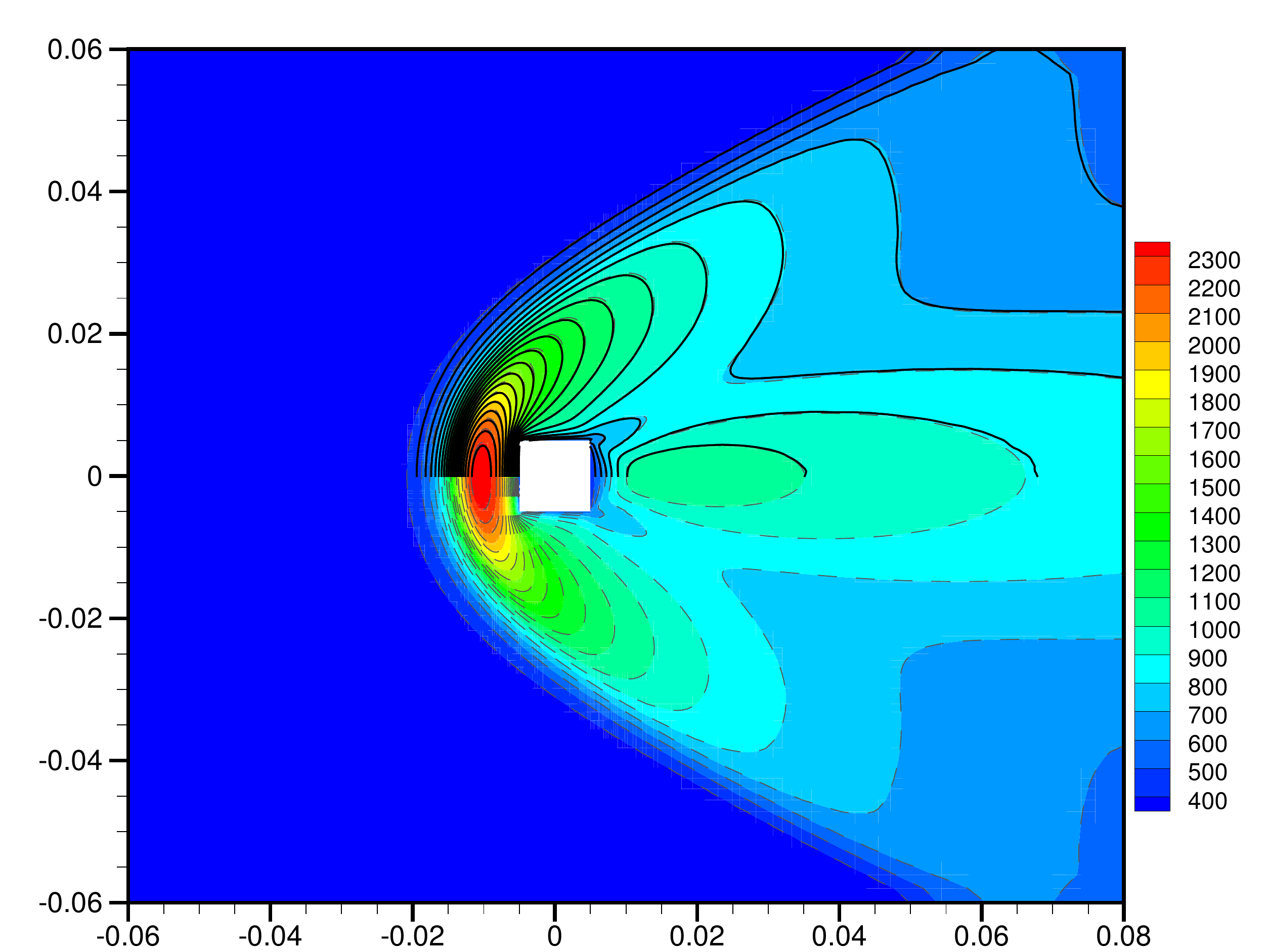}}\\
\subfigure[X velocity (m/s)]{\includegraphics[width=0.48\textwidth]{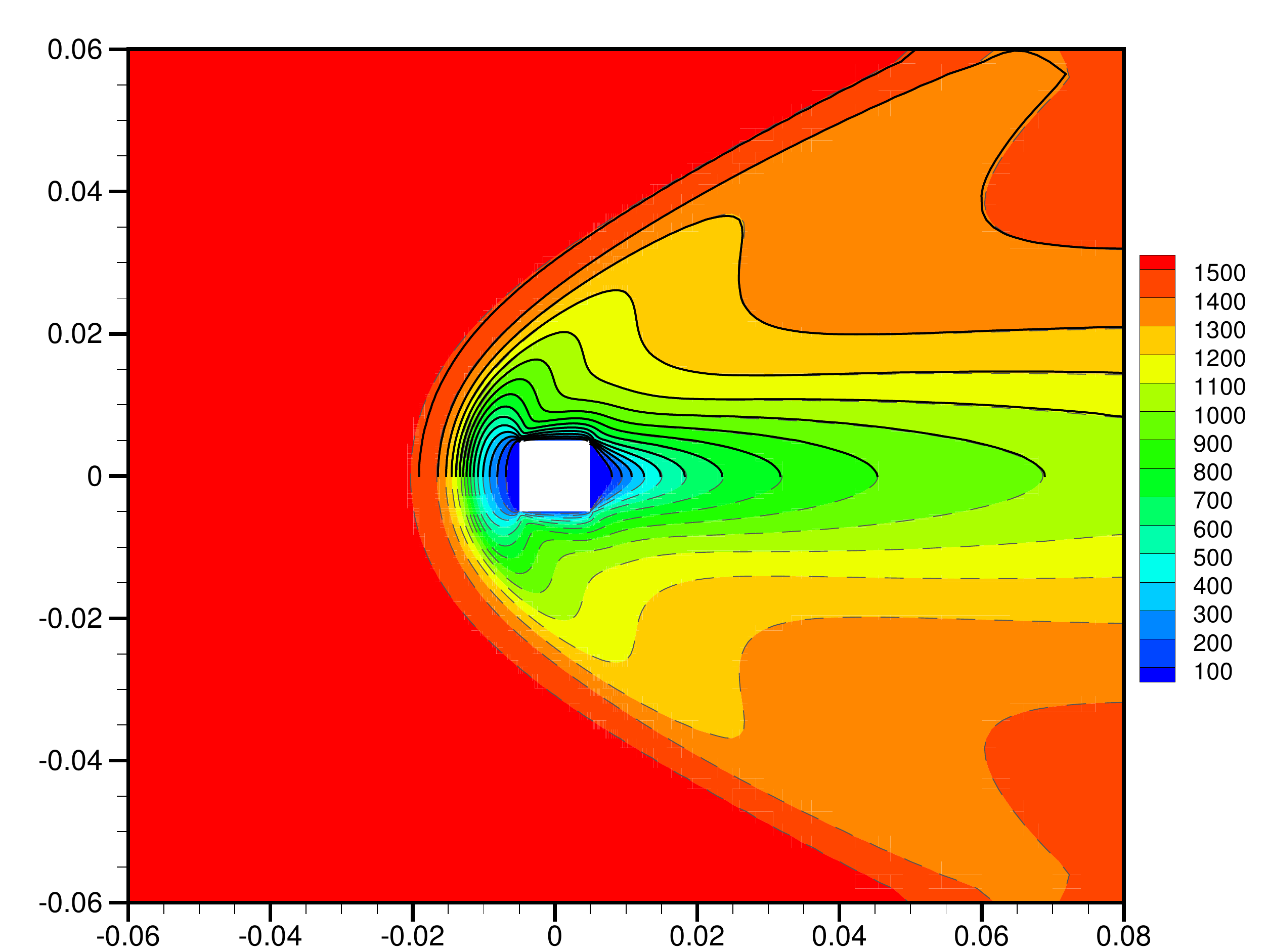}}	
\subfigure[Y velocity (m/s)]{\includegraphics[width=0.48\textwidth]{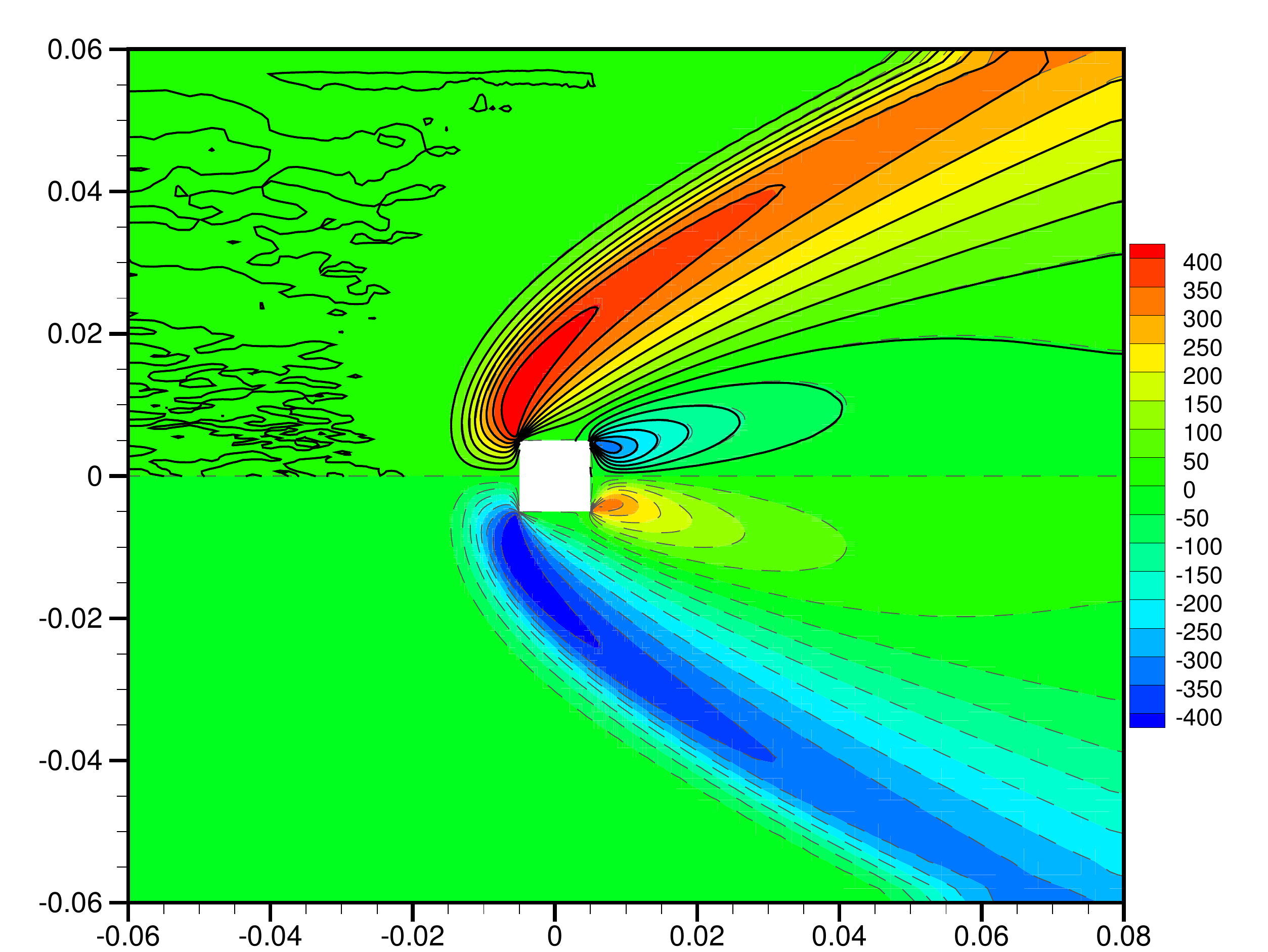}}
\caption{\label{fig:squareSteadyResults} Flow variable distributions at steady state for hypersonic flow past a square cylinder at $Ma=5$ and $Kn=0.1$. The background with dashes lines present the results from the IUGKS, and solid lines in the upper half domain denote the DSMC results.}
\end{figure}

\begin{figure}[H]
\centering
\subfigure[Pressure]{\includegraphics[width=0.32\textwidth]{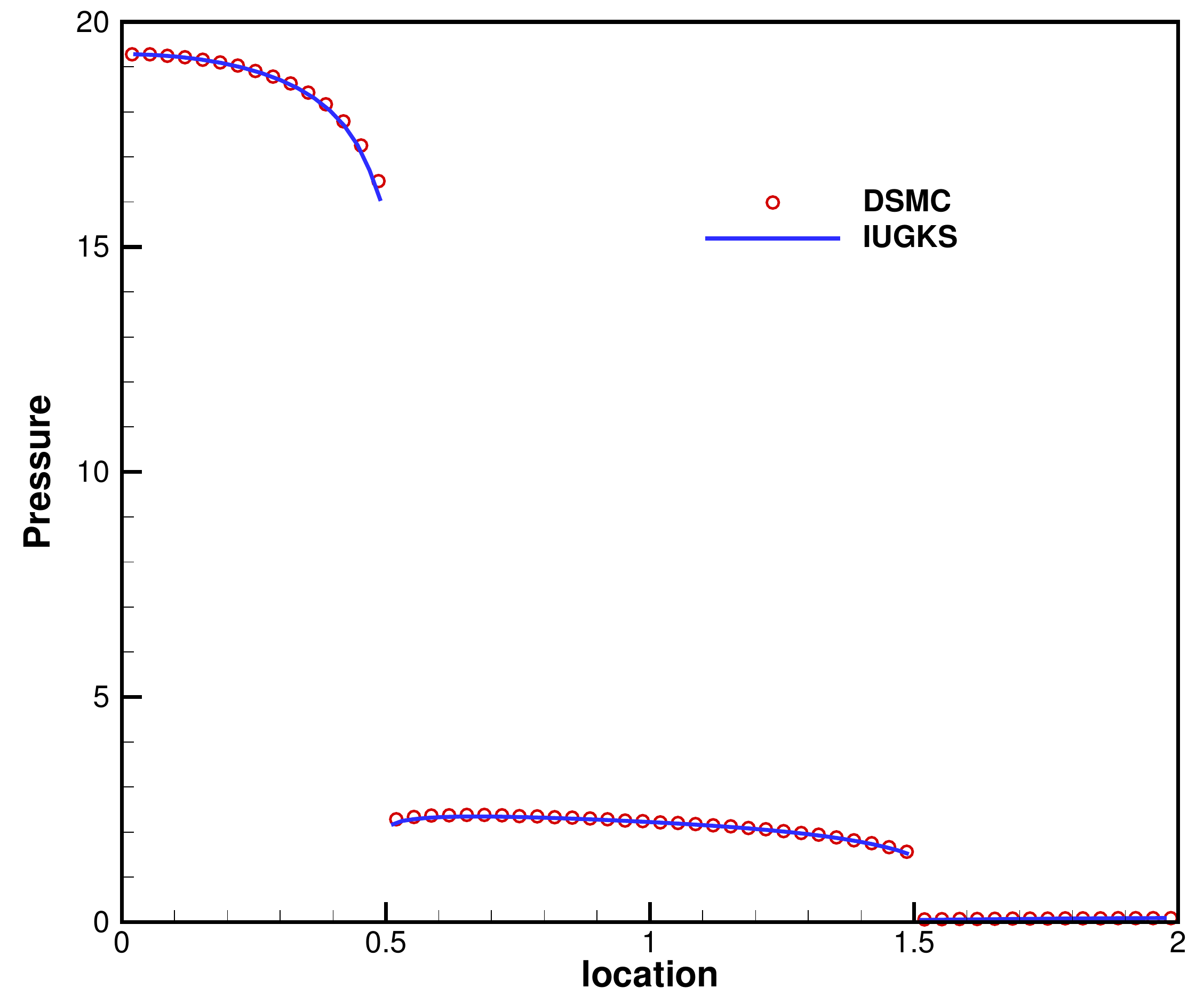}}
\subfigure[Shear stress]{\includegraphics[width=0.32\textwidth]{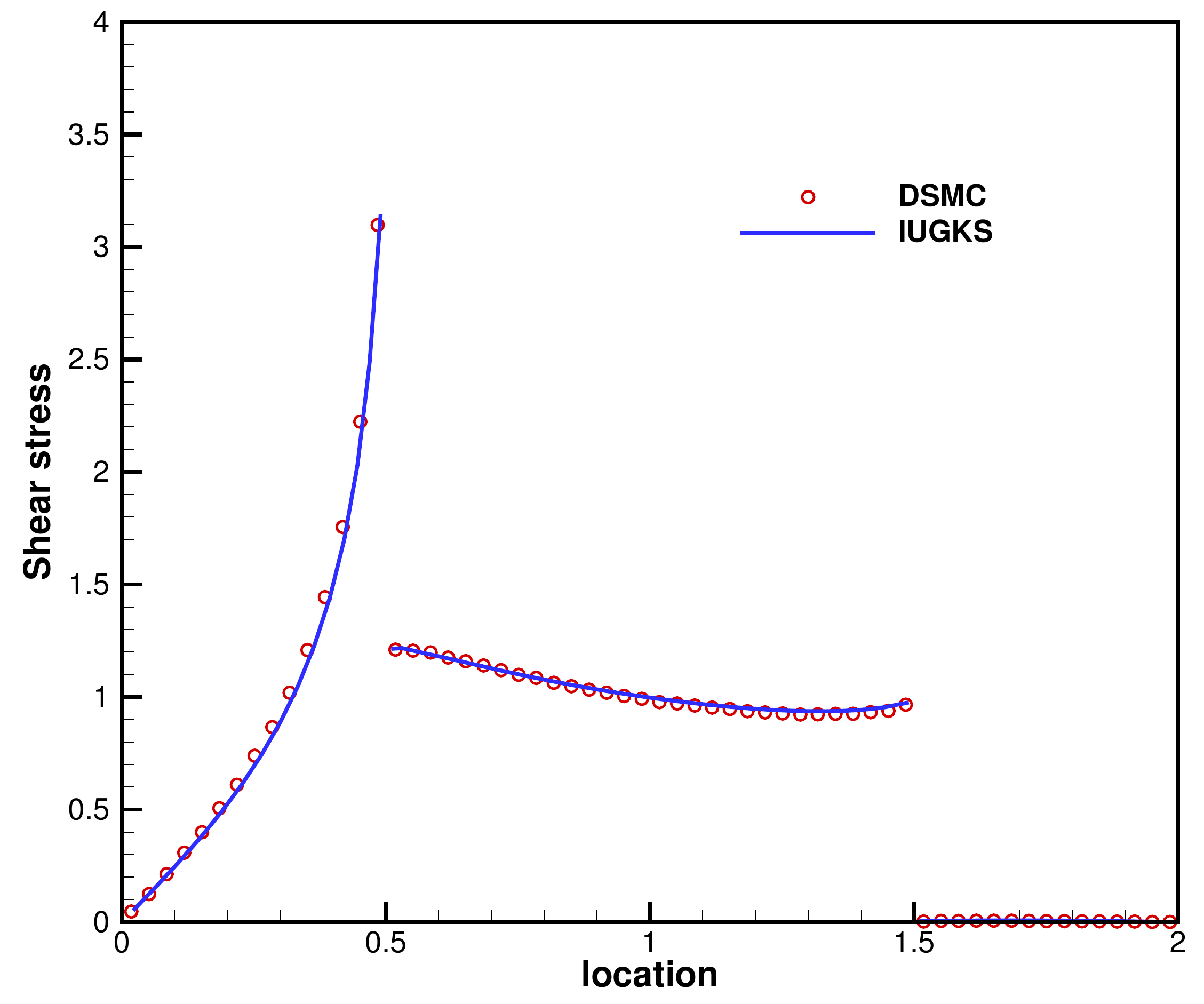}}
\subfigure[Heat flux]{\includegraphics[width=0.32\textwidth]{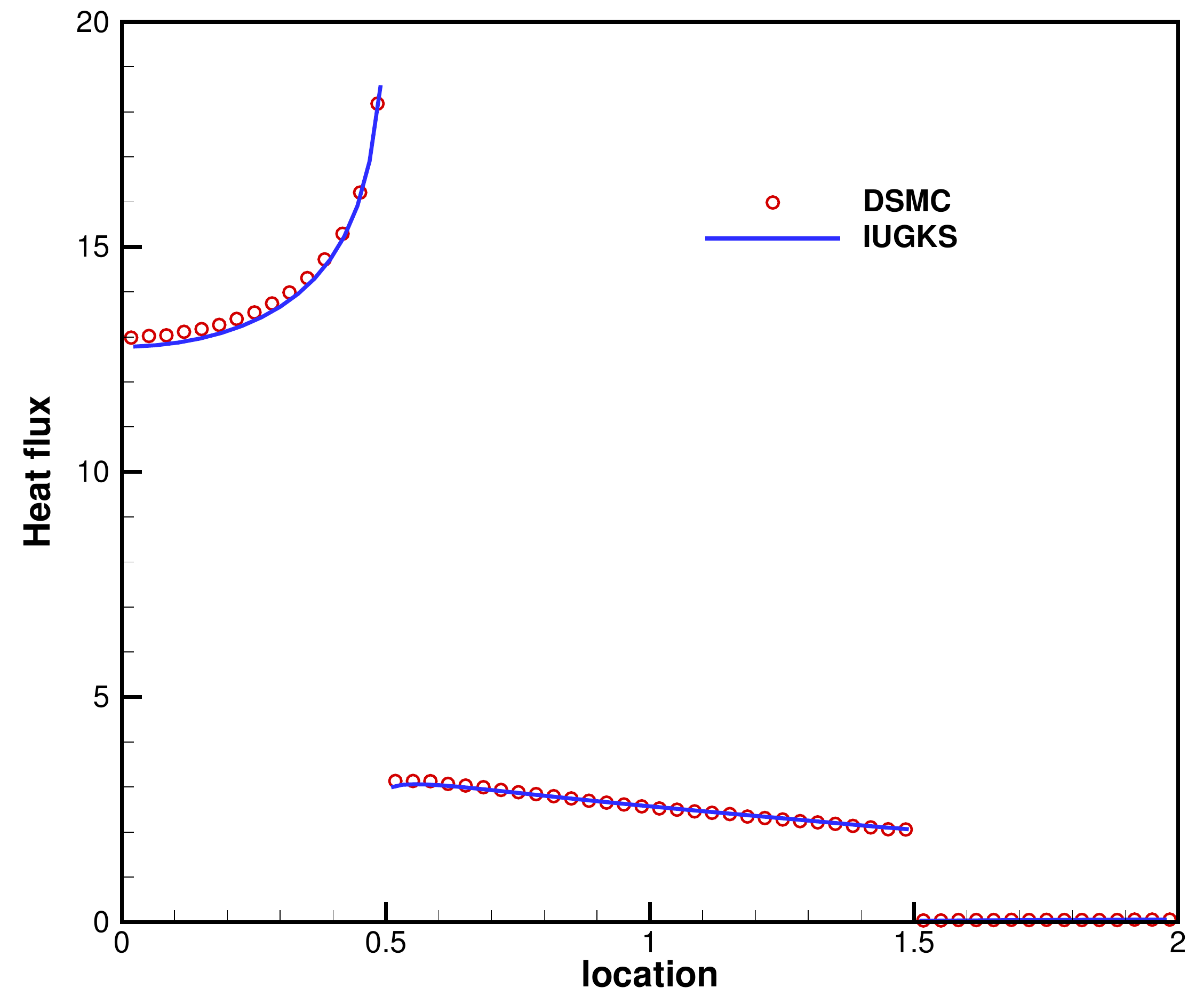}}
\caption{\label{fig:squareSurface}Surface quantities on the upper half walls of the square cylinder. The pressure and shear stress are normalized by $\rho_{0} C_{0}^2$ where $C_{0} = \sqrt{2 R T_{0}}$, and the heat flux is normalized by $\rho_{0} C_{0}^3$.}
\end{figure}

\begin{figure}[H]
	\centering
	\subfigure[]{\includegraphics[width=0.46\textwidth]{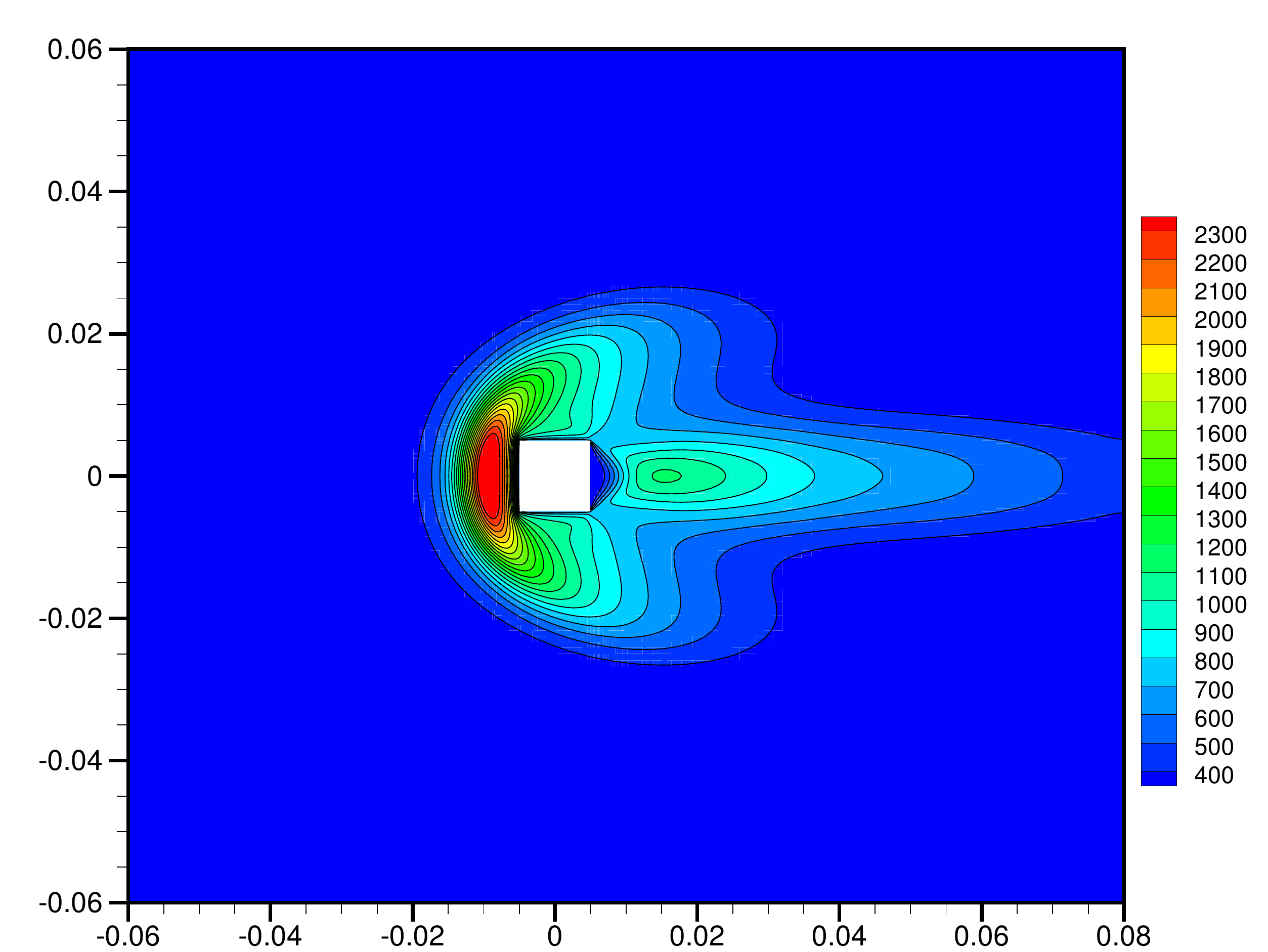}}
	\subfigure[]{\includegraphics[width=0.46\textwidth]{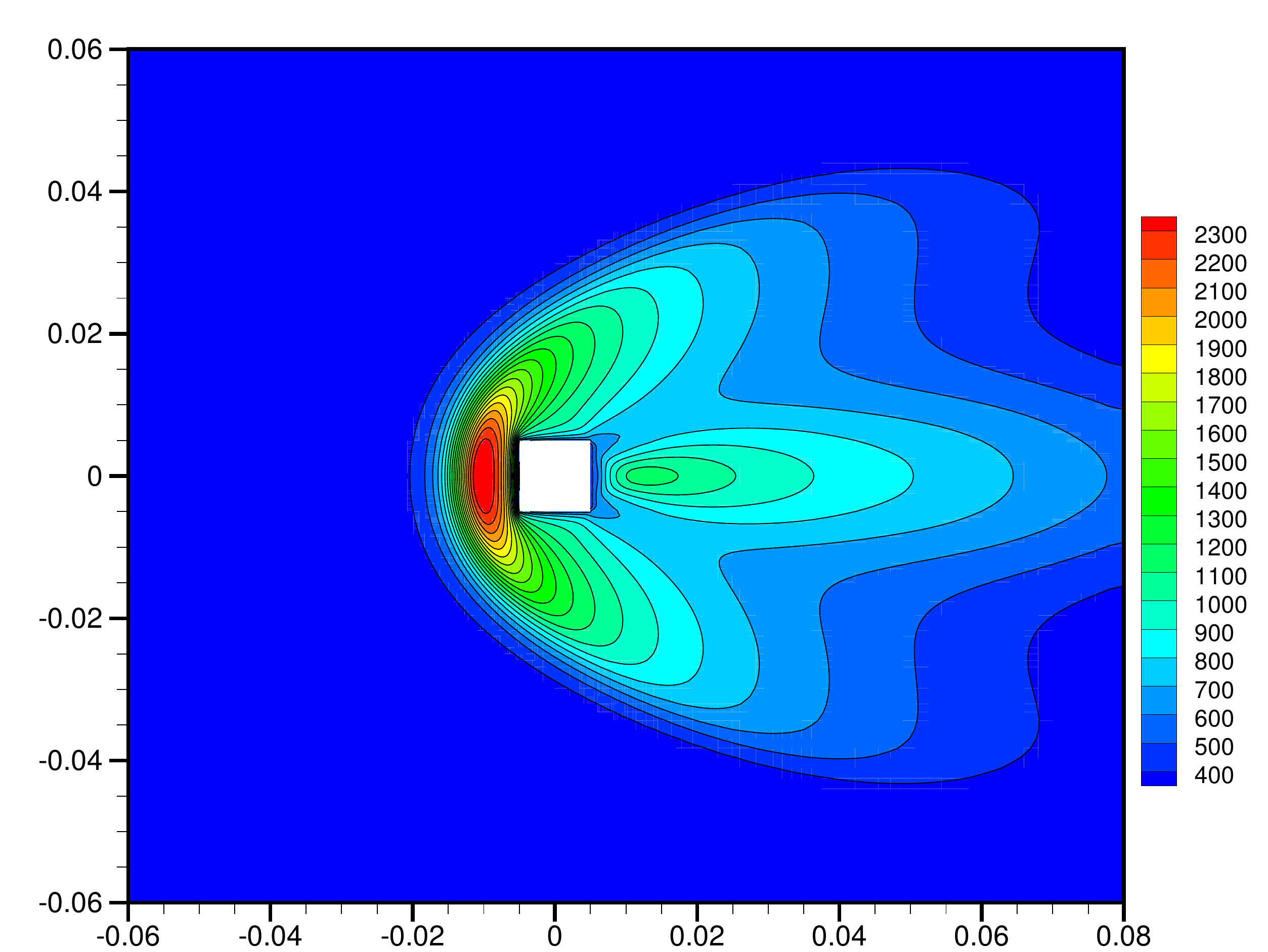}}\\
	\subfigure[]{\includegraphics[width=0.46\textwidth]{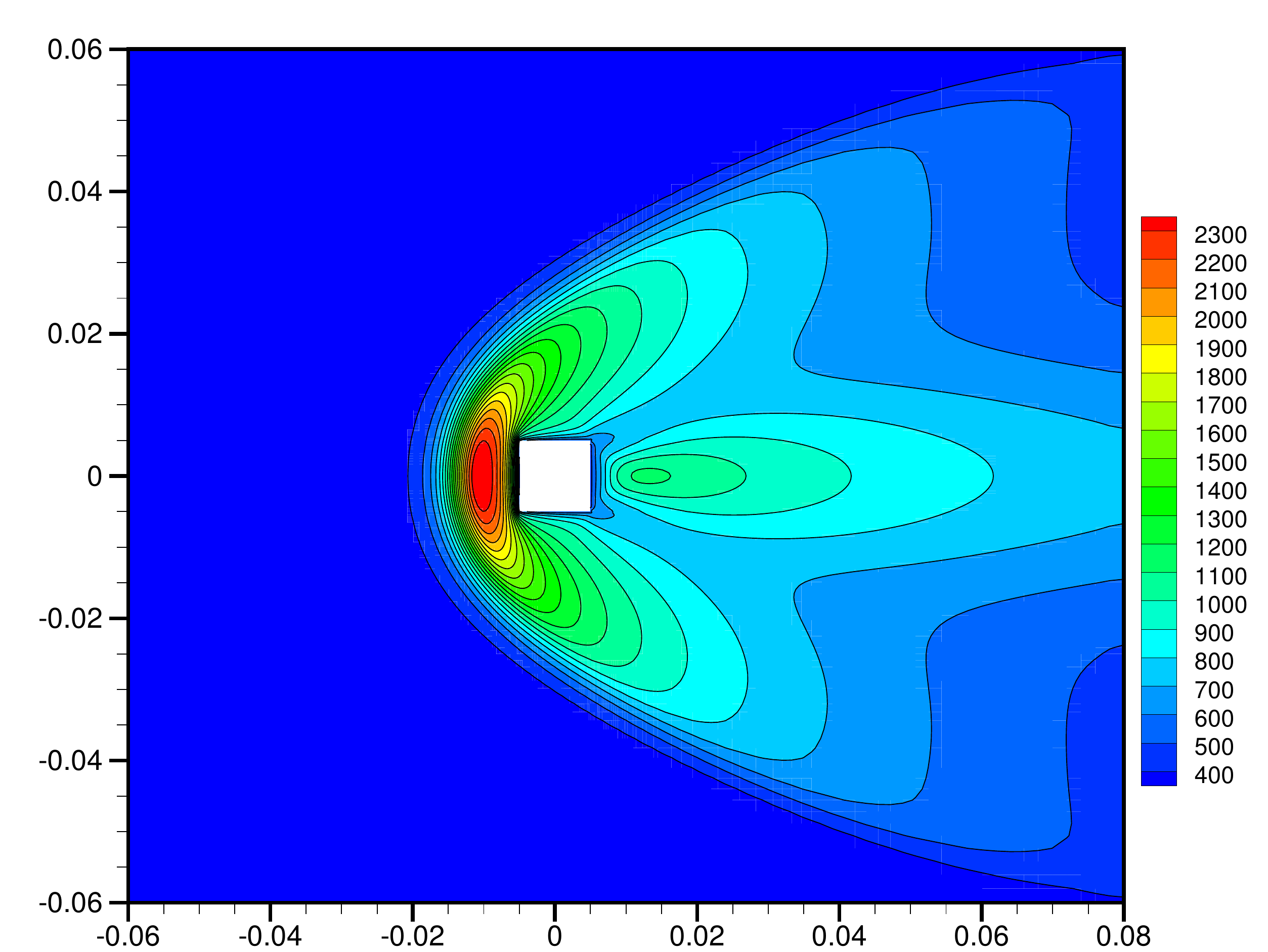}}
	\subfigure[]{\includegraphics[width=0.46\textwidth]{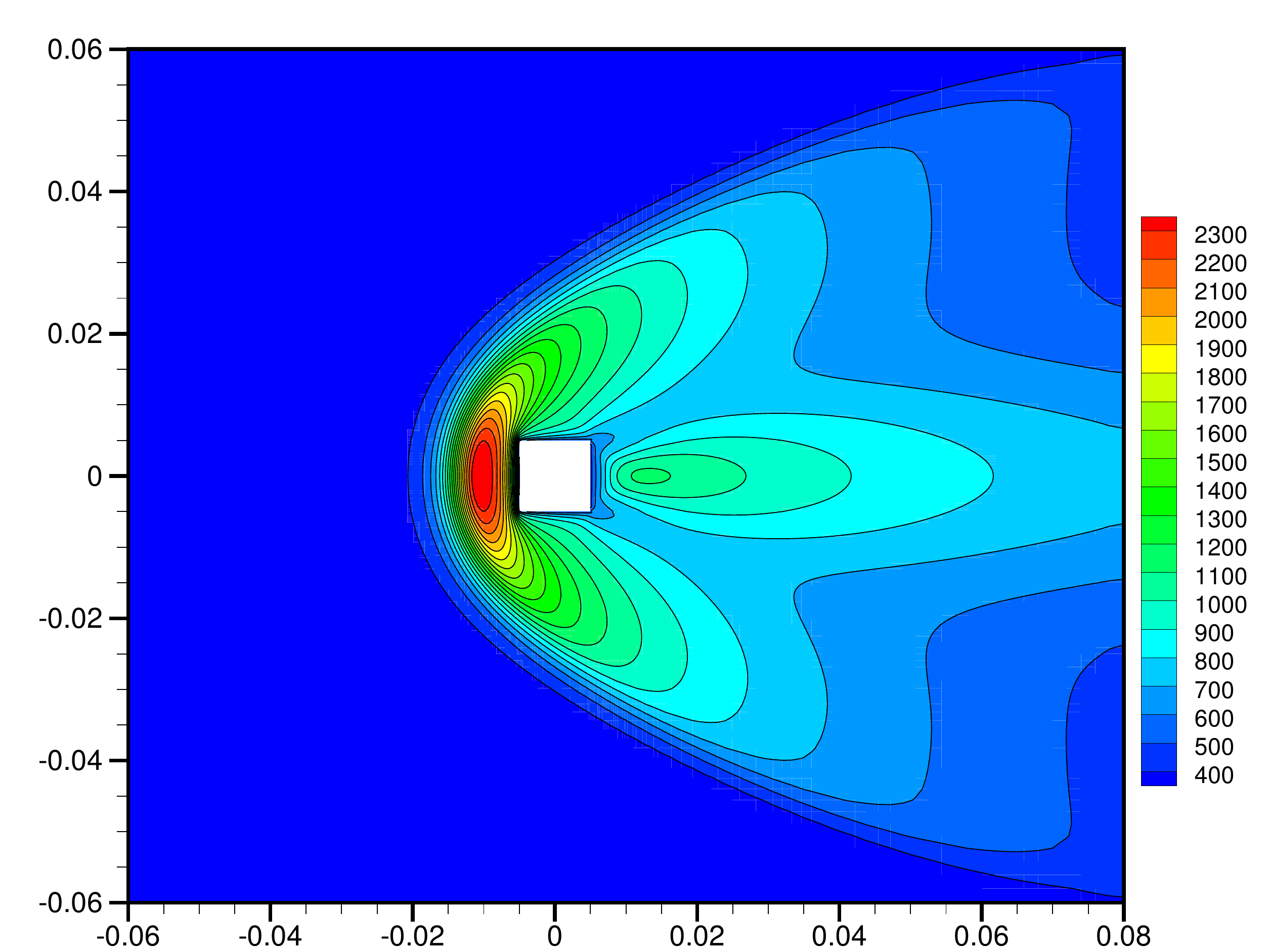}}
	\caption{\label{fig:squareInstantResults} Temperature distributions at different times for hypersonic flow past a square cylinder at $Ma=5$ and $Kn=0.1$; (a) $t=31.9{\rm \mu s}$, (b) $t=63.8 {\rm \mu s}$, (c) $t=95.7 {\rm \mu s}$ and (d) $t=127.6 {\rm \mu s}$.}
\end{figure}

The density, pressure and velocity distributions at the steady state are given in Fig.~\ref{fig:squareSteadyResults}, where the results obtained from the IUGKS are compared with those from the DSMC method. Generally, the IUGKS results agree well with the DSMC data. For the velocity along $y$-direction, the IUGKS gives a straight line on the symmetric plane for the zero contour line while the DSMC gives some noises in the freestream due to the statistic error. The surface quantities, such as pressure, shear stress, and heat flux on the upper half walls of the cylinder are given in Fig.~\ref{fig:squareSurface} and compared with the DSMC data. Good agreement has been obtained. For unsteady flow simulation, we show the time evolution of the temperature distribution at several instants in Fig.~\ref{fig:squareInstantResults} to illustrate the capability of the IUGKS for capturing unsteady flow evolution.

The final steady state solution is obtained at the physical time $t=319{\rm \mu s}$, which takes about $14$ hours using a serial computation.
In comparison with the DSMC method, for evolving the solution by the same amount of time, the IUGKS basically takes the same computational cost as DSMC, where the DSMC targets on the steady state solution and no sampling is conducted in the unsteady process.
The IUGKS has no noise and doesn't need any additional sampling for unsteady solution.
If only the final steady state solution is needed, a fully implicit with multigrid technique can be used in UGKS to get a much higher efficiency,
where the UGKS may become hundreds times faster than DSMC \cite{zhu2017multigrid}.

\section{Discussions and conclusions}
In this paper, an implicit unified gas-kinetic scheme (IUGKS) is proposed for unsteady flow simulations in all Knudsen number regimes. By solving the time-accurate implicit governing equations of the macroscopic variables and the gas distribution function within several inner iterations, the flow field can be updated with a large time step, which is no longer restricted by the CFL stability condition.

For unsteady flow simulation, non-uniform meshes are usually used in many practical engineering applications,
such as very fine mesh for viscous flow near solid boundaries and the mesh around the airfoil tip.
For the explicit scheme, due to the CFL condition the numerical global time step can become very small, which destroys the advantage of the multiscale nature of UGKS.
Since the UGKS employs the analytic solution of the kinetic relaxation model,
the numerical flux covers different flow regimes with the changing of the local ratio of the time step over the local mean collision time, i.e., $\Delta t_s / \tau$.
A uniform time step determined by the smallest cell size will make the UGKS flux to the kinetic scale everywhere,
such as $\Delta t_s / \tau \leq 1$. Consequently, a free transport dynamics will be used for the flux evaluation
even for these cells with much larger cell size away from the boundaries.
Therefore, the multi-scale property of the UGKS is not fully utilized.
In the current IUGKS, the implicit treatment releases the CFL time step constraint,
and the flow field can be updated by a numerical time step which is much larger than the global minimum one.
In other words, for the unsteady flow with a large variation of cell size the flow dynamics from IUGKS in different regions can be in different flow regimes.
In addition, according to the ratio between the numerical time step $\Delta t$ and the cell-size-determined time step $\Delta t_s$,
the temporal discretization for the implicit scheme has been improved.
As a result, when the global numerical time step in IUGKS becomes as small as the one used in the explicit UGKS,
the IUGKS can automatically recover the explicit method, where only one inner iteration for each step is needed.
Based on numerical tests, the IUGKS has been validated in terms of accuracy, stability, computational efficiency.
The IUGKS has second-order accuracy both in space and time in both continuum and rarefied flow regimes.
Generally, for unsteady flow the IUGKS is at least one order of magnitude more efficient than the explicit UGKS.
Even for the hypersonic rarefied flow as shown in this paper, the IUGKS can obtain the time accurate solutions as efficient as DSMC.

\section*{Acknowledgment}
The work of Zhu and Zhong was supported by the National Natural Science Foundation of China (Grant No. 11472219) and  the National Pre-Research Foundation of China, as well as the 111 Project of China (B17037).
The research work of Xu is supported by the Hong Kong research grant council (16206617,16207715,16211014) and NSFC (91530319,11772281).

\appendix

\section{Data for wall bounded Rayleigh flow}
The wall bounded Rayleigh flow tested in Section \ref{sec:validation} is a low-speed rarefied flow, which may become a benchmark for validation of numerical schemes. Here we list the data for at some points along the central lines and on solid walls at time $t=1.5$.

\newcolumntype{d}[1]{D{.}{.}{#1}}

\begin{table}[h]
	\centering
\caption{\label{tab:horizontal} Flow variables along the horizontal central line at $t=1.5$ for wall bounded Rayleigh flow at $Kn=0.05$.}
\begin{tabular*}{0.98\textwidth}{@{\extracolsep{\fill} } rrrrr}
	\toprule
 \multicolumn{1}{c}{$x$ (m)} & \multicolumn{1}{c}{$\rho / \rho_0$} & \multicolumn{1}{c}{$U$ (m/s)}
 & \multicolumn{1}{c}{$V$ (m/s)}& \multicolumn{1}{c}{$T$ (K)} \\
\hline
 $6.2111801e-03$  &  $8.18016026e-01$  &  $7.85891e-02$  &  $ 6.74227e+00$  &  $3.49874e+02$\\
 $1.3043478e-01$  &  $8.78648941e-01$  &  $1.86343e+00$  &  $ 2.03063e+00$  &  $3.24664e+02$\\
 $2.5465838e-01$  &  $9.25634041e-01$  &  $4.07566e+00$  &  $-3.24820e-01$  &  $3.08005e+02$\\
 $3.7888199e-01$  &  $9.63532054e-01$  &  $6.21682e+00$  &  $-1.29129e+00$  &  $2.96121e+02$\\
 $4.7826087e-01$  &  $9.86728280e-01$  &  $7.73843e+00$  &  $-1.46834e+00$  &  $2.89537e+02$\\
 $5.5279503e-01$  &  $1.00006841e+00$  &  $8.72092e+00$  &  $-1.41336e+00$  &  $2.86018e+02$\\
 $6.2732919e-01$  &  $1.01027031e+00$  &  $9.54374e+00$  &  $-1.27672e+00$  &  $2.83483e+02$\\
 $7.5155280e-01$  &  $1.02150242e+00$  &  $1.05300e+01$  &  $-9.86511e-01$  &  $2.80891e+02$\\
 $8.7577640e-01$  &  $1.02735866e+00$  &  $1.10240e+01$  &  $-7.09914e-01$  &  $2.79616e+02$\\
 $9.3788820e-01$  &  $1.02882990e+00$  &  $1.10878e+01$  &  $-5.90787e-01$  &  $2.79261e+02$\\
 $1.0372671e+00$  &  $1.02965450e+00$  &  $1.09349e+01$  &  $-4.30510e-01$  &  $2.78891e+02$\\
 $1.1242236e+00$  &  $1.02909392e+00$  &  $1.05386e+01$  &  $-3.19135e-01$  &  $2.78640e+02$\\
 $1.2484472e+00$  &  $1.02654205e+00$  &  $9.53798e+00$  &  $-1.99918e-01$  &  $2.78221e+02$\\
 $1.3726708e+00$  &  $1.02231600e+00$  &  $8.05902e+00$  &  $-1.18435e-01$  &  $2.77604e+02$\\
 $1.4968944e+00$  &  $1.01709842e+00$  &  $6.24463e+00$  &  $-6.56574e-02$  &  $2.76766e+02$\\
 $1.6211180e+00$  &  $1.01186368e+00$  &  $4.35078e+00$  &  $-3.38440e-02$  &  $2.75818e+02$\\
 $1.7453416e+00$  &  $1.00756617e+00$  &  $2.63935e+00$  &  $-1.63774e-02$  &  $2.74951e+02$\\
 $1.8695652e+00$  &  $1.00481128e+00$  &  $1.23354e+00$  &  $-8.04907e-03$  &  $2.74348e+02$\\
 $1.9937888e+00$  &  $1.00382149e+00$  &  $5.64568e-02$  &  $-5.51691e-03$  &  $2.74121e+02$\\
\bottomrule
\end{tabular*}
\end{table}

\begin{table}[h]
	\centering
	\caption{\label{tab:vertical} Flow variables along the vertical central line at $t=1.5$ for wall bounded Rayleigh flow at $Kn=0.05$.}
	\begin{tabular*}{0.98\textwidth}{@{\extracolsep{\fill} }rrrrr}
		\toprule
		\multicolumn{1}{c}{$y$ (m)} & \multicolumn{1}{c}{$\rho / \rho_0$} & \multicolumn{1}{c}{$U$ (m/s)}
	& \multicolumn{1}{c}{$V$ (m/s)} & \multicolumn{1}{c}{$T$ (K)} \\
		\hline
  $6.1728395e-03$  &  $1.04684423e+00$  &  $2.84048e+00$  &  $-1.02562e-02$  &  $2.74860e+02$ \\
  $1.2962963e-01$  &  $1.03715456e+00$  &  $7.87730e+00$  &  $-5.77711e-01$  &  $2.77250e+02$ \\
  $2.1604938e-01$  &  $1.03337308e+00$  &  $9.47163e+00$  &  $-8.65068e-01$  &  $2.78103e+02$ \\
  $2.9012346e-01$  &  $1.03133167e+00$  &  $1.02116e+01$  &  $-9.31617e-01$  &  $2.78548e+02$ \\
  $3.7654321e-01$  &  $1.02999141e+00$  &  $1.06908e+01$  &  $-8.34530e-01$  &  $2.78849e+02$ \\
  $4.7530864e-01$  &  $1.02951243e+00$  &  $1.09836e+01$  &  $-5.68704e-01$  &  $2.79000e+02$ \\
  $5.1234568e-01$  &  $1.02958102e+00$  &  $1.10483e+01$  &  $-4.43631e-01$  &  $2.79014e+02$ \\
  $5.8641975e-01$  &  $1.03011130e+00$  &  $1.11016e+01$  &  $-1.78284e-01$  &  $2.78972e+02$ \\
  $6.6049383e-01$  &  $1.03120727e+00$  &  $1.10072e+01$  &  $ 7.24708e-02$  &  $2.78822e+02$ \\
  $7.4691358e-01$  &  $1.03335834e+00$  &  $1.05529e+01$  &  $ 2.80216e-01$  &  $2.78453e+02$ \\
  $8.0864198e-01$  &  $1.03563223e+00$  &  $9.82787e+00$  &  $ 3.27270e-01$  &  $2.78005e+02$ \\
  $8.7037037e-01$  &  $1.03869504e+00$  &  $8.55994e+00$  &  $ 2.65045e-01$  &  $2.77341e+02$ \\
  $9.9382716e-01$  &  $1.04848214e+00$  &  $3.11158e+00$  &  $-3.31606e-03$  &  $2.74895e+02$ \\
	\bottomrule
	\end{tabular*}
\end{table}

\begin{table}[h]
\centering
\caption{\label{tab:upper} Surface quantities along the upper wall at time $t=1.5$ for wall bounded Rayleigh flow at $Kn=0.05$.}
\begin{tabular*}{0.98\textwidth}{@{\extracolsep{\fill} }rrrr}
\toprule
\multicolumn{1}{c}{$x$} & \multicolumn{1}{c}{$p_w / \rho_0 C_0^2 $} & \multicolumn{1}{c}{$\tau_w / \rho_0 C_0^2$}
  & \multicolumn{1}{c}{$ q_w / \rho_0 C_0^3$}  \\
\hline
$6.2111801e-03$  &  $5.3565e-01$  &  $1.2424e-02$  &  $7.4008e-02$ \\
$1.3043478e-01$  &  $5.2399e-01$  &  $8.0181e-03$  &  $3.8776e-02$ \\
$1.9254658e-01$  &  $5.2309e-01$  &  $7.0158e-03$  &  $3.0238e-02$ \\
$2.7950310e-01$  &  $5.2275e-01$  &  $6.0035e-03$  &  $2.1583e-02$ \\
$3.2919255e-01$  &  $5.2284e-01$  &  $5.5672e-03$  &  $1.7828e-02$ \\
$3.7888199e-01$  &  $5.2306e-01$  &  $5.2149e-03$  &  $1.4741e-02$ \\
$5.0310559e-01$  &  $5.2408e-01$  &  $4.6489e-03$  &  $9.3038e-03$ \\
$5.7763975e-01$  &  $5.2488e-01$  &  $4.4971e-03$  &  $7.2490e-03$ \\
$6.3975155e-01$  &  $5.2562e-01$  &  $4.4631e-03$  &  $6.0582e-03$ \\
$6.8944099e-01$  &  $5.2620e-01$  &  $4.4889e-03$  &  $5.3808e-03$ \\
$7.5155280e-01$  &  $5.2691e-01$  &  $4.5769e-03$  &  $4.8102e-03$ \\
$8.7577640e-01$  &  $5.2797e-01$  &  $4.8808e-03$  &  $4.3264e-03$ \\
$9.2546584e-01$  &  $5.2817e-01$  &  $5.0231e-03$  &  $4.2926e-03$ \\
$9.5031056e-01$  &  $5.2820e-01$  &  $5.0930e-03$  &  $4.2963e-03$ \\
$1.0000000e+00$  &  $5.2811e-01$  &  $5.2223e-03$  &  $4.3291e-03$ \\
$1.0745342e+00$  &  $5.2754e-01$  &  $5.3644e-03$  &  $4.4000e-03$ \\
$1.1366460e+00$  &  $5.2659e-01$  &  $5.4071e-03$  &  $4.4358e-03$ \\
$1.1490683e+00$  &  $5.2635e-01$  &  $5.4052e-03$  &  $4.4371e-03$ \\
$1.2484472e+00$  &  $5.2379e-01$  &  $5.2400e-03$  &  $4.3441e-03$ \\
$1.3726708e+00$  &  $5.1929e-01$  &  $4.6358e-03$  &  $3.9138e-03$ \\
$1.4968944e+00$  &  $5.1413e-01$  &  $3.6737e-03$  &  $3.1740e-03$ \\
$1.6211180e+00$  &  $5.0932e-01$  &  $2.5718e-03$  &  $2.3075e-03$ \\
$1.7453416e+00$  &  $5.0564e-01$  &  $1.5527e-03$  &  $1.5361e-03$ \\
$1.8695652e+00$  &  $5.0342e-01$  &  $7.2076e-04$  &  $1.0193e-03$ \\
$1.9937888e+00$  &  $5.0265e-01$  &  $3.2882e-05$  &  $8.2962e-04$ \\
\bottomrule
\end{tabular*}
\end{table}

\begin{table}[h]
	\centering
	\caption{\label{tab:lower} Surface quantities along the lower wall at time $t=1.5$ for wall bounded Rayleigh flow at $Kn=0.05$.}
	\begin{tabular*}{0.98\textwidth}{@{\extracolsep{\fill} }rrrr}
		\toprule
		\multicolumn{1}{c}{$x$} & \multicolumn{1}{c}{$p_w/\rho_0 C_0^2 $} & \multicolumn{1}{c}{$\tau_w / \rho_0 C_0^2$}
		  & \multicolumn{1}{c}{$ q_w / \rho_0 C_0^3$}  \\
		\hline
$6.2111801e-03$  &  $4.9029e-01$  &  $3.2843e-03$  &  $6.2854e-02$ \\
$1.3043478e-01$  &  $5.0831e-01$  &  $2.4156e-03$  &  $3.5665e-02$ \\
$1.8012422e-01$  &  $5.1159e-01$  &  $2.2900e-03$  &  $2.9628e-02$ \\
$2.2981366e-01$  &  $5.1413e-01$  &  $2.2386e-03$  &  $2.4683e-02$ \\
$3.1677019e-01$  &  $5.1747e-01$  &  $2.2807e-03$  &  $1.7943e-02$ \\
$3.7888199e-01$  &  $5.1931e-01$  &  $2.3864e-03$  &  $1.4290e-02$ \\
$5.0310559e-01$  &  $5.2218e-01$  &  $2.7265e-03$  &  $9.1862e-03$ \\
$6.2732919e-01$  &  $5.2442e-01$  &  $3.1866e-03$  &  $6.2751e-03$ \\
$7.5155280e-01$  &  $5.2617e-01$  &  $3.7280e-03$  &  $4.8470e-03$ \\
$8.7577640e-01$  &  $5.2730e-01$  &  $4.3004e-03$  &  $4.3391e-03$ \\
$9.5031056e-01$  &  $5.2754e-01$  &  $4.6209e-03$  &  $4.2856e-03$ \\
$1.0000000e+00$  &  $5.2746e-01$  &  $4.8070e-03$  &  $4.3023e-03$ \\
$1.0745342e+00$  &  $5.2689e-01$  &  $5.0179e-03$  &  $4.3514e-03$ \\
$1.1366460e+00$  &  $5.2597e-01$  &  $5.1075e-03$  &  $4.3727e-03$ \\
$1.1614907e+00$  &  $5.2548e-01$  &  $5.1165e-03$  &  $4.3684e-03$ \\
$1.2484472e+00$  &  $5.2325e-01$  &  $5.0104e-03$  &  $4.2670e-03$ \\
$1.3726708e+00$  &  $5.1887e-01$  &  $4.4720e-03$  &  $3.8390e-03$ \\
$1.4968944e+00$  &  $5.1384e-01$  &  $3.5658e-03$  &  $3.1145e-03$ \\
$1.6211180e+00$  &  $5.0915e-01$  &  $2.5077e-03$  &  $2.2672e-03$ \\
$1.7453416e+00$  &  $5.0555e-01$  &  $1.5192e-03$  &  $1.5121e-03$ \\
$1.8695652e+00$  &  $5.0337e-01$  &  $7.0676e-04$  &  $1.0052e-03$ \\
$1.9937888e+00$  &  $5.0261e-01$  &  $3.2270e-05$  &  $8.1898e-04$ \\
		\bottomrule
	\end{tabular*}
\end{table}

\begin{table}[h]
	\centering
	\caption{\label{tab:side} Surface quantities along side plate at time $t=1.5$ for wall bounded Rayleigh flow at $Kn=0.05$.}
	\begin{tabular*}{0.9\textwidth}{@{\extracolsep{\fill} }rrrr}
		\toprule
\multicolumn{1}{c}{$x$} & \multicolumn{1}{c}{$p_w/\rho_0 C_0^2$} & \multicolumn{1}{c}{$\tau_w / \rho_0 C_0^2$}
& \multicolumn{1}{c}{$ q_w / \rho_0 C_0^3$} \\
		\hline
$6.1728395e-03$  &  $5.2104e-01$  &  $1.4074e-02$  &  $7.8678e-02$ \\
$6.7901234e-02$  &  $5.2494e-01$  &  $9.6498e-03$  &  $6.4572e-02$ \\
$1.2962963e-01$  &  $5.2408e-01$  &  $7.5477e-03$  &  $5.7702e-02$ \\
$2.5308642e-01$  &  $5.2213e-01$  &  $5.5299e-03$  &  $5.0555e-02$ \\
$3.7654321e-01$  &  $5.2128e-01$  &  $4.8126e-03$  &  $4.7414e-02$ \\
$4.1358025e-01$  &  $5.2123e-01$  &  $4.7293e-03$  &  $4.6940e-02$ \\
$4.8765432e-01$  &  $5.2142e-01$  &  $4.6716e-03$  &  $4.6508e-02$ \\
$5.0000000e-01$  &  $5.2148e-01$  &  $4.6726e-03$  &  $4.6499e-02$ \\
$5.7407407e-01$  &  $5.2208e-01$  &  $4.7215e-03$  &  $4.6822e-02$ \\
$6.2345679e-01$  &  $5.2271e-01$  &  $4.7810e-03$  &  $4.7420e-02$ \\
$7.4691358e-01$  &  $5.2534e-01$  &  $4.9456e-03$  &  $5.0643e-02$ \\
$8.5802469e-01$  &  $5.2982e-01$  &  $5.0124e-03$  &  $5.7082e-02$ \\
$9.4444444e-01$  &  $5.3636e-01$  &  $4.9719e-03$  &  $6.7810e-02$ \\
$9.9382716e-01$  &  $5.4420e-01$  &  $5.0975e-03$  &  $8.2648e-02$ \\
		\bottomrule
	\end{tabular*}
\end{table}

\clearpage
\newpage


\bibliographystyle{elsarticle-num}
\biboptions{numbers,sort&compress}
\bibliography{unsteadyUGKS}

\end{document}